# THÈSE

Pour obtenir le grade de

## DOCTEUR DE L'UNIVERSITÉ DE GRENOBLE

Spécialité : **Mathématiques, Sciences et Technologies de l'Information, Informatique**

Arrêté ministériel : 7 août 2006

Présentée par

# Frank Meyer

Thèse dirigée par **Eric Gaussier** et
codirigée par **Fabrice Clerot**

préparée au sein du **Laboratoire LIG**
dans **l'École Doctorale MSTII**

# Recommender systems in industrial contexts

Thèse soutenue publiquement le **25 janvier 2012**
devant le jury composé de :

**Mme Anne Boyer**
Professeur de l'Université Nancy II — Rapporteur
**M. Fabrice Clerot**
Responsable de l'Unité R&D PROF à Orange Labs — Co-encadrant
**Mme Nathalie Denos**
Maître de conférence de l'Université Grenoble II — Membre
**M. Marc El-Bèze**
Professeur de l'Université d'Avignon et Pays de Vaucluse — Rapporteur
**M. Eric Gaussier**
Professeur de l'Université Grenoble I — Directeur
**Mme Isabelle Tellier**
Professeur de l'Université Paris 3 — Membre

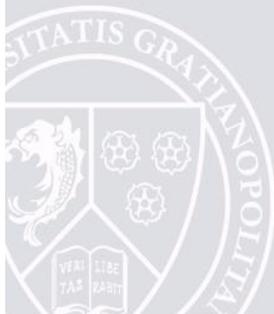

**Version 3.30.**
**May, 2012.**

*"Discovery is when something wonderful that you didn't know existed, or didn't know how to ask for, finds you"*

Fortune Magazine.

Cited by Greg Linden, creator of Amazon's recommender engine.

# Acknowledgments

First of all, I would like to thank all members of the Jury for having kindly agreed to evaluate my work. I am honored that Anne Boyer and Marc El-Bèze have agreed to report on my thesis.

I warmly thank my supervisor, Eric Gaussier, for the confidence he has shown me during the last four years. His regular supports, despite the distance between Grenoble and Lannion, his advice and his encouragement have been a key driver for my work.

I warmly thank my co-supervisor of PhD, Fabrice Clerot for his support, his numerous re-readings, his advice and for many constructive discussions which enabled me to achieve my overall analysis protocol that I present in this thesis.

I thank the laboratory ASAP Orange Labs Lannion for allowing me to undertake my thesis during my research project: for this reason I thank once again Fabrice Clerot, Head of R&D Unit PROF and Patrice Soyer, former Head of the laboratory ASAP, who made this project possible.

The final phase of drafting this document lasted several months during 2011. Without the facilities I received this year in the "Object of Research Media Search" this work would have been very difficult. As such, I thank Henri Sanson, Head of the "O.R. Media Search" and Sylvie Gruson, Head of Development Projects "Recommendation" for their kindness.

I also want to thank the team of the of R&D Unit PROF for their help, advice, or just for their expressions of sympathy. In alphabetical order, thank you to Anaïs, Aurélie, Barbara, Bruno, Carine, Christophe, Dhafer, Dominique, Françoise, Marc, Nicolas, Pascal G., Philippe, Raphaël, Romain G., Romain T., Sami, Sylvie, Tanguy and Vincent.

Several people participated in the development of the versions of Reperio or part of its benchmark during the last four years: Laurent Candillier (co-works on similarity measures, core Reperio version 1), Stéphane Geney (optimizations, DBMS, API, industrialization), Pascal Gouzien (platform tests, web services), Patrick Losquin (optimization and engine API, Web services), Julien Schluth (API engine and thematic part) and Damien Hembert (rewrote version of Reperio E). I thank them greatly for their technical contributions which were often essential.

Finally, many thanks to my English teacher, Steve Hill, of LCI, for the time we spent to read and improve my English in many passages, and for all the time spent talking on many subjects, including the automatic recommender systems. I am not yet expert in English, it is a large project, but no doubt that Steve is very familiar now with automatic recommender systems!

Dedicated to my family.

Page 8

# Executive Summary

This work presents the results of a research and development project dealing with automated recommendation systems in an industrial context, at Orange company (France Telecom Group).

Chapter 1 of this thesis provides a formal framework to the automatic recommendation, and provides standardization and formalization of core functions. We list the possible modes of recommendation, crossing the main objects or attributes manipulated by a recommendation system: items, metadata about items, users, socio-demographic data on users. We identify the various types of data sources used for a recommendation engine. We end up with four types of recommendations (item-to-items, item-to-users, user-to-users, user-to-items). Then we propose four core features of the automatic recommendation:

**Help to Decide**: predicting a rating for a user for an item
**Help to Compare** : rank a list of items in a personalized way for a user
**Help to Discover**: provide a user with unknown items that will be appreciated
**Help to Explore**: give items similar to a given target item

The first two features require a personalized scoring primitive function of an item for a user, the last feature requires a similarity function between objects. The "Help to Explore" feature requires the use of a primitive function of similarity between two objects. The "Help to Discover" feature can be implemented by one primitive or the other.

We propose a classification of Recommender Systems (R.S.) not based on the data source, but based on the primitive functions: a scoring engine, a similarity engine, and, for the last emerging category, a tags engine.

In Chapter 2, the state of the art, we identify the main techniques used by the automatic recommendation systems. Three features attracted our attention: the association rules, the K-Nearest-Neighbor (KNN) methods and the fast matrix factorization methods, especially a method called Gravity. We consider that the KNN methods correspond to a generalization of the association rules method and are more flexible. We therefore focus on the presentation of KNN and fast matrix factorization algorithms. We give an overview of different systems to show the diversity and richness of recommender engines' technology, while using as magnifying glass the first chapter to analyze it.

The second stage of this work was the creation of a recommendation system to meet industrial needs. Chapter 3 describes an automatic recommendation system designed and developed in Orange, the Reperio engine. This is an automatic recommendation engine using a KNN method as kernel. Its first version was developed in 2007 and was operational in the same year. Since then it has been used in operational services of France Telecom.

The recommendation engine Reperio allowed us to study the key factors that influence the performance of systems based on KNN methods.

In chapter 3 we show that the KNN methods are very competitive as long as you use good similarity measures to define neighborhoods.

We introduce an appropriate modification of the Pearson similarity and demonstrate its efficiency.

We show that the KNN methods are fully adapted to the functions and requirements defined in Chapter 1: KNN methods allow us to build industrial engines recommendation both generic and flexible.

We also show that hybridization techniques, and content-based recommendations are well adapted to cold start issue especially if user profiles are small. However, with profiles of sufficient size (over 20 ratings), a purely collaborative system becomes after a few hundred users more efficient. These results are of course related to a given size catalog (around 20,000 items).

The third stage of this work is the proposal of an evaluation process of automatic recommendation systems. This methodology is based on the combination of the four core functions tasks of a recommendation system:
- Help to Decide,
- Help to Compare,
- Help to Discover,
- Help to Explore,

taking into account the different behaviors of the following segments:
- heavy users and popular items,
- heavy users and unpopular items,
- light users and popular items,
- light users and unpopular items.

We experience this method by comparing four recommendation algorithms: two major methods of the state of the art: KNN and Gravity and two simple methods: a random predictor, and a default predictor based on the average users' and items' ratings.

We show that the two major methods are rather complementary as they perform differently across the different segments. We show that we could use the two algorithms together, for instance to produce a KNN matrix based on the dimension reduction done by Gravity, with still a correct quality. Gravity could be used to implement a "Help to Explore" function, and Gravity could be used as a component for fast KNN search.

We introduce a new measure, the Average Measure of Impact, to deal with the usefulness and the trust of the recommendations. Many algorithms and systems are evaluated only using the Root Mean Squared Error (RMSE) or Mean Absolute Error (MAE), and assuming that the lower the RMSE the higher the quality of the recommendation. We show that there is no clear correlation between RMSE measures and other quality measures of the recommendation. This fact strongly justifies our protocol, which independently evaluate the four core functions.

We have demonstrated the utility of our protocol as it may change
- the classical vision of the recommendation evaluation, often focused on the RMSE/MAE measures as they are assumed correlated with the system overall performances,
- and the way to design the recommender systems to achieve their tasks.

When designing a recommender engine 's general recommendation strategy, we have to think about the impact of the recommender: recommending popular items to heavy users might be not so useful. On the other hand, it can be illusory to make personalized recommendations of unpopular and unknown items to light and unknown users.

A possible simple strategy could be:
- Use personalized algorithms to recommend popular items to light users.
- Use personalized algorithms to recommend unpopular items of the long tails for heavy users.

*Note*: This thesis focuses on the automatic recommendation systems: for example, we excluded from this study editorial recommendation systems that can be managed by tools such as Content Management System (CMS) and the so-called recommendations to my friends, which can be a messaging system, or discussion forums coupled with a product catalog. We focus on automated systems, although the strict distinctions with other forms of more human recommendations is not perceived for people outside the community, particularly for marketing services.

# Présentation générale

Ce travail présente les résultats d'un projet de recherche et développement sur les systèmes de recommandation automatiques dans un contexte industriel, l'entreprise Orange (France Telecom Group).

Le Chapitre 1 de cette thèse fournit un cadre formel à la recommandation automatique, et propose une standardisation et une formalisation des fonctions cœur. Nous listons les modes de recommandation possibles en croisant les principaux objets ou attributs manipulés par un système de recommandation : items, métadonnées sur les items, utilisateurs, données socio-démographiques sur les utilisateurs. Nous inventorions les différents types de sources de données utilisables pour un moteur de recommandation. Nous aboutissons à 4 types de recommandation (item-to-items, item-to-users, user-to-users, user-to-items). Puis nous proposons 4 fonctionnalités cœur de la recommandation automatique :

**Aide à la Décision**: prédire une note d'un utilisateur pour un item
**Aide à la Comparaison** : classer une liste d'items de manière personnalisée pour un utilisateur
**Aide à la Découverte** : fournir à un utilisateur des items inconnus qu'il va apprécier
**Aide à l'Exploration** : donner des items similaires à un item cible donné

Les deux premières fonctionnalités demandent une primitive de scoring (méthode de prédiction de note) personnalisée d'un item pour un utilisateur. La fonctionnalité "Aide à l'Exploration" demande une fonction primitive de similarité entre deux objets. La fonction "Aide à la Découverte" peut être implémentée par l'une ou l'autre de ces deux primitives.

Nous proposons une classification des systèmes de recommandation au delà de la source des données, basée sur les fonctions source : un moteur de scoring, un moteur de similarité, ou, pour la dernière forme émergeante, un moteur de tags.

Dans le chapitre 2, l'état de l'art, nous inventorions rapidement les principales techniques utilisées par les systèmes de recommandation automatiques. Trois fonctions retiennent notre attention: les règles d'associations, les méthodes à K-plus-proches-voisins (KPPV) et les méthodes de factorisation rapides de matrices (principalement, l'algorithme Gravity). Nous considérons que les méthodes à K-plus-proches voisins correspondent à une généralisation des méthodes à règles d'association et qu'elles sont plus flexibles. Nous concentrons donc la présentation des algorithmes sur les méthodes KPPV et sur les méthodes de factorisation rapide de matrice.

La deuxième étape de ce travail a été la réalisation d'un système de recommandation répondant à des besoins industriels. Le chapitre 3 détaille un système de recommandation automatique conçu et développé à Orange : le moteur de recommandation Reperio. Il s'agit d'un moteur de recommandation automatique hybride générique utilisant une méthode de type K-plus-proches voisins. Sa première version a été développée début 2007 et a été opérationnelle courant 2007. Reperio est déjà utilisé dans des services opérationnels de France Télécom.

Le moteur de recommandation Reperio nous a permis d'étudier les facteurs clés qui influent sur les performances d'un systèmes basé sur les KPPV.

Dans le chapitre 3, nous montrons que les méthodes de type K-plus-proches voisins sont très compétitives à condition de bien spécifier les mesures de similarité utilisées pour définir les voisinages.

Nous introduisons une modification de la mesure de similarité de Pearson et nous montrons son efficacité.

Nous montrons que les technologies de KPPV sont totalement adaptées aux fonctions et pré-requis définis au chapitre 1 : les méthodes à KPPV permettent de réaliser des moteurs de recommandation industriels génériques et flexibles.

La 3ème étape de ce travail est la proposition d'un processus d'évaluation des systèmes de recommandation. Cette méthodologie est fondée sur le croisement des 4 tâches clés d'un système de recommandation :
- aide à la Décision,
- aide à la Comparaison,
- aide à la Découverte,
- aide à l'Exploration,

en prenant en compte le comportement de 4 segments clés des utilisateurs et items :
- gros utilisateurs et items populaires,
- gros utilisateurs et items peu fréquents,
- petits utilisateurs et items populaires,
- petits utilisateurs et items peu fréquents.

Nous expérimentons cette méthodologie en comparant quatre algorithmes de recommandation : deux méthodes principales de l'état de l'art : KPPV et Gravity, et deux méthodes simples: un prédicteur aléatoire, et un prédicteur par défaut basé sur la moyenne des notes des utilisateurs et des items.

Nous montrons que les deux principales méthodes, Gravity et KPPV sont complémentaires car elles se comportent différemment dans les différents segments que nous avions définis. Nous montrons que nous pouvons utiliser ces deux algorithmes ensemble, par exemple pour produire une matrice KNN de qualité correcte en nous basant sur la réduction de dimension produite par Gravity. Gravity pourrait être utilisée pour mettre en œuvre la fonction "Aide à l'Exploration", et Gravity pourrait être utilisée comme composant pour un processus de recherche rapide de KPPV.

Nous proposons une nouvelle mesure, la mesure moyenne de l'impact (AMI), pour gérer les notions d'utilité et de confiance des recommandations. De nombreux algorithmes et systèmes sont évalués en utilisant uniquement la Root Mean Squared Error (RMSE) ou la Mean Aboslute Error (MAE), et en supposant que minimiser la RMSE revient à augmenter la qualité de la recommandation. Nous montrons qu'il n'y a pas de lien strict entre la mesure et la RMSE autres mesures de qualité de la recommandation. Ce fait justifie fortement notre protocole qui évalue indépendamment les quatre fonctions coeur.

Nous démontrons l'utilité de notre protocole qui pourrait changer
- la vision classique de l'évaluation de la recommandation, souvent axée sur les mesures RMSE / MAE car elles sont supposées être corrélées avec les performances globales du système,
- et la façon de spécifier les systèmes de recommandation pour atteindre leurs tâches.

Lors de la conception de la stratégie générale de recommandation d'un moteur de recommandation automatique, nous devons penser à l'impact des recommandations : recommander des articles populaires pour les gros utilisateurs pourrait ne pas être très utile en terme d'augmentation d'usages. Mais d'un autre côté, il peut être illusoire de faire des recommandations personnalisées d'éléments impopulaires et inconnus à de petits utilisateurs sur lesquels on dispose de peu de connaissances.

Une stratégie simple pour augmenter les usages pourrait être :
• d'utiliser des algorithmes personnalisés pour recommander des articles populaires pour les petits utilisateurs,
• d'utiliser des algorithmes personnalisés pour recommander des items peu connus de la longue traine aux seuls gros utilisateurs "connaisseurs".

*Note* : Cette thèse est centrée sur les systèmes de recommandation automatique : nous avons par exemple exclu de ce travail les systèmes de recommandation éditoriaux, qui peuvent être gérés par des outils de Content Management System (CMS) et ce qu'on appelle la recommandation par les amis, qui correspond à un système de messagerie, ou à un ensemble de forums de discussion couplés à un catalogue de produits. Nous nous concentrons sur les systèmes automatiques, bien que la distinction stricte avec les autres formes de recommandations plus "humaines" n'aille pas de soi pour des personnes externes à la communauté, pour les services marketing notamment.

**Industrial contributions**

The Reperio engine is a flexible and generic hybrid recommender system designed during the work of this thesis. The Reperio engine was presented to the Executive Committee of France Telecom Group in June 2009. This prototype engine is currently used in the following services: recommendation of radio stations (service LiveRadio), recommendation of books and readers (service Lecteur.com). It was also used as a first prototype of the service VideoParty Orange. Reperio is entering a process of industrialization in France Telecom through its subsidiary IT&Labs, for use in future projects for operational services.

Two patent applications were made after a proof of concept achieved through Reperio.

**Contributions industrielles**

Le moteur de recommandation Reperio est un système de recommandation hybride, flexible et générique conçu durant ce travail de thèse. Le moteur Reperio fut présenté au Comité Exécutif du groupe France Telecom en juin 2009. Ce moteur prototype est actuellement utilisé dans les services suivants : recommandation de stations radio sur le service opérationnel LiveRadio, recommandation d'utilisateurs et de livres sur le service Lecteur.com. Il a aussi été utilisé en tant que 1er prototype sur le service VideoParty d'Orange. Reperio est en cours d'industrialisation à Orange France par sa filiale IT&Labs, pour utilisation dans des projets de services opérationnels ultérieurs.

Deux demandes de brevets ont été effectuées suite à une preuve de concept réalisée grâce à Reperio.

**Publications**

*French Conference*

- Frank Meyer, Éric Gaussier, Fabrice Clérot, Julien Schluth: Apport des données thématiques dans les systèmes de recommandation : hybridation et démarrage à froid. EGC 2011.

*International book chapter*

- Frank Meyer, Damien Poirier, Isabelle Tellier, Françoise Fessant: "REPERIO: a flexible architecture for recommendation in an industrial context", Intelligent Techniques in Recommendation Systems: Contextual Advancements and New Methods. Publisher IGI Global (2011).
- Laurent Candillier, Kris Jack, Françoise Fessant, Frank Meyer, "State-of-the-Art Recommender Systems". In Collaborative and Social Information Retrieval and Access: Techniques for Improved User Modeling, chapter 1. Publisher: IGI Global (2008).

*International Conference*

- Frank Meyer, Françoise Fessant: "Reperio: a generic and flexible recommender system", IEEE/WIC/ACM conference on web intelligence, 2011.
- Laurent Candillier, Frank Meyer, Françoise Fessant: "Designing Specific Weighted Similarity Measures to Improve Collaborative Filtering Systems". ICDM 2008.
- Laurent Candillier, Frank Meyer, Marc Boulle, "Comparing state-of-the-art collaborative filtering systems". 5th International Conference on Machine Learning and Data Mining (MLDM), Leipzig 2007.

**Patents**

Frank Meyer, Julien Schluth, Damien Hembert. "Device and a method for updating a user profile", US12/780,343, date of application: 2010.

Frank Meyer, Stéphane Geney. "Device and a method for predicting comments associated with a product", US12/782,396. date of application: 2010.

## Notations

Unless otherwise specified, we will use the following notations:

| | |
|---|---|
| **M** | number of users |
| **N** | number of items |
| $u, v \in \{1,..,N\}$ | indexes for the users |
| $i, j \in \{1,..,M\}$ | indexes for the items |
| **I** | set of the items |
| **U** | set of the users |
| $S_u \subseteq I$ | the set of items rated by **u** |
| $T_i \subseteq U$ | the set of users who have rated **i** |
| **\|S\|** | the cardinality of the set **S** (the number of elements of **S**) |
| $r_{u,i}$ | the rating of the user **u** for the item **i** |
| $\hat{r}_{u,i}$ | the predicted rating of the user **u** for the item **i**<br>the symbol "hat" denotes information predicted, following a model |
| **R** | the matrix of the logs of ratings $r_{u,i}$ |
| **Q**$^\text{T}$ | the transpose of a matrix **Q** |
| **x.y** | the scalar product of vectors **x** and **y** |

# Table of Contents

# List of figures

# List of tables

# Introduction

*"It is capital mistake to theorize before one has data"*
(Sherlock Holmes to Dr Watson in "A scandal in Bohemia".)

Sir Arthur Conan Doyle.

This thesis deals with automatic recommendation systems. Automatic recommendation systems are systems that allow, through for example data mining techniques, to automatically recommend to users, based on their past consumption, items that may interest them. These systems allow for example to increase sales on e-commerce websites: the Amazon site has a marketing strategy mainly based on the recommendation. Amazon has popularized the use of automatic recommendation based on the recommendation function that we call item-to-items, or "Help to Explore": the famous "people who have seen / bought this product have also seen / bought these articles".

The central contribution of this thesis is to analyze the recommendation systems in the industrial context, including marketing needs, and to cross this analysis with academic works. We will try to answer four main questions:

1. What are the main features of an automatic recommender system, in an industrial point of view?
2. What are the main techniques involved in automatic recommender systems?
3. What are the technical implications and the expected performances of recommender systems, if we choose a K-Nearest Neighbor technique as kernel?
4. What is the expected added value of recommender systems, from an industrial and marketing point of view?

This thesis consists of four parts:

**Chapter 1**
An analysis of the core functions and the prerequisites for recommender systems in an industrial context. We identify four core functions for recommendation systems: Help do Decide, Help to Compare, Help to Explore, Help to Discover. The implementation of these functions has implications for the choices at the heart of algorithmic recommender systems.

We show that a system only based on a rating prediction function is not sufficient to implement the "Help to Explore" function, while a similarity function between items can be the basis for the implementation of all the core functions.

**Chapter 2**
A state of the art, which deals with the main techniques used in automated recommendation systems. The two most commonly used algorithmic methods, the K-Nearest-Neighbor methods (KNN) and the fast factorization methods are detailed. This state of the art also presents purely content-based methods, hybridization techniques, and the classical performance metrics used to evaluate recommender systems. This state of the art gives also an overview of several systems, both from academia and industry (Amazon, Google ...).

**Chapter 3**

An analysis of the performances and implications of a recommendation system developed during this thesis. This system, called Reperio, is a hybrid recommender engine using KNN methods. We study the performance of the KNN methods, including the impact of similarity functions used. Then we study the performance of the KNN method in critical uses cases such as cold start situation.

We show that when implementing KNN methods, the similarity functions chosen to define the neighborhood in the KNN methods are often underestimated.

We analyze the performances of KNN methods with content-based data and using data hybridization techniques.

Concerning the cold start evaluation, we find that content-based methods and hybrid methods are interesting, when usage data are still too sparse to allow for proper profiling of users.

**Chapter 4**

A methodology for analyzing the performance of recommender systems in industrial context. This methodology aims to assess the added value of algorithmic strategies and recommendation systems according to its core functions. For this we take the four functions we have defined in Chapter 1 : Help do Decide, Help to Compare, Help to Explore, Help to Discover, and we cross with four key segments of the performance analysis of the recommendation systems : heavy users and popular items, heavy users and unpopular items, light users and popular items, light users and unpopular items.

We apply our methodology to analyze the performance of two state of the art methods for the implementation of a recommender system: KNN methods and fast matrix factorization methods.

We define a measure of the impact of the recommendation: we show that the positive impacts of the recommendations seem more important with KNN-type method than with fast matrix factorization-type method.

We show that the assumption deeply rooted in the community that optimizing the score prediction error (RMSE) is equivalent to optimize the quality of recommendation may be false: there is no clear correlation between RMSE and precision, or between RMSE and the impact of the recommendations.

We finally show that recommendation systems should redefine their challenges: it is unrealistic to recommend personalized unpopular items to users who are not known, for example. To increase usage, effective strategies are rather: recommend unpopular items only to heavy users, recommend to light users only the popular items. The long tail paradigm, both for infrequent items and infrequent users, should be revisited.

# 1 Industrial context of recommender systems

*"Making a thousand decisions, even the wise will make a mistake"*

Chinese proverb.

The purpose of this chapter is:

- to define the core functions of recommender systems in an industrial context,
- to define the prerequisites of recommender systems,
- to make a first technical check of the relevance of a technique based on item-item similarity matrices to build a recommender system.

# 1.1 Introduction

This chapter discusses the key points of automatic recommender systems, from an industrial point of view.

In a first step, we will list the different classifications of automatic recommendation systems. We show that the usual types can be specified and extended, particularly through a functional typology. We identify four core functions of recommendation systems.

In a second step, we will list the features required for a recommendation system. Then we shall discuss the technical implications, in industrial settings, of these requirements.

This chapter will allow us to focus on the state of the art of some specific techniques in relation to the industrial context. Then in Chapter 4, it will allow us to define a new evaluation protocol for industrial recommender engines.

# 1.2 A fast emerging interdisciplinary area

A standard catalog of an online video on demand service such as Netflix can have more than 100,000 titles. The e-commerce site Amazon now contains several million product references. URLs of web sites today are counted in billions or tens of billions. We are entering an era of huge catalogs and databases where one person cannot consider himself/herself to have an overview of what is available and what might be of interest to him/her.

Two types of systems were developed in parallel with the rapid inflation of available contents: search engines and automatic recommendation systems (Table 1-1). Search engines are useful for people who know what they want and who will perform a search query. Automatic recommendation systems are often used as a support system for discovery and navigation or as a support system for decision making. Aid to discovery issues personal recommendations. Navigation aid provides a contextual help such as similar products to the product being viewed. The support for decision making predicts, for a user, the potential value a product may have for him/her, for instance a personalized rating for this product.

|  | Search System | Recommendation System |
|---|---|---|
| The user knows what she wants | Yes | No |
| The system needs a request from a user | Yes<br>It's mandatory<br>The request is explicit | No<br>It's optional,<br>or the request is implicit |
| The main technique is: | Content-based | Collaborative-based |
| The system analyzes the usage of the users | Optional | Mandatory |
| The system might make rating predictions for the returned items | No | Yes |
| The technique for Related Search is | Content-based only | Collaborative-based and/or Content-based |
| The system pushes items to users | No | Yes<br>Generally it is the main task |

**Table 1-1: Brief comparison between Search and Recommendation**

For about 20 years we have witnessed a rise of interest for automatic recommendation systems (Goldberg et al., 1992) (Resnick et al., 1994) (Herlocker et al., 2000) (Adomavicius and Tuzhilin, 2005) (Rao and Talwar, 2008) (Su and Khoshgoftaar, 2009). The general goal of automatic recommender systems is to help users to find products (items) that should interest them, from large catalogs. Items are defined as any object that can be consumed, bought, read, viewed... Automatic recommendation systems exist for items as diverse as Web pages, movies, TV programs, books, restaurants, humorous jokes, songs, people within a social network, etc. (Rao and Talwar, 2008) identified 96 recommendation systems on various subjects, academic or industrial.

## 1.3 Main typologies of recommender systems

Several recommender system typologies are known in the literature (see Figure 1-2):
- the classical typology, with collaborative filtering, content-based filtering and hybrid filtering, for instance used in (Adomavicius and Tuzhilin, 2005),
- the typology of (Su and Khoshgoftaar, 2009), restricted to collaborative filtering,
- the typology of (Rao and Talwar, 2008), depending on the data sources used by the systems.

### 1.3.1 The classical typology

This typology identified 3 types of recommender systems:
1. the collaborative filtering systems,
2. the content-based filtering systems, also known as thematic filtering,
3. the hybrid systems, using both collaborative and content-based techniques.

This typology is based on the type of information used as inputs for the models. Collaborative filtering (CF) uses logs of users, generally user ratings on items, with dates. Content-based filtering (CBF) uses item metadata (i.e. intrinsic characteristics of the items, described by text, structural data - or information extracted from the items in some cases: acoustic features from songs for instance). In CF logs of users are compared in some way in order to find usage correlations. In CBF metadata are compared to explicit or implicit user preferences. The difference between CF and CBF is thus easy: CF is called "content agnostic": this means that a CF does not need data describing the items which will be recommended: an "item ID" is enough. The hybrid filtering on the other hand is a combination of CF and CBF.

### 1.3.2 Su and Khoshgoftaar

For purely collaborative systems (Su and Khoshgoftaar, 2009) proposes a sub-classification: it includes the hybrid techniques (which are necessarily collaborative) to classify them in hybrid collaborative methods. It classifies CF in 3 categories:

1. Memory-based CF approaches for type K-nearest-neighbors,
2. Model-based CF approaches encompassing a variety of techniques such as: clustering, Bayesian networks, matrix factorization, Markov decision processes,
3. Hybrid CF, which combines a CF technique recommendation with one or several other methods, which can be CBF approaches of other CF approaches. In this second case "hybrid" does not mean a combination of CF and CBF: it is only a hybridization of several CF methods.

### 1.3.3 Rao and Talwa

(Rao and Talwar, 2008) propose a classification depending on the source of information used to recommend items:

- CBF, also known as thematic filtering. These systems use item correlations to generate recommendations, and/or matching between item features and a user profile to calculate a score for each item,
- CF: recommender systems only based on logs of user usages on items,
- Demographic filtering: recommender systems using a a priori knowledge on groups of users. This knowledge is used to build stereotypes (with information such as age, location,..) linked with particular lists of items to recommend,
- Hybrid filtering makes a use of CF and CBF.

Without further details, (Rao and Talwar 2008) also cite two other systems, one based on a priori knowledge of relationships between items and users, said "knowledge-based recommender system", and another one based on a notion of utility calculated for a user and an item, called "utility-based recommender systems".

### 1.3.4 Data sources

An automatic recommendation system handles 4 basic types of objects (see Figure 1-1) that are:
1. Users,
2. Items,
3. Users' features: socio-demographic data: age, sex, location,
4. Items characteristics called metadata, or descriptors. In the folksonomy domain, user-generated metadata are called tags.

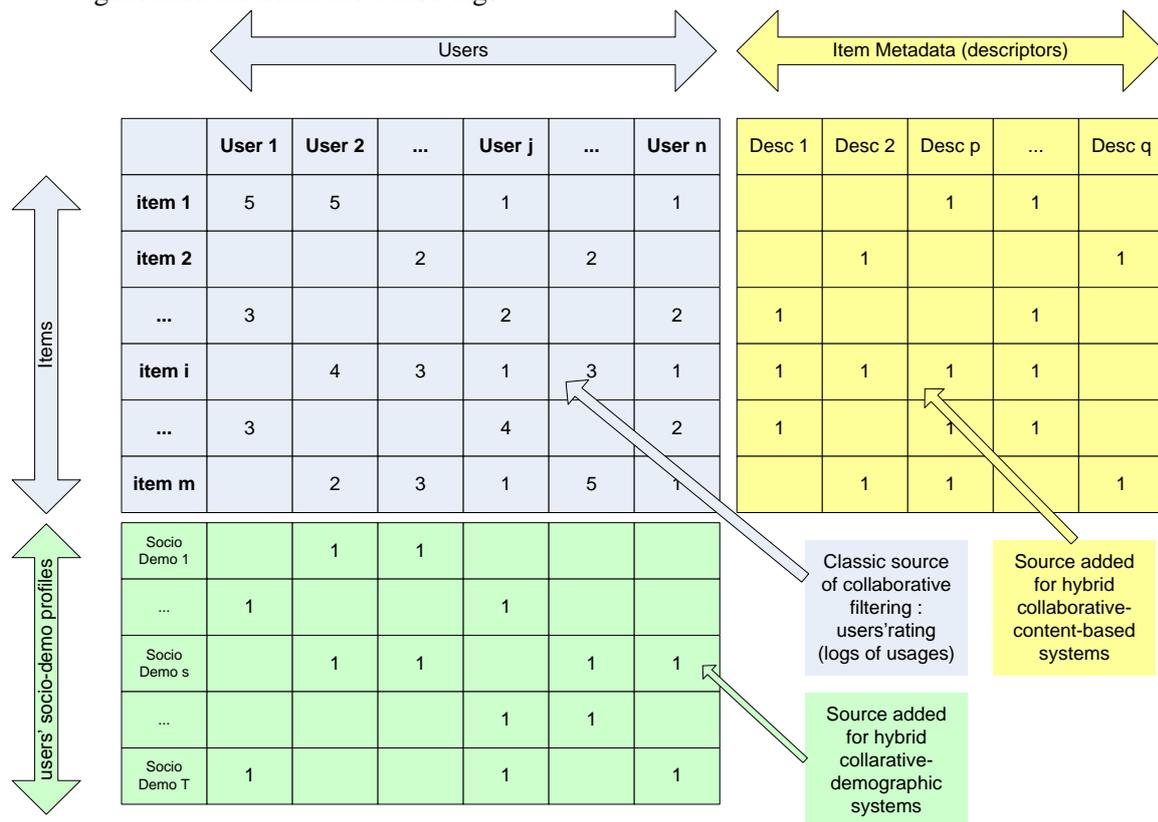

**Figure 1-1 : Possible input data of a recommender system or a hybrid filtering system**

Each cross-reference between two types of objects can have a value which can be a Boolean, a number, a tag, a text... Each cross-reference between an object type and another one gives a data table. For example, the crossing between users and items gives a matrix of uses (shopping, browsing or notes) on the items. Similarly, a cross-reference between items and metadata provides an array of catalog items. Since there are **n=4** object types, inventory of all possible cross-references gives **n×(n +1) / 2 = 10** cross-references, listed in the Figure 1-2. For each data source, the object "item" is described either by rows or by columns of the matrix.

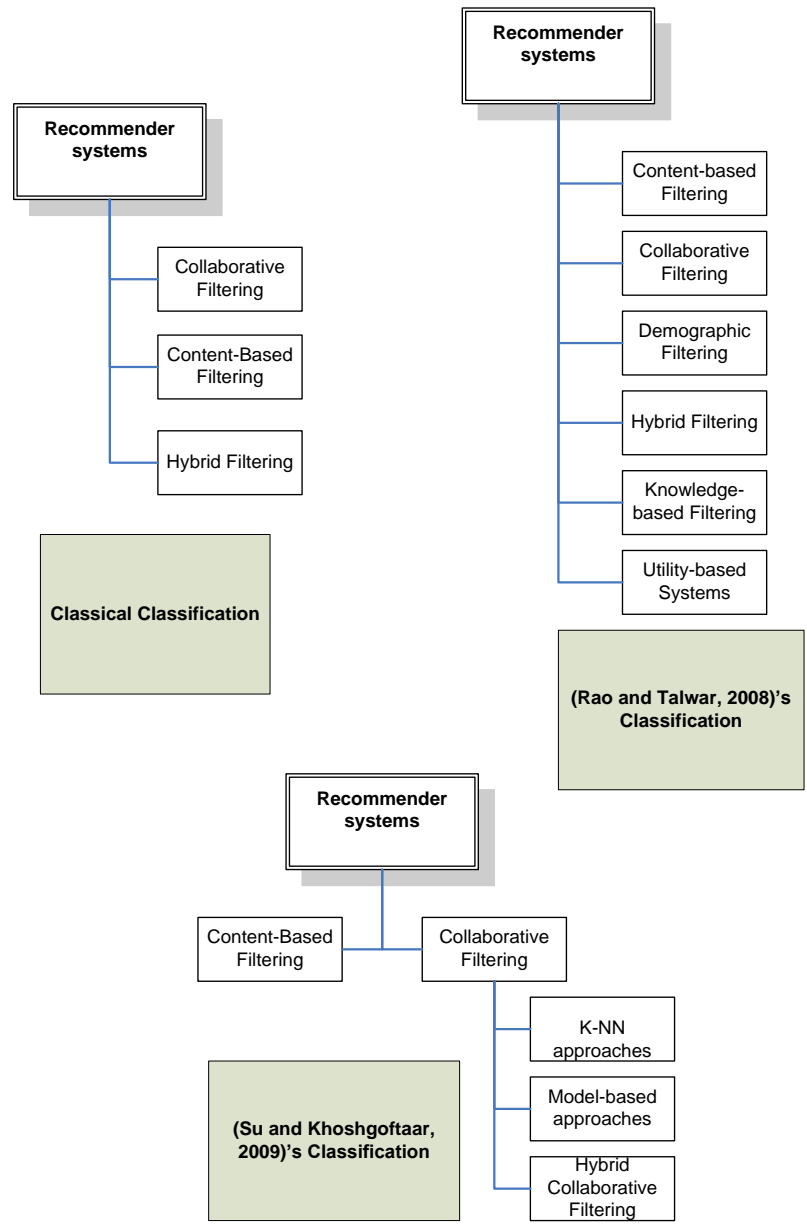

**Figure 1-2 Main classifications of recommender systems**

For each of 10 tables of data, we can design functions such as:
- Prediction (scoring) of some unknown values, such as rating of interest for an item, or for a metadata (characteristic of an item),
- Personalized recommendations of items, or users, or characteristics (metadata): the system can recommend books, or give advice for a community of readers, or recommend themes or genres of books, etc,
- Cross-Recommendation: "People who like this actor also like that genre or that director", for example,
- Recommendation of socio-demographic characteristics (location, employment) to link people according to their socio-style.

|    | Data source: input matrices | Examples in industrial contexts |
|----|------------------------------|---------------------------------|
| 1  | items × users | Logs (ratings, purchases, browsing…), with scalar or Boolean or Tag data. Discussion forums of users on items, with text data. |
| 2  | users × user's socio-demographic data | Client files. Geo-marketing files. |
| 3  | users × items' metadata | Declared users' preferences on item characteristics or tags. |
| 4  | users × users | Declared friends in a social network. Edges between users in a graph of messages or calls. |
| 5  | items × items | Links between HTML pages. Items linked by a "related search" device. |
| 6  | items × metadata | Catalog of items. Documents x keywords, for text applications. Tagged items. |
| 7  | items × user's socio-demographic data | Specific marketing segmentation for a campaign for instance. |
| 8  | items metadata × user's socio-demographic data | Generic marketing segmentation for a campaign, for instance. |
| 9  | items metadata × items metadata | Thesaurus giving links between genre, types, keywords. |
| 10 | user's socio-demographic data × user's socio-demographic data | A priori segmentation for a meeting website, for instance. |

**Table 1-2: Inventory of possible matrix data sources for a recommendation system**

Table 1-2 gives the list of the possible simplest data sources. Those data sources should also be, in fact recommenders' productions, as we'll see below. Note that binary data sources are more usual than real-valued data sources: descriptors on items, user's socio-demographics data are often coded with discrete values then binary coded. Links, Tags, marketing segmentations, social networks,... are also generally simply binary coded. And in the industry, logs of navigation and logs of purchases are most frequent than logs of ratings.

### 1.3.5 Recommendation target types

A main part of recommender systems focuses only on items and users. Meta data suggestion is generally done via social tagging systems whereas any catalog-based system can support this function. Socio-demographic suggestion is, may be culturally, not really used, and restricted to very specific areas like meeting websites. Although the types of recommendations can be almost symmetric with respect to the kind of data-sources listed in the previous table, some features are more common than others. The 4 cross-references between users and items described in the Table 1-3 are the most seen in the industry. Possible other types of recommendations may be for instance user-to-metadata (i.e. you should like movies with the characteristic "Director=Stanley Kubrick"), or item-to-metadata ("if you like this item **i**, you should like items with this characteristic c"), already implemented in search engines as non-personalized "related search".

| Type of recommendation | Key example | Usage |
|---|---|---|
| item-to-items | "People who have bought/viewed/liked this item have also bought/viewed/liked these items." | It is the famous recommendation mode, popularized by Amazon[TM]. |
| user-to-items | "Here is a list of items according to your profile ..." | This recommendation assumes that it knows the profile of the user. Used by Amazon for instance, for its mail-based recommendations on a thematic profile. |
| user-to-users | "Here is a list of people with similar tastes to you (profile)" | Emerging usages on social networks services. |
| item-to-users | "Here is a list of targeted people who may be interested by this item." | For each item: find the users who will be targeted to sell this item. Used in marketing campaigns. |

**Table 1-3: Main recommendation types in the industry**

## 1.4 Main functionalities of recommender systems

Another approach to the typology of recommendation systems could rely on the functionalities the system offers. Although the literature focuses heavily on the only feature of score prediction (the rating values generally), there are actually several classic functionalities that we formalize in this section.

We propose 4 essential core functions that are:
1. rating prediction,
2. ranking prediction, ordering of items by relevance criterion,
3. contextual, non-personalized recommendation for an anonymous website user,
4. recommendation of personalized items.

These features are essential because they cover the four essential needs of users facing huge catalogs of items: Decide, Compare, Explore, Discover. This is summarized in the following table:

| Essential function | Description | Core recommender function |
|---|---|---|
| **Help to Decide** | Given 1 item **i**, user **u** wants to know if he will appreciate **i** | Rating prediction |
| **Help to Compare** | Given **n** items, user **u** wants to know what item to choose | Ranking prediction |
| **Help to Explore** | Given **1** item **i**, user **u** wants to know what are the related k items | Item-to-items recommendation |
| **Help to Discover** | Given a huge catalog of items, user u wants to find k new interesting items | Personalized recommendation |

**Table 1-4: The four essential functions of the automatic recommendation for users**

Two functions are emerging with social networks:
- anonymous tag (or property) labeling for each item,
- personalized tag (or property) labeling for each item.

These functions help users in different ways using tags. The Tag-to-Tag recommendation may help users to navigate, not directly on items, but on item's tags. Giving for each item some tags likely to represent well the items' characteristics may help users to Decide or to Compare. Using tags to refine personalized recommendation may help users to Discover new contents. We believe that tags can be managed as items, or as items' metadata. Some other tag-related functions exist such as tag-suggestion when tagging items, but they are more used to organize the terms of the folksonomies and are out of the scope of this thesis.

There are also several requirements for recommender systems:
- compliance with many data sources as industrial systems have to be deployed on different services with different kinds of data,
- robustness to noisy or possibly corrupted data, as some data sources may be more or less reliable,
- management of the cold start at the time of the recommender's launch: the cold-start is the problem for the system to start prediction on users or items without enough data: the predictions may be poor, if not impossible,
- scalability as operational systems have to deal with volumetric,
- reactivity of the system, as response time and on-line learning time issues are also crucial,
- trusted relationship, based on transparency and explainability: the transparency of a recommendation system is an important factor for the acceptance of recommendations,
- the management of the "long tail", that is to say the management of all the items that are not often bought/seen but that we want to promote.

Although the academic literature focuses on the score prediction, industry prioritization of the functionalities is very different: the generated recommendations themselves, and their utility, are more important. Automatic recommender systems are often used on e-commerce websites. These systems work in conjunction with a search engine for assistance in catalog browsing to help users find contents. Many users of e-commerce websites are anonymous, and therefore the main feature is the contextual recommendation of item, for anonymous users. The purpose of these systems is also increasing uses (the audience of a site) or increasing sales, so the recommendation itself is more important than the score prediction. Moreover, prioritizing a list of items on a display page is a more important functionality than the prediction of a rating.

## 1.4.1 Rating prediction

For an item **i**, and a user profile $S_u$ represented by a set of items rated by user **u**, we can provide a predictive rating the user u would give to the item **i**. This is the rating prediction functionality of the recommendation. This is the most studied functionality since the challenge Netflix (Netflix Prize, 2007). Numerous publications exclusively deal with this issue (Su and Khoshgoftaar, 2009) However, we cannot reduce the function of the recommendation to the rating prediction as we shall see later.

**Definition 1.1: "The rating prediction problem"**

Given a set of users **U**, a set of items **I**, a rating matrix **R** on a subset of $U \times I$, and a cost function **Q**, find a predictive model as a function $\hat{r}: U \times I \to [a, b]$ such that $\sum_{r_{u,i} \in R} Q(\hat{r}_{u,i} - r_{u,i})$ is minimized (**a** and **b** being respectively the min and the max values of the ratings).

Conventionally we use the Mean Absolute Error (MAE) and Root Mean Squared Error (RMSE) as cost functions (see state-of-the-art chapter).

## 1.4.2 Ranking prediction

The items for which the predicted ratings are highest are not necessarily the most useful items for a user. Indeed, we must distinguish the fact that a user likes an item and the fact that an item is already known by the user (Herlocker et al., 2004). To recommend a movie the recommendation system must find an item that would probably be well rated by the user but also an item that the user probably does not know.

The ranking can incorporate a notion of item utility and is not necessarily derived from the rating prediction score. It is a function of its own. For a profile $S_u$ and a list $L$ of items, a recommendation system can provide a relevance score which can sort the items in the list $L$. The ranking is an important feature allowing implementing filtering services coupled with a dedicated search engine.

**Definition 1.2 "The ranking prediction problem"**

Given a user-item matrix **R**, **U** a set of users, **I** a set of items, and **C** a set of constraints represented by 3-uples **(i,j,u)** such that $i, j \in I$ and $u \in U$ and $r_{u,i} < r_{u,j}$, find a ranking function $h$ maximizing

$$\sum_{(i,j,u) \in C} [\![h(i) < h(j)]\!]$$

with $[\![\ ]\!]$ equals 1 if and only if the bracket expression is true.

Note that this ranking function gives higher values to high rated items: the rank is in decreasing order, the best items with the highest indexes.

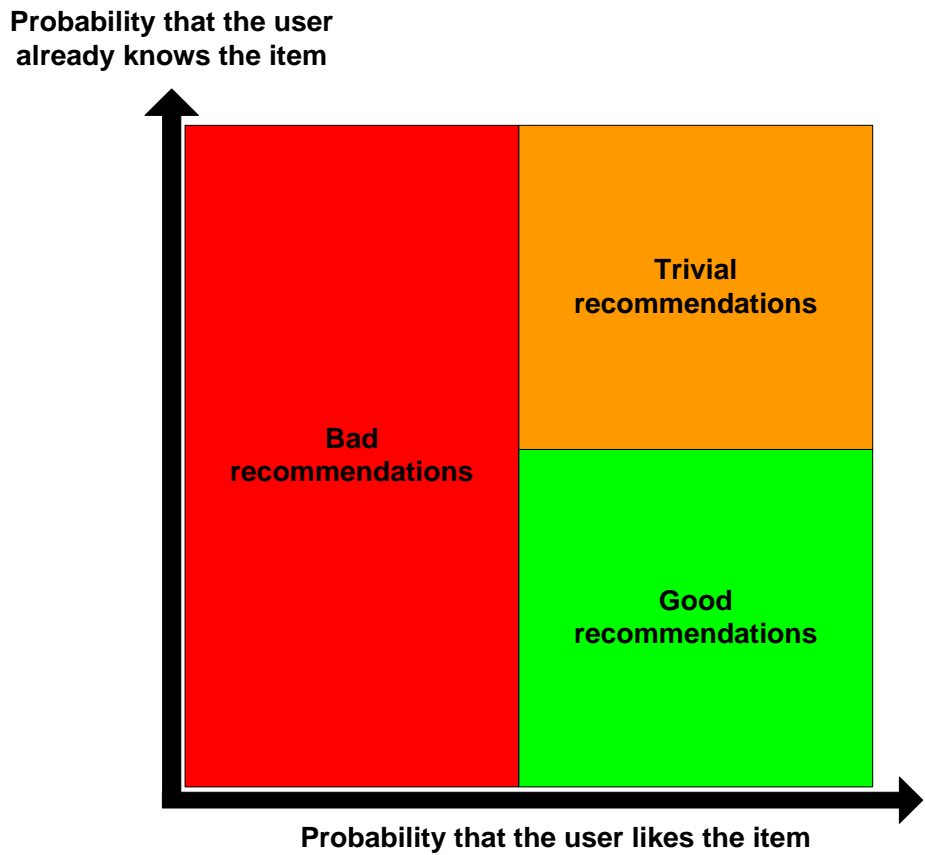

**Figure 1-3: A good recommender system should avoid bad and trivial recommendations**

### 1.4.3 Item-to-item recommendation to anonymous users

This function is sometimes called "contextual item-to-item" or "contextual recommendation". We restrict ourselves here to item-based context-dependent impersonalized recommendation. It should be distinguished from the next function, the personalized recommendation

For an item **i**, we can give a list of similar items **L(i)**: it is the classic item-to-item-recommendation, popularized by the e-commerce website Amazon (www.amazon.com). This feature has an immediate implementation if we have an item-item similarity matrix **M(i, j)**, which gives each item the list of similar items and their associated similarity weight. This mode is widely used in industry because it can make recommendations for anonymous users, based on the item he/she consults. This browsing aid in a catalog is simple and very effective. This functionality is so important that by itself it legitimizes the use of techniques based on item-item matrix (Linden et al., 2003) (Koren, 2010).

Tests exist to assess the rating prediction function and even the personalized recommendation function. Impersonalized recommendation is more difficult to assess: the rate of transformation (that is to say, the rated of selection / purchase of the recommended items) cannot be measured except in a real service.

**Definition 1.3: "The Item-to-Item recommendation problem"**

Given a user-item matrix **R** on a set of identified users **U** and a set of items **I**, a number of round **n**[1], a set of anonymous users **A** one of them managed during one round **r**, a number of items k to be recommended at each round, at each round r a context item $i_r$ being viewed by an anonymous user $u_r \in A$, find **n** ordered sets of items $X_i=\{x_1,…,x_k\}$ such that $i \notin X_i$ maximizing a utility function **f**. The utility function must use the feedback of the anonymous user **u** on each item **x**.

Find **n** sets $X_i=\{x_{i,1},…,x_{i,k}\}$ and a ranking function **h(x)** such that

$$\sum_{r=1}^{n} \sum_{x \in X_{i_r}} f_{u_r}(x) \times h(x) \text{ is maximized}$$

where $f_u(x)$ is the feedback of user **u** to item **x**, and **h(x)** the order index of **x**.

Note that the ranking function h(x) is in decreasing order, the best items with the highest indexes. Many types of $f_u(x)$ are possible, for instance:

| | |
|---|---|
| click rate | $f_u(x)= 0$ if the item is not selected by **u** |
| | $f_u(x)=1$ if the item is selected |
| purchase | $f_u(x) =0$ if item **x** is not purchased |
| | $f_u(x) = 1$ if item **x** is purchased |
| cash flow | $f_u(x) =0$ if item **x** is not purchased |
| | $f_u(x)=$ cash flow of **x**, if item **x** is purchased |

**Table 1-5: Examples of utility functions**

Note that personalized item-to-item recommendation can also be considered in two ways:
1. using a context item, then generating a list of similar items, then post-filtering the similar items for instance using a personalized ranking system.
2. using a user profile as a list of context items, to generate directly personalized recommendation.

### 1.4.4 Personalized recommendation of items

This function is usually called "Item-based Top-N recommendation" (Karypis, 2001) (Deshpande and Karypis, 2004) in the literature. As the ranking must be well differentiated from rating prediction, the personalized recommendation of items must be seen as a specific function which is not necessarily deduced from the others. Here one wants to provide a user **u** with a shortlist **L(u)** of **p** items from a catalog of items **C**.

---
[1] the purpose of the n rounds (n high enough) is to give a statistically reliable measure of the performance

A naive idea for personal recommendation would be to predict the item ratings, or even better to give a utility score to all items in the catalog **C** and then take the top **p**. However in practice this solution may be infeasible, if the catalog is very large. Rather, we use pre-computed approaches. For a user profile **u** represented by a set of rated items, and noted $S_u$, one can provide a list of items similar to items in the profile **u**, denoted $L(S_u)$: For this function we can use the principle of the Item-to-item recommendation applied to the entire profile, plus a post-filtering step: at least a post-filtering process systematically deletes the items already known by **u** (that is to say in the profile of the user **u**).

**Definition 1.4: "The Top-N personalized recommendation problem"**

Given a set of users **U**, a set of items **I**, and a set of items $S_u \subset I$ (purchased or rated) for each user $u \in U$, find an ordered set $X_u \subset I$ such that $|X_u| \leq N$ and $X_u \cap S_u = \emptyset$ maximizing a utility function $f_u$. The utility function must use the feedback of **u** on each item **x**.

Find $|U|$ sets $X_u = \{X_{u_1}, \ldots, X_{u_N}\}$ and the function $h(x)$ such that

$$\sum_{u \in U} \sum_{x \in X_u} f_u(x) \times h(x) \text{ is maximized}$$

where $f_u(x)$ is the feedback of user **u** to item **x**, and $h(x)$ the order index of **x**.
Here also many utility functions can be considered.

## 1.4.5 Emerging new classification

With the development of social network services on the Internet, new services like collective classification by tags are emerging as a new way to recommend items, users or tag-based information.
In fact, two main functions can be identified:
1. Anonymous tag (or property) labelling for items,
2. Personalized tag (or property) labelling for items.

**Anonymous tag (or property) labeling**

This issue, generally called "folksonomy", mainly focuses on finding shared vocabularies, reinforcing meaningful tags, leading to a static shared representation: each item is finally viewed with the same tags, for any user. This is for example addressed in (Heymann, 2008). The specific problem of anonymous tag labelling can be formulated as follows: given a set of items **I**, a set of users **U**, a set of tags **T** and a set $M \subset I \times U \times T$ of the user-generated tags on the items, derive a global and static item representation of each item, for any user, $M' \subset I \times T$.

**Personalized tag (or property) labeling**

Following the collaborative and personalized framework to give personalized predicted ratings for users given an item and their profile, some systems give personalized tags as suggestions (when tagging an item, for instance to avoid synonymy or to ensure completeness) or as information (when requesting specific items). These systems are also used for personalized ranking for user requests. The FolkRank algorithm (Jaschke et al., 2008), a PageRank-like algorithm (Brin et al., 1998), is an example of such a system. The specific problem of personalized tag labelling can be formulated as follows: Given a set of items **I**, a set of users **U**, a set of tags **T** and a set $M \subset I \times U \times T$ of user-generated tags on the items, derive a set of views $V_{u,i}$ giving for any couple $(u, i) \in U \times I$ a personalized view of the probabilities of each tag **t** to be perceived by **u** for **i**: $V_{u,i} \subset T \times [0; 1]$.

Other tag-based functions for tag-based recommenders exist but are out of the scope of this thesis. The Figure 1-4 gives a possible new classification of recommender engines, using the aforementioned list of functions. In this classification, 3 main techniques, score oriented, similarity oriented and tag oriented lead to different functionalities. The score-oriented methods use a predicted score function to score items in a personalized or not personalized way. They are well adapted for rating prediction and ranking prediction tasks. Note that carrying out the ranking function, they are able to generate recommendations, but they are not adapted to item-to-item contextual recommendation for instance. The similarity-oriented methods use the similarity between 2 items to generate directly recommendations. The recommendations can be contextual and anonymous if one of the item is a contextual item such as the item currently watched during catalog navigation. Using the similarity links principle applied to a user's profile, the similarity-oriented methods can generate recommendations, pushing items similar to those already enjoyed by the user. The tag-oriented methods can generate tags for anonymous users, or in a personalized way. And then this tags can be used to query a search engine for instance, to generate recommendations, which can also be anonymous or personalized.

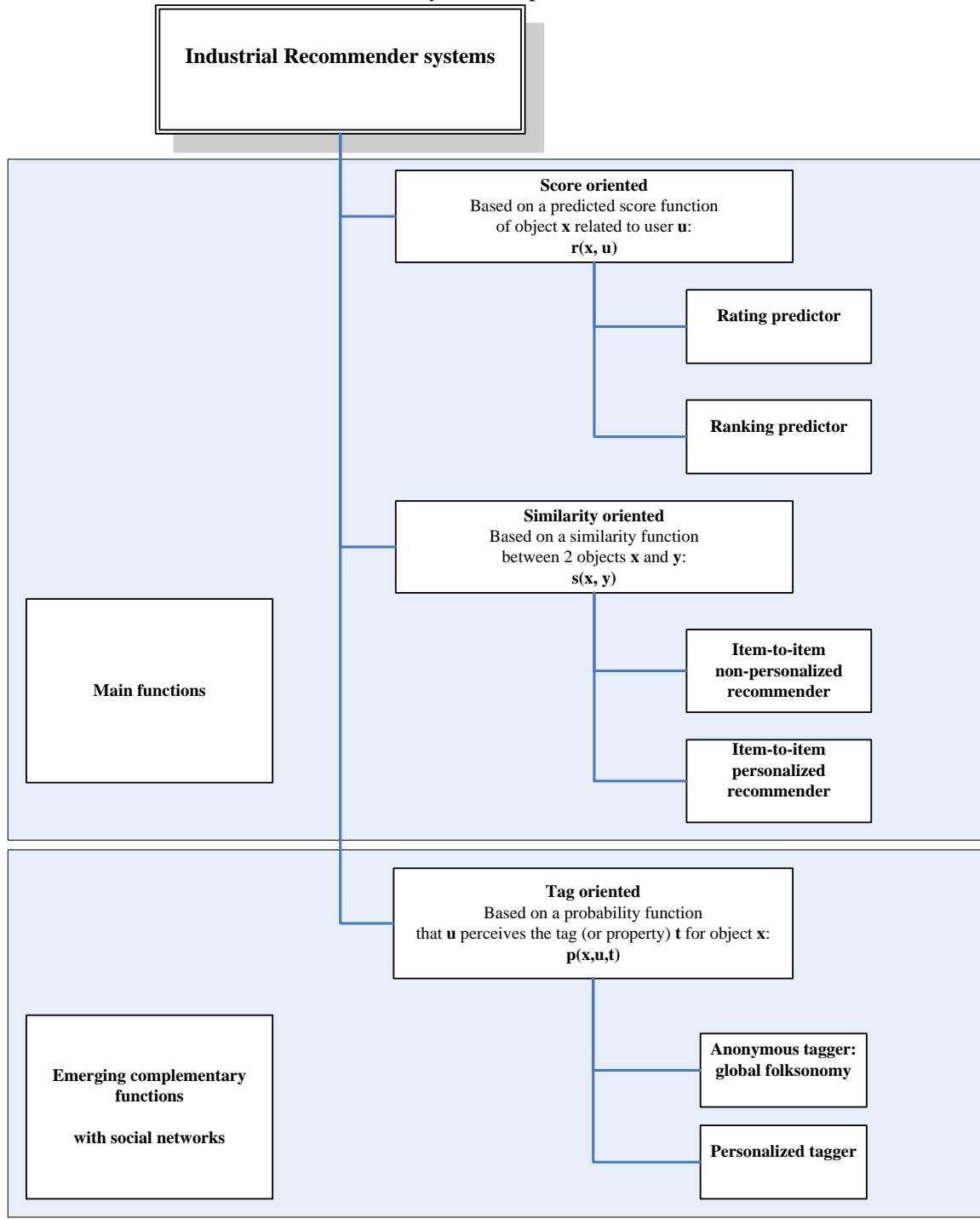

**Figure 1-4: A functional classification of industrial recommender systems**

## 1.5 Requirements for recommender systems

A recommendation system must provide the main functionalities aforementioned but will also have other mandatory operational functions/requirements.

### 1.5.1 Ability to use multiple data sources

This requirement has a main reason: an industrial system may be used in different contexts. Especially in a big company, a recommender service may be launched with logs of purchases in one case and with logs of ratings in another case, with or without access to rich items metadata, etc. Moreover the context can change over time: users can just select item, then with a new service component users may be able to rate items. As mentioned by (Burke, 2007), this ability is also useful to address the cold-start problem (see 1.5.2). To launch a new service, a recommender system should use external data, for example:
- from an internet online database, to build a good representation of the items for content-based recommendation,
- from another service with the same items but other users, to build a collaborative model, item-based (Poirier et al. 2010),
- from another service with the same users but with other items, to build a collaborative model, user-based.

Ideally the system should natively support the use of multiple data sources: this will avoid manually switching from one data source to another.

### 1.5.2 Management of the cold start and of new events

At the opening of a service, the system must be able to make useful recommendations, even if the usage data are not yet available. This well-known issue is called the cold-start problem (Adomavicius and Tuzhilin, 2005) (Su and Khoshgoftaar, 2009). To deal with the cold-start problem, the recurrent choices are the use of metadata with a content-based recommender or the use of socio-demographic data to infer profiles (Nguyen et al., 2007). Another interesting choice is the use of external data, for instance other known usages on the items to recommend (Poirier et al., 2010). The use of external data will be necessary until the service has enough user data (profiles, explicit and / or default profile).

The problem of the cold-start is very close to the problem of new events: a new event is a new user or a new item (Adomavicius and Tuzhilin, 2005) (Su and Khoshgoftaar, 2009) that arrives in the system with no more information than a rating (first log of a user/item in a collaborative case) or a list of descriptors (new entry in the catalog in a content-based case).

### 1.5.3 Robustness to noisy or corrupted data

As the recommender system will have to deal with different kinds of data sources, the quality of these sources will be sometimes questionable. This can be true for both collaborative-oriented sources and content-based-oriented data sources.

For collaborative data, purchase logs are the most reliable data sources because each user has to pay the item to generate each log entry. Indeed, purchase data are in general the most informative about users and the most precious for predictive models[2]. On the contrary, public rating logs can be corrupted, for example by companies wanting to promote their products on websites or wanting to strike against their competitors. This phenomenon can disturb a lot a recommender system (Lam and Riedl 2004, Mobasher et al., 2005). Social networks and e-commerce websites may have many fake users whose profiles should be ignored by automatic data mining systems.

---

[2] Based on personal knowledge

In content-based filtering one can face the same problem of poor-data-quality. For example, if the item metadata are gathered on the Internet by automatic tools (specific aggregator engines), many problems may occur: duplicate item keys due to differences in item spelling, duplicate item descriptors due to synonymies, possibly wrong cross references, etc. Even large catalogs from content providers are not 100% error free, when they evolve quickly with many updates per week.

### 1.5.4  Scalability

The scalability is a key feature for industrial systems. Industrial systems are working on millions of users and tens of thousands of items or more. The scalability of a system is its ability to deal with a lot of data and a lot of users, without decreasing the quality of service. Although this property is well taken into account in the literature for batch model building (Breese et al., 1998), (Rashid et al., 2006) (Takacs et al., 2009), few articles deal with the reactivity of the recommender systems *online* (Sarwar et al., 2001).

### 1.5.5  Reactivity

A system must be able to take into account a change of a user profile in real time in the case of online recommendation. For example, a user adding a rated item in his profile, or changing a rating to an item already rated, must be able to immediately see the impact that this new information has on the recommendation system. This is even more important if the user reacts to a wrong recommendation indicating that a particular item that the system has recommended does not please him. Otherwise the interaction with the system is severely degraded, and users are much less inclined to provide feedback. (Sarwar et al., 2001).

### 1.5.6  Trust: Transparency, Explanations and Confidence

The transparency of a recommender system is its ability to explain how it works to give a confidence index for its recommendation. The system's ability to be able to give a *confidence index* (Basu et al., 1998) of its recommendation and an explanation are important factors for acceptance by users (Herlocker et al. 2000), (Bilgic, 2004). For an industrial system, *explaining the recommendation* is important for the user, but also for the service manager (marketing…) for validation and traceability of the service.

### 1.5.7  The management of the "long tail"

It is essential for a commercial application to fully exploit its catalog, including its "long tail", the long tail corresponding to the items not very often purchased (Anderson, 2006), (Celma, 2008). The concept of "coverage" deals with this aspect: the coverage is the rate of the catalog items that are actually recommended by the system. The "help to explore" function of recommender systems also assumes that the system does not confine itself to recommend a "Top N" most viewed or most purchased items: The system's ability to make a recommendation, in a relevant way, for all items in the catalog is very important.

## 1.6 A first conclusion

### 1.6.1  Is a scoring-based system enough?

Many publications in the field of automatic recommendation focused exclusively on scoring-based systems, that is to say systems capable of predicting an appetence score of a user for a given item. This emphasis on scoring systems is the result of several factors we discuss below.

The first factor is the scarcity of public data sources. MovieLens and Netflix databases are the best known and both deal of logs of anonymous users' ratings on items.

The second factor is methodological. The evaluation of a scoring system is simple and well known in the community. Several measures exist (Root Mean Squared error, Mean Average Error) and are well accepted. Instead it is difficult to establish the added value of a recommender system issuing recommendations in another way than by an online service.

Of course, logs allow using a protocol with 2 sets, Learning and Test, to simulate recommendations based on profiles only known in the Learning set and to check if they correspond to existing items (case logs of usage) or appreciated items (if logs of ratings) in the Test set. (Deshpande and Karypis, 2004) for instance describe this type of protocol. But these principles of simulation present some problems of interpretation:
- each recommendation (user, item) which does not match an identical couple of (user, item) existing in the Test logs cannot be assessed,
- they do not distinguish which of the recommendations are really useful, those that lead to a purchase that would not have happened without the recommendation,
- they prefer simple systems that recommend the most viewed or rated items, at the expense of systems performing risky but useful recommendations.

On-line experiments on real users, using measurements like conversion rate (percentage of recommendations triggering an act of consumption) or LifeTime Value (measure of the value generated by each client) via a A/B testing protocol (analysis of a marketing campaign with a criterion varying over two different samples of users) can actually evaluate the real value of the pushed recommendations. But generally academic studies do not report on such tests, often confidential, except for rare exceptions like (Davidson et al., 2010).

A third factor is that the scoring function appears to be sufficient to produce the other functionalities of a recommendation system. An implicit assumption is that a scoring system can make recommendations: simply browse the items in the catalog, scoring each item for a user and then return the result. This view, however, has very significant limitations:
- if there is no notion of similarity between two items, the item recommendation for anonymous users will not be possible: this recommendation is based only on the items data (characteristics or usages characteristics) and require a similarity function to link items.
- for personalized recommendation, the system would eventually not scale: consider an application for web pages recommendation for users whose profiles are changing every day: if every profile change implies to browse the entire catalog (the web sites!) to keep the items with the highest score, the system will quickly collapse.
- the prediction of the highest rated item is not necessarily the most useful recommendation (Cremonesi et al., 2010). For instance, the item with the highest predicted rating will most likely be already known by the user.

### 1.6.2 The need for a notion of similarity between items

A system based on a single rating prediction algorithm will therefore be incomplete. As we discuss below, in an industrial context, systems based on similarity matrices between items are best suited to meet the necessary functional spectrum of automatic recommendation. We define as an item-item matrix-based system any predictive system using a matrix (usually pre-calculated) linking pairs of items according to a similarity measure. These systems compare pairs of items. For each item, a list of similar items, with a similarity index, is associated.

# 1.7 General principles of recommender systems based on item-item matrix

We present here the general principles that ensure the functional coverage of industrial systems, described above. It is only an intuitive layout. The presentation will be made in detail in a separate section.

## 1.7.1 Rating predictions

Given a target item **i** and an identified user **u,** one wants to predict the rating of **u** for **i**. $S_u$ is the user **u**'s profile, consisting of pairs (item, rating) entered by the user. The profile is of the form $(i_1, r_1), (i_2, r_2),...,(i_m, r_m)$ ... The more an item **i** is similar to a high rated item of $S_u$ (and not similar to low-rated items of $S_u$), the higher the rating predicted for **i**. A common technique is to perform the weighted average of the $S_u$ using the similarity measure of each item $(i_1,...,i_m)$ with the target item **i**. A matrix of similarities calculated between all items can implement this function (Sarwar, 2001). Naturally all scoring systems, such as those based on matrix factorization (Takacs, 2009) also provide this functionality.

## 1.7.2 Ranking

The ranking is similar in principle to the prediction of a rating. Here we compare a target list of **n** items $L=(j_1, j_2,..., j_n)$ with a profile $S_u=\{(i_1, r1), (i_2, r_2),..., (i_m, r_m)\}$. One seeks to assign a rank to each target item, from **1** to **n**. For example, a recoding by rank (from the 1st to the last) of the items rated by **u** can put in a situation where an average weighted by the similarity of each target item yields a score of rank. Again a matrix of similarity between items may be the basis to the function implementation. Here also any scoring oriented system can provide the feature.

## 1.7.3 Item-to-item recommendation to anonymous users

Given an anonymous user using a context item **i**: **i** is being viewed, or **i** is being purchased, we wish to make relevant recommendations of items related to **i**. Clearly, a pre-calculated similarity matrix item-item can implement this feature immediately. The main methods for making anonymous recommendations based on a contextual item are:
- Neighborhood methods which build a matrix of similarity between items,
- Association rule methods, which can also generate matrices associating an item with other items by calculating co-occurrences.

The scoring-oriented methods such as factorial methods, neural networks... seem inadequate to this task. One might use a similarity based on the provided score of these methods, but the relevance of such similarity is questionable.

## 1.7.4 Personalized recommendation of items

$S_u$ is the user profile, consisting of pairs (item, score) entered by the user. The profile is of the form $(i_1, r_1), (i_2, r_2), ..., (i_m, r_m)$. A simple heuristic for recommending **N** items customized to a user according to his profile is to apply the item-to-item contextual recommendation principle to the user's profile. The principle should be as follows:
- Keep only **t** seed items $s_1, s_2, s_3$ ... well rated by **u,** eventually using a heuristic to choose the seeds (most recent items, best rated items …)
- For each of these **t** seed $s_x$ generate a recommendation list $L_x$ (using contextual item-to-item functionality): Find the **k** most similar items.
- Concatenate the **t** lists $L_x$ of items obtained as a basis to recommend *L*.
- Sort the final list *L* using ranking and eventually another heuristic (diversity…)
- Select the first **N** items of *L*.

Improvements of this principle might be: post-filtering by removing any item already in the profile of **u**, removing items close to items rated poorly by **u**, etc.

The item-item matrix items are very well suited to the personal recommendation of items. Scoring methods can also address this problem but with the severe limitations outlined above.

### 1.7.5 Ability to manage multiple sources

Building an item-item matrix can be done by any space-vector representation of the items. Thus items can be represented in an [item x user] matrix where each item vector is composed of user ratings in the space of all the users. This principle is retained in the vast majority of systems using collaborative techniques such as K-Nearest-Neighbours (KNN).

Items can also be represented in a [item $\times$ metadata] matrix and therefore each item vector is then represented in the space of its descriptors. In general, the descriptors are Boolean attributes encoding pairs (attribute, value) of the item catalog. For example, in the field of movies, Star Wars will be represented by couples (genre, science fiction) (actor, Harrison Ford) (actor, Carry Fisher), (Director, George Lucas)... Hybridization for KNN methods can be done for instance in 2 simple modes:
- Concatenation of space vectors from the user ratings on the one hand and space metadata on the other hand, (feature selection, or feature weighting or feature construction can be used first).
- Creation of two similarity matrices, one for each data source and then merge the 2 matrices obtained, for example by weighting each similarity matrix.

The hybridization of sources is also possible with other methods (matrix factorization...).

### 1.7.6 Management of the cold start

One very interesting property of item-item similarity matrices is that they can be built from any source of data related to items. The itemIDs of external data can be joined after using specific item-matching techniques (not detailed here).

For a new service, for instance of a DVD rental, with almost no user, an item-item similarity matrix can be built from several sources such as:
- catalog of the DVDs, using items' metadata
- purchase logs on the same DVD, but from another service (with other users)
- social networks such as Flixster using automatic opinion classification and itemID matching: see (Poirier et al. 2010) for instance.

Item-item similarity matrices are well suited to the cold-start management problem.

### 1.7.7 Robustness to noisy or corrupted data

This requirement is often managed by combining different approaches. Concerning fake users, a classic approach is to make fake users more expensive to create. In services where profiling is done only from the purchase history, fake users are very unlikely. In other services, automatic controls are generally carried out to prevent automatic "robots" to register in place of real users: complex visual validation codes, specific email confirmation, etc.

We'll see that similarity-based systems can handle noisy data, by choosing good similarity measures and good model combinations such as hybrid methods. Moreover hybrid methods are known to be effective techniques against several attacks (see Mobasher et al. 2007a).

The similarity-matrix-based recommender systems can be compliant with the robustness and corrupted data requirement if we choose a good similarity measure (possibly including feature selection or feature weighting schemes) and/or good model combinations.

### 1.7.8 Scalability

The scalability may be a requirement difficult to hold for a system based on a similarity matrix as:
- Direct KNN search methods are quadratic in the number of items to compare,
- Closely-related methods as association rules are also considered of high complexity.

Matrix factorization methods such as Gravity (Takacs et al., 2008) and other scoring-oriented methods are considered more scalable than methods based on item-item matrix.

In fact, several heuristics and algorithms exist to ensure proper scalability to KNN based methods. The KNN search can be significantly accelerated by clustering methods or dedicated methods like locally sensitive hashing (Gionis et al., 1999). Association rules exploiting the scarcity of data and / or operating on time slices can provide a very good scalability. Amazon (Linden et al., 2003) or YouTube (Davidson et al., 2010), operate according to these principles.

### 1.7.9 Reactivity

Scoring-oriented methods using a compiled predicted model will have limitations in terms of reactivity: consider a user **u** who updates her rating for item **i** from rating 5 (very good) to rating 1 (very bad). Such methods will need to recalculate the predictive model so as to take this modification into account.

On the contrary, to recommend items for an updated profile $S_u$ with re-rated items, a model based on a similarity matrix can reapply the principle of the similarity-based personalized recommendation: It consists in using the item-to-item strategy on the updated user profile with the pre-computed similarity matrix. Few changes in updated items of $S_u$ will have negligible effect on the similarity matrix.

Item-item matrix methods are well adapted to responsiveness constraints. This point is especially noted in (Koren, 2010).

### 1.7.10 Trust: Transparency, Explanation and Confidence

Transparency is another area where systems based on matrices of item-item similarity have an advantage over pure scoring methods.

The generation of a confidence score is possible using the similarity information between items. A simple heuristic is that the more an item **i** will be similar to highly rated items, without being similar to low-rated items, the higher the confidence index will be.

The explanation of a personal recommendation is also easy and intuitive: you can use for each item **i** recommended to a user **u**, the nearest similar item **j** which was in the profile of **u** and which contributed most to the recommendation of **i**. This type of explanation of a personalized recommendation works with logs data, so on a collaborative system. It has no equivalent with scoring-oriented systems for collaborative data as a similarity function is required. This type of explanation can also be done using an item-item similarity matrix built on item's metadata, so we can use a generic explanation scheme both in collaborative and thematic modes.

For contextual recommendations, in collaborative mode, item-item matrices allow the famous "people who have seen / bought this item also seen / bought these items" popularized by Amazon which is a transparent, neutral, and explicit way to explain these recommendations.

### 1.7.11 Management of the "long tail"

The long tail management is in fact part of the recommendation generation strategy. This will be detailed in the chapter dedicated to Reperio. Items belonging to the long tail must be identified by a simple statistic process: items not very often purchased, viewed or rated can be classified in the long tail. Item-item similarity matrices can be used to link every item of the long tail with "better known items" for instance. This "better known items" can be seen as triggers for the recommendation of the corresponding long tail items. We can use this principle to recommend items from the long tail both in the contextual and personalized recommendation frameworks.

It is worth saying that this simple pre-linked principle, using a similarity matrix, insures scalability and responsiveness of the system even with a huge catalog of items.

# 1.8 Conclusion

The traditional classification of automatic recommendation systems is based on the nature of the source data used as input: roughly speaking, usage logs for collaborative filtering systems, metadata catalog coupled to a user profile for thematic filtering systems, and if we use several sources hybrid filtering systems.

Some authors detail the data sources, beyond the simple dichotomy metadata / usage logs, which leads to more types of systems. If we make an inventory of all possible data sources, it actually comes to 10 different possibilities. We believe that another possible classification of the recommender systems, based on the types of available features, rather than on data sources is also possible. This typology identifies two current main systems, those scoring-oriented, and those similarity-based -oriented, plus an emerging new category, tag-oriented.

The functions of automatic recommendation cannot be reduced to the simple score prediction for a given item **i** and a given user **u**, even if this function is the best known and most widely covered by the community. Specific functions of item-to-item recommendation, and personalized recommendations should for example be taken into account when designing a recommendation system.

We identified 4 core functions for automatic recommender systems:

1. **Help to Decide**: predicting a rating for a user for an item
2. **Help to Compare**: rank a list of items in a personalized way for a user
3. **Help to Discover**: provide a user unknown items that will be appreciated
4. **Help to Explorer**: give items similar to a given target item

Other pre-requisites are necessary for an operational system: adaptation to diverse sources of data, robustness, cold-start management, scalability, reactivity, trust relationship, and long tail management. We have introduced for each requirement some principles compliant with similarity-matrix-based systems. These principles will be addressed later in this thesis.

# 2 State of the art

*"Advice is like snow: the softer it falls, the longer it dwells upon, and the deeper it sinks into the mind."*

Samuel Taylor Coleridge.

The purpose of this chapter is to present the main techniques used in the automatic recommender systems' community, and to give an overview of some recommender systems' implementations, both in the academic field and in the industry.

# 2.1 Introduction

This chapter deals with the techniques and methods used in automatic recommender systems. We discuss the algorithms and methods, principles of performance measurement, and concrete examples of real systems from academia or industry.

Two types of algorithmic methods are distinguished from others because they are widely used both in academia and industry: K-Nearest Neighbors methods and Matrix Factorization methods. These two methods are discussed in more detail. Then content-based methods, and hybridization methods are quickly made, then the traditional metrics used in evaluation.

Overall the variety of existing techniques makes it difficult to make a state of the art that is both comprehensive and thorough. We prefer to develop several approaches in the context of an overview of existing recommendation systems, which illustrates the diversity of concrete principles that can be implemented. This overview examines the principles and algorithms used to implement systems and also methods used for performance evaluation: several industrial systems are included in this study, including those of major players of the Internet (Amazon, Google... ).

# 2.2 Main algorithm techniques

## 2.2.1 Association rules

Association rules mining are very popular in data-mining for marketing. This class of algorithms extracts rules that predict the occurrence of an item based on the presence of other items in a transaction. For instance, the Apriori algorithm (Agrawal and Srikant, 1994) is a very famous association rule algorithm, widely used in market basket analysis. It finds groups of items frequently occurring together in the transactions. These groups are called frequent item sets. The notion of frequency is central: a threshold of minimum frequency, called minimum support, must be given to Apriori before it searches for frequent item sets. A minimum confidence threshold must also be given to select the most reliable rules.

For recommender systems, association rules have been studied for instance by (Mobasher et al. 2001) and (Lin et al. 2002). Association rule algorithms have to be adapted to the specific field of recommenders: recommender systems need to be able to recommend any item of a catalog: the association rule algorithm must be able to make associations for any item even those which have a small support. But on the other hand, setting a global support threshold too small leads to an explosive combinatory of possible rules. Traditional approaches use heuristics techniques such as adaptive support (Mobasher et al. 2001) and sliding windows (Mobasher et al., 2001) (Davidson et al., 2010) to insure the scalability of the method.

The real performances of association rules in term of quality still need to be investigated. For instance, in the performances reported both by (Mobasher et al. 2001) and (Lin et al., 2002) the precision of the recommendations is generally under 50% for the former and under 75% for the second, which represent not excellent precision. In the Netflix challenge (Netflix Prize, 2007), no association rule algorithms have emerged clearly. Note also that these algorithms are not adapted to real-valued data and therefore need data preprocessing for logs of ratings and post-processing for rating predictions.

Finally this approach is, in fact, close to item-item similarity-based collaborative filtering techniques. In pure association rule frameworks co-occurrences are analyzed within transactions whereas item-item collaborative filtering systems analyze co-occurrences through all the users, thanks to a similarity measure, on possible real-valued data and without needing a concept of transaction. Item-item similarity-based collaborative filtering techniques such as KNN item-item similarity matrix can generalize simple association rules techniques in a more flexible way.

### 2.2.2 Bayesian classifiers

Bayesian classifiers (Friedman et al., 1997) consider all features and classes of a learning problem as random continuous or discrete variables. Using conditional probabilities and the Bayes' Theorem, the goal of a Bayesian classifier is to maximize the posterior probability of the class of any item to classify given data. Using ratings as classes with discrete values, a Bayesian classifier can be applied to real-valued rating data.

The naïve Bayesian approach assumes that the features (the users, or the items) are independent given a class, the class being for instance a rating taking a discrete value. With this simplification, the probability of the class given all features can be computed very efficiently. Of course, as the assumption of naïve Bayesian approaches is the inverse of the collaborative filtering assumption - the users, and the items, are not independent - the accuracy of such a model is definitively suboptimal. But as a default predictor coupled with a most powerful model, the naïve Bayesian approach may help. (Candillier et al., 2007) used a naïve Bayesian model approach as default predictor with a main KNN-based algorithm, with global good results in term of accuracy.Using more complex Bayesian approaches one needs to consider that attributes are not independent, using Bayesian Networks for instance. But Bayesian Networks learning process becomes in this case complex (Breese et al., 1998).

### 2.2.3 Neural networks approaches

Neural Networks (NN) approaches (or Artificial Neural Networks) are very common in machine learning, signal processing and data mining applications (Bishop, 1995) (Bishop, 2006). Many Artificial Neural Networks exist, the most famous class being the multilayer perceptron using the backpropagation algorithm.

Neural Networks are not often used in recommender systems. There might be several reasons:
- Classical large neural networks in high dimensional problems are known to learn (converge) very slowly,
- For classification tasks in content-based systems, maybe there is no need to complex non-linear classifiers, as the experiment of (Pazzani and Billsus, 1997) seems to demonstrate,
- NN have a black box effect, that is to say the output of a NN cannot easily be interpreted. Even if some explanation methods were proposed, for instance (Feraud and Clerot, 2002), the perceived complexity of NN is still an issue in e-commerce applications.

On the other hand, some systems were proposed for instance in (Jennings and Higuchi, 1993) or (Hsu et al., 2007), and fast matrix factorization techniques such as the Gravity Algorithm (Takacs et al., 2008) can be viewed as specific Neural Network systems.

### 2.2.4 K-Nearest Neighbor approaches

KNN approaches, also called memory-based approaches, are a mainstream in the recommendation system field (Su and Khoshgoftaar, 2009), (Adomavicius and Tuzhilin, 2005). They generalize the association rule principles to compare either items or users globally, on binary of real-valued data. They are well adapted to the 4 core functions of recommenders. They can easily be adapted for regression so they can provide ranking and rating predictions. They are also well-suited for item-to-item recommendation. Their only weakness is the lack of scalability when the number of objects to compare during the KNN search increases. The exact algorithm is intrinsically quadratic: the time to build the model is proportional to the square of the number of objects to compare.

The KNN approaches are more detailed in this chapter. Clustering techniques, closely related to KNN techniques, are presented at the same time.

### 2.2.5 Matrix factorization techniques

Matrix factorization techniques gained popularity during the Netflix challenge (Bell and Koren, 2007c), because they are fast and accurate. In fact, to optimize the error of a rating prediction task, fast matrix factorization techniques such as the Gravity algorithm (Takacks et al., 2008) currently represent state-of-the-art. This class of algorithms is fast and accurate and easy to implement. It proved to be very effective during the Netflix Challenge (Netflix, 2007). On the other hand it seems to be not adapted to item-to-item recommendation as it is a pure scoring method. The matrix factorization techniques are more detailed in this chapter.

### 2.2.6 Other techniques

Several techniques used in the Information Retrieval domain are also often used in Recommender Systems when items can be represented by text documents. For instance, the Rocchio classifier (Rocchio, 1970) is a widely used method to refine search query with user's relevance feedback. The Latent Semantic Analysis (Deerwester et al.,1990) and its probabilistic variant pLSA (Hoffman, 1999) are well known techniques of semantic representation of documents. The LSA framework is based on a matrix factorisation technique and the pLSA algorithm is closely related to non-negative matrix factorisation (Gaussier and Goutte, 2005).

Other techniques from other fields may include for instance: Markov Decision Process (Shani et al., 2005), sequential association rules and related k-order markov models (Brun and Boyer, 2009), Boltzmann machines (Salakhutdinov et al., 2007), Random Walk systems (Jamali and Ester, 2009).

## 2.3 The K-nearest neighbor approaches

For the implementation of recommendation systems, K-Nearest Neighbor (KNN) methods are very popular. They were used from the onset of collaborative systems (Resnick et al., 1994), have been extensively studied (Deshpande and Karypis, 2004) (Sarwar et al., 2001), are used in industry (Linden et al., 2003) and until recently have been widely used in combination with other methods in the Netflix challenge, among the most competitive methods (Bell et al., 2007) (Koren, 2010).

We distinguish methods based on similar users and methods based on similar items. All KNN methods for collaborative filtering are defined by:
- a similarity function (or a distance) that can associate to any pair of objects a similarity index,
- a neighborhood, based on the similarity function, giving for any object its list of similar objects,
- and a method for combining the rating of a set of objects together.

### 2.3.1 K-Nearest Neighbor user-based approaches

For user-based approaches (Resnick et al. 1994; Shardanand and Maes, 1995), the prediction of a user rating of the user **u** on an item **i** is based on the ratings, on that item **i**, of the nearest neighbors of the user **u**. So a similarity measure between users needs to be defined. Then a set of nearest neighbors is selected. And finally, a method for combining the ratings of those neighbors on the target item needs to be chosen.

The way the similarity between users is computed will be discussed in the next sub-section. Let **sim(a, u)** be that similarity measure between users **a** and **u**. The number of neighbors considered is often set by a system parameter that we denote by **K**. So the set of neighbors of a given user **a**, denoted by $T_a$, is made of the **K** users that maximize their similarity to user **a**.

We denote by $S_u$ the set of items rated by user **u**, and $r_{u,i}$ a rating of user **u** on item **i**. A possible way to predict the rating of the user **a** on the item **i** is then to use the weighted sum of the ratings of the nearest neighbors $u \in T_a$ who have already rated the item **i**:

$$\hat{r}_{a,i} = \frac{\sum_{\{u \in T_a | i \in S_u\}} sim(a,u) \times r_{u,i}}{\sum_{\{u \in T_a | i \in S_u\}} |sim(a,u)|} \qquad (2\text{-}1)$$

In order to take into account the difference in use of the rating scale by different users, predictions based on deviations from the mean ratings have been proposed. In that case, $r_{a,i}$ is computed using the sum of the user mean rating and the weighted sum of deviations from their mean rating of the neighbors that have rated the item **i**:

$$\hat{r}_{a,i} = \bar{r}_a + \frac{\sum_{\{u \in T_a | i \in S_u\}} sim(a,u) \times (r_{u,i} - \bar{r}_u)}{\sum_{\{u \in T_a | i \in S_u\}} |sim(a,u)|} \qquad (2\text{-}2)$$

$\bar{r}_u$ represents the mean rating of user **u**:

$$\bar{r}_u = \frac{\sum_{\{i \in S_u\}} r_{u,i}}{|S_u|} \qquad (2\text{-}3)$$

Let us suppose a user rates 4 a movie she likes and 1 a movie she dislikes while another user rates 5 a movie she likes and 2 a movie she dislikes. Then using deviations from their mean rating will reduce the effect of such a difference in use of the rating scale. The time complexity of user-based approaches is **O(n² × m × K)** for the neighborhood model construction, it is **O(K)** for one rating prediction, and the space complexity is **O(n × K)**, with **n** the number of user and **m** the number of items.

### 2.3.2 K-Nearest Neighbor item-based approaches

For item-based approaches (Sarwar et al., 2001), the prediction of a user rating of the user **u** on an item **i** is based on the previous ratings of **u** on items similar to **i**. Item-based approaches have gradually replaced the user-based methods (Sarwar et al. 2001; Karypis, 2001; Linden et al. 2003; Deshpande and Karypis, 2004). The first argument was considerations of computational complexity: the computation of a matrix of exact neighborhood is a quadratic function of the number of objects to compare. The number of users (several hundred thousand to several million) was often greater than the number of items from a catalog (a few tens of thousands in general). It was then more advantageous to build an item-item similarity matrix. This advantage would be less important today as many user-generated catalogs (user-generated video on Youtube,...) or recommender applications to the web or to the music can lead to very huge catalogs with more items than users.

(Linden et al, 2003) showed that functionally, the item-based neighborhood methods provide an advantage over other methods, user-based neighborhood or even using other types of models: in many industrial systems, most of the recommendation is made by the contextual recommendation to anonymous users, as on the Amazon's website (www.amazon.com). This recommendation is called item-to-item recommendation, and generally includes the archetype associated message "people who have seen / bought this item also viewed / purchased these items. This recommendation is directly based on an item-item similarity matrix on navigation logs or purchase logs calculated in batch mode.

We note here $T_i$ the neighborhood of the item **i**. Symmetrically to the user-based approach, two natural ways to predict the rating of a user **a** to an item **i** are:

the use of the weighted sum:

$$\hat{r}_{a,i} = \frac{\sum_{\{j \in S_a \cap T_i\}} sim(i,j) \times r_{a,j}}{\sum_{\{j \in S_a \cap T_i\}} |sim(i,j)|} \quad (2\text{-}4)$$

the use of the weighted sum of deviation from the mean of the ratings:

$$\hat{r}_{a,i} = \bar{r}_i + \frac{\sum_{\{j \in S_a \cap T_i\}} sim(i,j) \times (r_{a,j} - \bar{r}_j)}{\sum_{\{j \in S_a \cap T_i\}} |sim(i,j)|} \quad (2\text{-}5)$$

$\bar{r}_x$ being the mean of ratings of the item **x** (on all the users):

$$\bar{r}_x = \frac{\sum_{\{u \in U | x \in S_u\}} r_{u,x}}{|\{u \in U | x \in S_u\}|} \quad (2\text{-}6)$$

The time complexity of item-based approaches is **O(m² × n × K)** for the neighborhood model construction, it is **O(K)** for one rating prediction, and the space complexity is **O(m × K)**, with **n** the number of users and **m** the number of items.

### 2.3.3 Performance of item-item matrices compared to user-user matrices

(Deshpande and Karypis, 2004) indicate that models based on user-user matrices give better predictive performance than models based on item-item matrix. This is not the conclusion of (Sarwar et al., 2001). More recently, particularly through the Netflix challenge (2006 - 2009) contributions have shown the superiority of item-item models compared to user-user models (Bell et al., 2007) (Takacs et al., 2007). (Koren, 2010) points out that, outside the best predictive performance verified on methods based on item-item matrix, the 2 main advantages of models based on item-item matrix are:

- **Easy explanation**: the system recommends an item **j** to user **u** because the user **u** gave a good rating to an item **i** which is similar to **j**. This is easy to explain automatically. The transparency of the explanation is typical to this type of method. The matrix factorization methods are less easy to interpret, for instance.
- **Easy management of the new ratings**: some news ratings of a user do not significantly modify the item-item similarity matrix. Therefore it is not necessary to recalculate the matrix when a user rates a few additional items. As at the same time the prediction formula based on the item-item similarity matrix takes into account the immediate change of a user profile, it is very easy to manage new users (with one rating) or user's profile change. Again, models such as matrix factorization do not offer the same properties, a learning phase for each user feedback being necessary to adapt the rating prediction scheme.

### 2.3.4 Similarity measures

The similarity between 2 defined items or 2 users is the crucial factor in systems based on neighborhood methods. The first to propose a similarity measure were (Resnick et al., 1994) with the Pearson correlation coefficient. Other similarities, such as the simple cosine, distance of Manhattan, ... are also classics.

$$pearson(i,j) = \frac{\sum_{u \in T_i \cap T_j}(r_{u,i} - \bar{r}_i)(r_{u,j} - \bar{r}_j)}{\sqrt{\sum_{u \in T_i \cap T_j}(r_{u,i} - \bar{r}_i)^2 \sum_{u \in T_i \cap T_j}(r_{u,j} - \bar{r}_j)^2}} \quad (2\text{-}7)$$

**(classic Pearson measure between 2 items)**

$$cosine(i,j) = \frac{\sum_{\{u \in T_i \cap T_j\}} r_{u,i} \times r_{u,j}}{\sqrt{\sum_{u \in T_i \cap T_j} r_{u,i}^2 \sum_{u \in T_i \cap T_j} r_{u,j}^2}} \quad (2\text{-}8)$$

**(classic Cosine measure between 2 items)**

To use the scarcity nature of the data, the Pearson measure is only applied to the attributes common to two items (or two users) to compare. We find the same principle applied to the cosine. It is a very traditional practice (Resnick et al., 1994) (Sarwar et al., 2001) (Adomavicius and Tuzhilin, 2005) (Rao and Talwar, 2008). The classic formula for Pearson, however, suffers from a flaw: 2 items which have a common user, who noted these two items the same way, will have a maximum similarity (and conversely with 2 users, if one uses a user-user matrix). For example, in the field of cinema, consider two users, a fan of science fiction, the other amateur of comedy. Initially, these users have nothing in common so their similarity is zero. Then suppose that these two users both rate "Men In Black", a comedy of science fiction. According to Pearson's formula, applied only to the movies rated in common, these two users will now have a maximum similarity. The problem of the classical Pearson measure is particularly identified in (Deshpande and Karypis, 2004) and (Breese et al., 1998).

The Jaccard similarity does not have this limitation since it measures the proportion of common attributes of 2 vectors, considered as 2 sets of elements. On the other hand, the Jaccard similarity does not take into account differences in ratings, only events related to some features present or not. If 2 users have rated exactly the same movies, but for every movie with ratings totally opposite the Jaccard similarity, however, returns a maximum similarity. For 2 items **i** and **j** respectively represented by the set of the users' ratings $T_i$ and $T_j$, the Jaccard similarity is defined by :

$$Jaccard(i,j) = \frac{|\{T_i \cap T_j\}|}{|\{T_i \cup T_j\}|} \qquad (2\text{-}9)$$

When the Jaccard similarity is combined with other measures of similarity, the resulting similarity can benefit from their complementarities. For example, the product of the Jaccard similarity with the Pearson similarity, denoted wPearson, produces better results than each similarity taken individually (Candillier et al., 2007).

### 2.3.5 Clustering techniques for KNN

Many clustering techniques have been used to improve the scalability of KNN-based methods. The clusters are used to reduce the KNN search of a vector **v** (an item or user representation) to the cluster of the vector **v**. The clustering phase can be done offline as a preprocessing phase. Many clustering techniques are known to be efficient both in terms of accuracy and in term of speed: K-means, Top-down hierarchical algorithm like Birch...

(Sarwar et al., 2002) used bisecting K-means to cluster users. They found that the cluster-based methods has lower performance in Mean Average Error (MAE, see next section) than the direct neighborhood method, but still correct. They also found that increasing the number of clusters has a bad effect to the performances, as the best performances were for 10 clusters: increasing the number of clusters, in fact, decreases their size, and reducing the search of the neighbors in small clusters leads to poor performances.

ClustKNN (Rashid et al., 2006) used a similar approach and leaded to the same results. An empirical analysis showed that the percentage of the top **N** real nearest neighbors found in a cluster decreased rapidly and is already under 16% with only 20 clusters of users on the MovieLens 1M database (containing 6000 users and 3700 items). On the other hand, the precision of the cluster-based method is less impacted.

The clustering of items was tried for instance by (O'Connor and Herlocker, 1999). They tried different algorithms, both on the rating space and on the metadata space of the little MovieLens 100,000 rating database. Again the accuracy in term of MAE is worse with the cluster-based methods, and as the number of clusters increases and their size decreases the MAE increases.

Clustering methods seem to be necessary viewed as a trade-off between speed and accuracy.

### 2.3.6 Limitations of KNN-based predictors and recent extensions

The K-nearest neighbor methods have some limitations (Koren 2010). First, these methods have scalability problems in limit cases, because the computational complexity of the search for neighbors is quadratic. Then, these methods generally yield results slightly worse than the matrix factorization methods, in terms of rating prediction. The challenge Netflix, just based on the rating prediction has shown on many contributions, the superiority of matrix factorization methods like Gravity.

New ideas for the expansion of KNN methods have emerged during the Netflix challenge. We will mention two that are representative, the KNN based on a pretreatment algorithm such as a Boltzmann machine (Salakhutdinov et al. 2007), and the use of a KNN approach in a matrix factorization framework (Tackacs et al. 2008, Koren 2010) that will be presented in the next section.

## 2.4 Matrix factorization techniques

Matrix factorization techniques are methods to transform a given matrix **M** into simpler matrices, typically 3 simpler matrices. The main method is Singular Value Decomposition, SVD.

The general formalization of the SVD is as follows:

Given a **m × n** matrix **R**, with rank **r**, the SVD of **R** is defined as:

$$\mathbf{SVD(R)} = \mathbf{U} \times \mathbf{\Sigma} \times \mathbf{V^T}$$

where **U** is a **m × m** matrix, **Σ** a **m × n** matrix, and **V** a **n × n** matrix. **Σ** is a diagonal matrix with **r** nonzero entries ($s_1, s_2,...s_r$), and by convention with the property $s_i >= s_{i+1}$. The $s_i$ entries are called the singular value of **R**, and are positive or null. **U** and **V** are unitary matrices: $\mathbf{UU^T}=\mathbf{I}$ and $\mathbf{V^TV}=\mathbf{I}$.

The SVD of **R** gives the best linear approximation of **R** when selecting the **k** first columns of **U**, the **k** first singular values of **Σ**, and the **k** first rows of **V**.

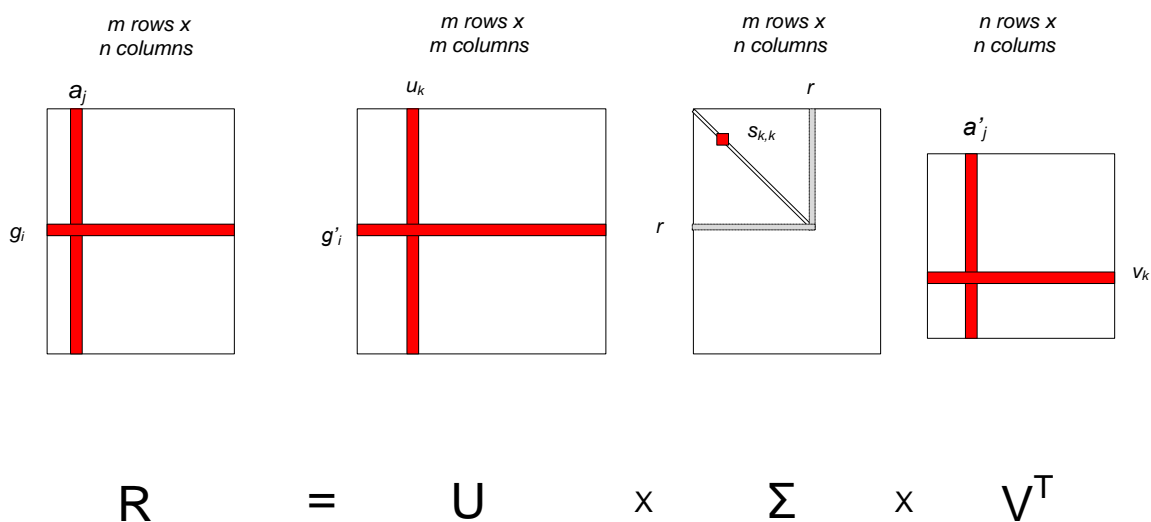

**Figure 2-1. Principle of the Singular Value Decomposition**

The great interest of the SVD is that the approximation of k-factor-based **U × Σ ×V**, with **k << m** and **k << n**, can be very good. In this case it makes a significant compression of information associated with a data denoising.

The SVD has been adapted to automatic recommendation systems. The sparsity nature of the rating matrix, and its size, require special adaptations. Indeed, the SVD is not defined in the case of matrices with unknown information, so sparse matrices must be well defined: empty values should be specified by default values.

The use of matrix factorization in collaborative filtering is not trivial. Compared to methods used in information retrieval, data entries include many missing values (more than 98% of missing values in the Netflix Challenge data for example). Methods for Singular Value Decomposition must handle the case of incomplete data. The first SVD approaches were therefore first to replace the missing values of the rating matrix (Sarwar et al., 2000B). However the application of the replacement of missing data leads to very large dense matrices, with risks of poor estimation of missing data and scalability problems. Subsequently approaches working directly with missing data, using parameters and regularization to avoid over-learning have been developed (Takacs et al., 2008).

We present here very briefly the classical SVD then a simple algorithm, well-known and efficient to perform fast matrix factorization: the algorithm Gravity.

## 2.4.1 Classical SVD applied to collaborative filtering

The **R** matrix of user ratings on items is first filled in: for example, each $r_{u,i}$ missing is replaced by the mean ratings of **u**.

One then searches for the decomposition of the matrix **R**:

$$\widehat{R} = U \times \Sigma \times V^T \qquad (2\text{-}10)$$

**U** is a **m × k** matrix, **V** is a **k × n** matrix, and **Σ** is a **k × k** diagonal matrix with the first **k** singular values.

The rating prediction for a user u for an item **i** is given by the scalar product of the pseudo user $(U \times \sqrt{\Sigma})[u]$ with the pseudo item $(\sqrt{\Sigma} \times V^T)[i]$:

$$\hat{r}_{u,i} = (U \times \sqrt{\Sigma})[u] \times (\sqrt{\Sigma} \times V^T)[i] \qquad (2\text{-}11)$$

using the notation**:** **(M)[j]** is the j-th row vector of the matrix **M**. $\sqrt{\Sigma}$ represents the diagonal matrix of the square roots of the singular values.

The SVD algorithm for a **m × n** matrix complexity has a cost in **O(mn²)** operations, (Thefether et al.,1997) assuming that **m > n**. This is very expensive for big matrices so approximation heuristics are needed for real applications.

## 2.4.2 The Gravity approach

The Gravity algorithm (Takacs et al., 2008) tries to directly approximate the matrix of ratings **R** rating by the product of a matrix **P** of the factors of the users with a matrix **Q** of the factors of the items.

$$\widehat{R} = P^T Q \qquad (2\text{-}12)$$

**Q** is a **K × n** matrix, with n the number of items, and **P** is a **K × m** matrix, with m the number of users. **R** is the matrix of the rating logs with **m** users (rows) and **n** items (colums), with **K** the number of factors (**K** is the main parameter of the model)

We take the following notations: $q_{k,i}$ is the **k**-th factor of the **i**-th item and $p_{k,n}$ is the **k**-th factor of the user **u**.

The predicted rating is given by:

$$\hat{r}_{u,i} = \sum_{k=1}^{K} p_{k,u}{}^T q_{k,i} \tag{2-13}$$

The predictive error is given by:

$$e_{u,i} = r_{u,i} - \hat{r}_{u,i} \tag{2-14}$$

Gravity's algorithm is based on a gradient descent using the back-propagation of the error. The error of the predictive model is given by:

$$E = \sum_{(u,i)\in R} e_{u,i}^2 = \sum_{(u,i)\in R} \left(r_{u,i} - \sum_{k=1}^{K} p_{k,u}{}^T q_{k,i}\right)^2 \tag{2-15}$$

We deduce the gradients of the error:

$$\frac{\partial}{\partial p_{u,k}} e_{u,i}^2 = -2 e_{u,i} \cdot q_{k,i} \,, \frac{\partial}{\partial q_{k,i}} e_{u,i}^2 = -2 e_{u,i} \cdot p_{k,u}{}^T \tag{2-16}$$

The algorithm tries to minimize the overall error on the training set iteratively. It takes as input the list of the logs, and for each triple **(u, i, r)** performs the predictive ratings $\hat{r}_{u,i}$, the error $e_{u,i}$, and then back-propagates the error's gradient to the weights of the matrix **P** and **Q**.

$$p_{k,u} \leftarrow p_{k,u} + \alpha \cdot e_{u,i} \cdot q_{k,i}$$
$$q_{k,i} \leftarrow q_{k,i} + \alpha \cdot e_{u,i} \cdot p_{k,u}$$

Both to prevent over fitting and divergence of the weights $p_{k,u}$ and $q_{k,i}$, a regularization principle is applied at this step. High weights are penalized by introducing in the value $e_{u,i}$ to minimize a value proportional to the square of their Euclidian norm :

$$e'_{u,i} = e_{u,i}^2 + \lambda(\|p_u\|^2 + \|q_i\|^2)$$

$$e'_{u,i} = e_{u,i}^2 + \lambda \left(\sum_{k=1}^{K} p_{k,u}^2 + \sum_{k=1}^{K} q_{k,i}^2\right)$$

Then we want now to minimize $E' = \sum_{u,i \in R} e'_{u,i}$

We compute the gradient of $e'_{u,i}$ according to each factor $p_{k,u}$ and $q_{k,i}$

$$\frac{\partial}{\partial p_{u,k}} e'_{u,i} = -2 e_{u,i} \cdot q_{k,i} + 2\lambda p_{k,u}$$

$$\frac{\partial}{\partial q_{k,i}} e'_{u,i} = -2 e_{u,i} \cdot p_{k,u} + 2\lambda q_{k,i}$$

To optimize this solution, we update the weights in the direction opposite to the gradient, so we obtain the following update rules for the algorithm *Gravity*:

$$p_{k,u} \leftarrow p_{k,u} + \alpha.(e_{u,i}.q_{k,i} - \lambda p_{k,u}) \qquad (2\text{-}17)$$

$$q_{k,i} \leftarrow q_{k,i} + \alpha.(e_{u,i}.p_{k,u} - \lambda q_{k,i}) \qquad (2\text{-}18)$$

Where $\alpha$ and $\lambda$ are respectively the learning rate and the regularization parameters.

The Figure 2-2 gives a simple overview of the principle of the Gravity Algorithm.

Gravity works exclusively on the observed values: the empty values are assumed to be predicted. This is a natural handling for empty values.

Gravity is efficient: it converges quickly, within several passes on the logs, and gives good results for the rating prediction task (Takacs et al., 2008). However it is very sensitive to its 2 parameters: the learning rate $\alpha$, and the regularization parameter $\lambda$. Ideally the regularization parameter needs to be calculated by cross-validation, usually for ratings between 1 and 5, a value of 0.05 is good. The learning rate needs to be set by trials/errors, usually small values of 0.01 to 0.001 are well.

The complexity of Gravity is linear according to the number of the nonzero matrix's entries; in the case of collaborative filtering this is equivalent to the number of logs, **L**. The number of passes is generally of several tens. The global cost of Gravity is **O(LC)**, with **L** the number of logs, and **C** the number of passes required to insure convergence

### 2.4.3 Adding bias

The regularization parameter forces Gravity to work with factors as small as possible, leading to $\hat{r}_{ui}$ as small as possible though according to the RMSE minimization goal. This would assume that there's no global tendency for specific users or specific items. This is not the case in real life applications. Statistical analysis of usage logs of MovieLens or Netflix databases, for instance, show important differences between users' behavior or items' properties. Some users rate higher that the mean, some items have higher mean that others. The global ratings' mean should also be taken into account. To capture specific information about items and users, biases are added to the original equation:

a bias for the global ratings' mean: $\mu$
a bias for the item's popularity: $b_i$
a bias for the user's behavior: $b_u$

The Gravity's equation becomes:

$$\hat{r}_{ui} = \mu + b_i + b_u + p_u^T q_i \qquad (2\text{-}19)$$

To learn this new model (see first algorithm in Figure 2-2), the biases are learnt simultaneously on a second step:

$$b_i = b_i + \alpha \left(r_{ij} - \lambda_2 (b_i + b_u - \mu)\right) \qquad (2\text{-}20)$$
$$b_u = b_u + \alpha \left(r_{ij} - \lambda_2 (b_u + b_{ui} - \mu)\right) \qquad (2\text{-}21)$$

where $\lambda_2$ is another regularization parameter.

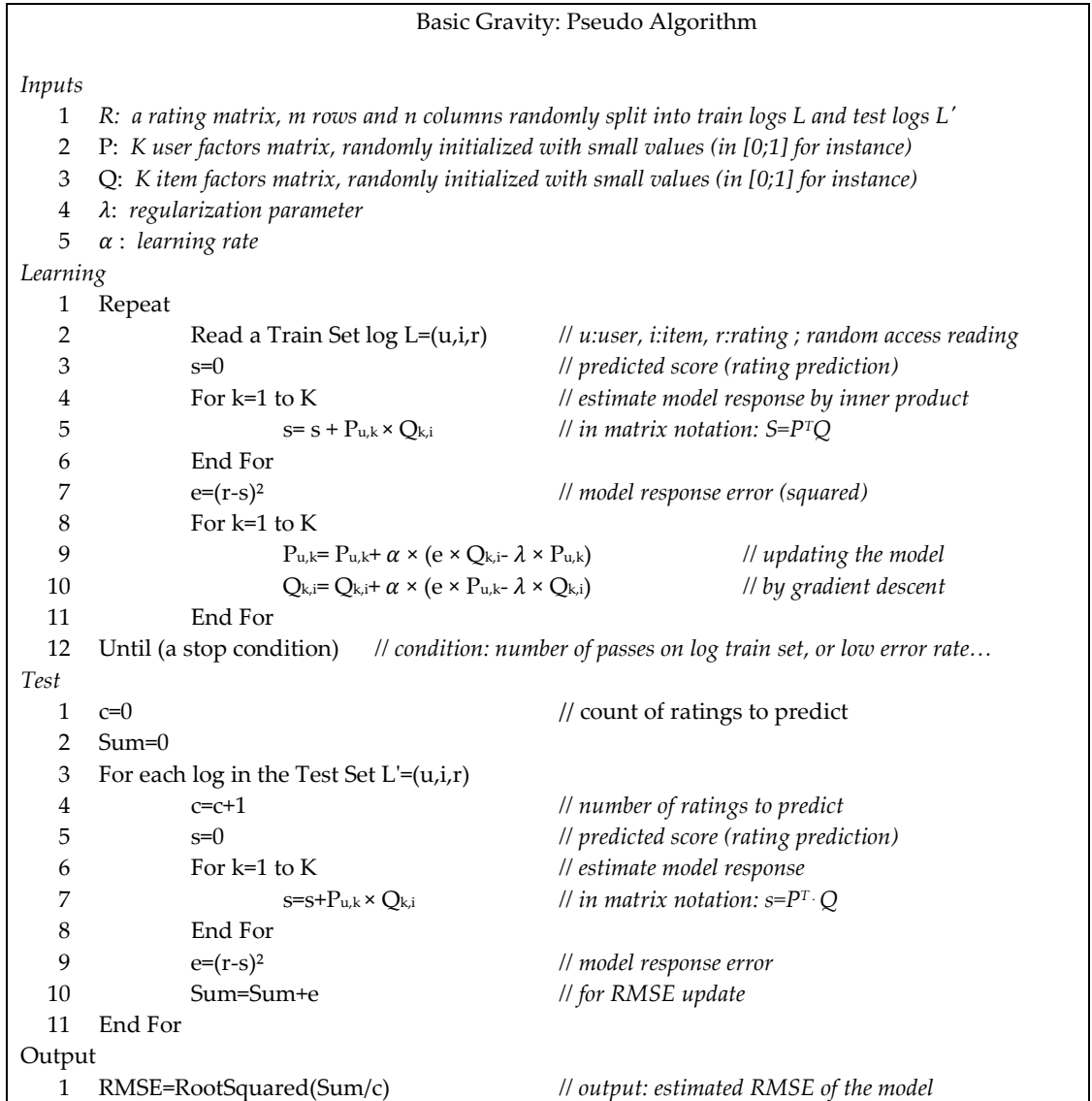

**Figure 2-2: Pseudo-algorithm: Principle of the basic Gravity method**

Another way to simply introduce bias features into Gravity is to fix to 1 the value of the 1st factor of the **P** matrix (users factors) and of the 2nd factor of the **Q** matrix (items factors): they will stand for **b$_u$** and **b$_i$**. (Takacs et al, 2008).

## 2.4.4 Extensions

(Takacs et al., 2008) propose to use the matrix factorization scheme into the conventional KNN-scoring scheme:

$$\hat{r}_{ui} = p_u^T q_i + \gamma \frac{\sum_{j \in S_u} w_{ij}(p_u^T q_j - r_{uj})}{\sum_{j \in S_u} w_{ij}} \qquad (2\text{-}22)$$

$\gamma$ is a parameter to optimize by cross-validation.
$w_{ij}$ are the similarities between items **i** and items **j**
$p_u^T q_i$ stands for the bias for **u** and **i**; the equivalent pure KNN-based rating equation use the user mean.

The authors propose 2 types of similarity weights for the $w_{ij}$, the normalized scalar product or the normalized Euclidian distance. The $w_{ij}$ are computed on the **Q** matrix of items' factors so this computation is faster than with conventional KNN techniques which uses the original vector of items. The training is similar to the regular Gravity's algorithm training.

(Koren, 2008b) also proposed a factorized neighborhood model, which is directly trained. But the model contains many parameters (8 parameters plus the neighborhood's size) that could make it difficult to deploy.

Gravity can also be extended to deal with both collaborative data and content-based data (Pilazci et al., 2011). The main idea is to use specific factors of users to be modified by item metadata.

### 2.4.5 Connection with Clustering and Dictionary learning

Suppose **X** is a matrix composed of **n** rows of items represented in the space of **p** user's ratings. Then the factorization algorithm approximating a matrix **X** by $\hat{X} = UV^T$ can be interpreted as a clustering process called dictionary learning: see Figure 2-3.

$V^T$ is a dictionary of **k** virtual items represented in the space of the **p** users' ratings (indeed the items are represented in dimension m), and **Q** is the decomposition coefficient representation of the matrix **R** using the dictionary. The rows of the $V^T$ matrix are not strictly speaking clusters but element of a (possibly over complete) basis. The encoding generally adds constraints of sparse coding, that is to say that many of the **Q** coefficients are null, so any original item vector of **R** can be approximated by a linear combination of few vectors of $V^T$. Those techniques of dictionary learning for sparse representations are used in many fields of signal processing (denoising), image processing (compression) and statistics. See for instance (Elad, 2006).

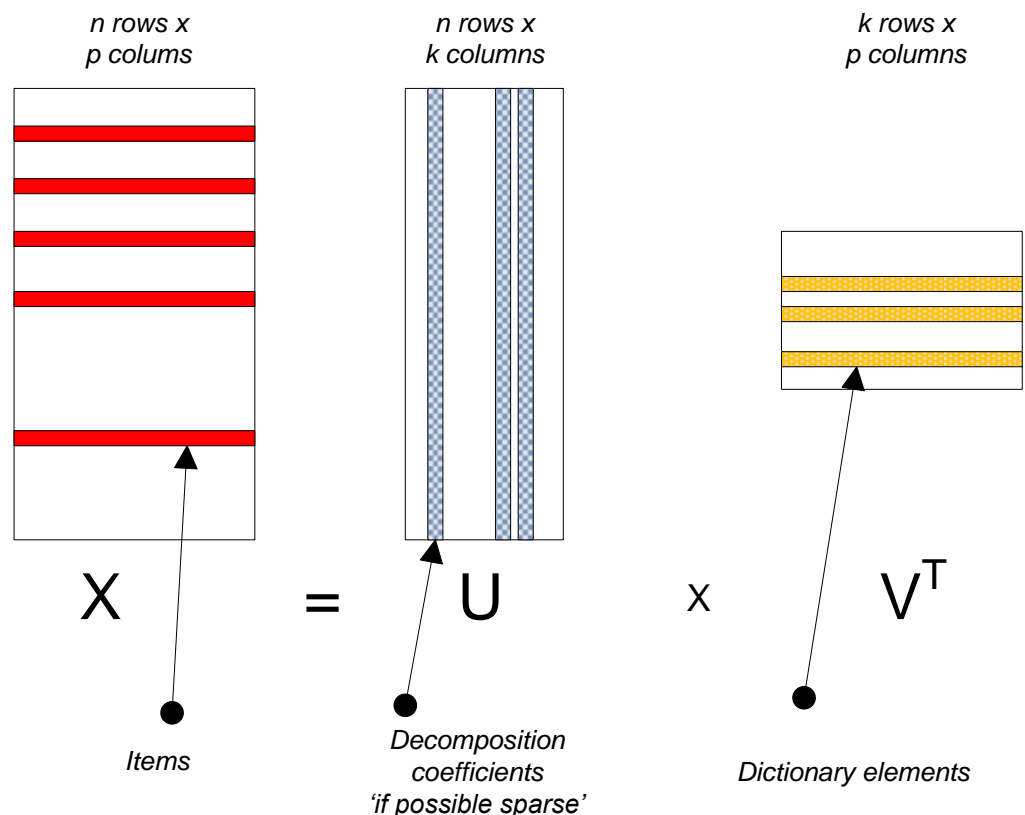

**Figure 2-3: Matrix factorization viewed as a dictionary learning process**

# 2.5 Content-based recommender systems

## 2.5.1 Introduction

Content-based filtering methods originate from the field of Information Retrieval (IR). For this reason, publications about these techniques are rarely listed in recommendation systems, where the articles deal principally with collaborative filtering, or at least hybrid methods (Su and Khoshgoftaar 2009; Adomavicius 2005). The methods known as content-based filtering, also known as thematic filtering, recommend an item to a user based on the description of both the item and the user profile in terms of item characteristics. These characteristics are called metadata. Content-based recommender systems need:
- item's metadata to describe the items to recommend,
- a representation of user profiles,
- and a method for comparing the user profile with the item to score.

A restrictive definition: a pure content-based system is possible. A purely content-based recommender system is a system working with one user at a time and building a profile based on the characteristics of items she consulted, using the feedback from the user. This definition of (Balabanovic and Shoham, 1997) has the advantage of precision as it is opposed to those of collaborative system which can only work with multiple users.

Items' metadata can be for example the genre of a movie, or the location of a restaurant, according to the type of items to be recommended. Finally, the items most similar to the preferences of the user will be recommended.

The user profile can be constructed implicitly from the ratings or from other previous actions on items (eg research, stop watching, buying,...) or explicitly based on questionnaires, for example by ratings the general characteristics she likes / dislikes.

A user model can be learned implicitly by a learning method, using the descriptions of items for a supervised algorithm, producing the ratings of the user as output. The user profiles are often represented by vectors of metadata's weight on the descriptions of items. Other user models can be considered. For example, if a rule induction algorithm is used, then a model of a user could be:

"IF genre IS action AND actor IS Stallone THEN film IS liked"

Following (Balabanovic, 1997) we can define a content-based filtering system on documents (such as Web pages, descriptions of products catalog, news from a newspaper,...) with 4 elements:

1. a method to represent the documents, for instance as a vector **d**
2. a method to represent the user's interest, for instance as a vector **u**
3. a function **p(d,u)** which gives the relevance of the document **d** for the profile of the user **u**
4. a function **v(d,u,f)** which gives an updated version of the user's profile, given her feedback **f** on the new document **d**.

Items are often represented as vectors within a space of their characteristics. The vector representation is well suited to treatments such as classifications or regressions underlying content-based filtering systems. In the vector-space model, documents and queries are vectors in a space of words. Each document is represented by a vector **v** where each component of the vector $v_i$ is the weight of the word $d_i$ in this document. If the document does not contain $d_i$, then $v_i = 0$. Classic pretreatments, such as those of Fab (Balabanovic, 1997), are to take the stems rather than the words (their radical), ignoring a predefined list of empty words (articles,...) then calculating the weight of words by a Term Frequency-Inverse Document Frequency (TF-IDF) technique that weights each term in proportion to its frequency in a document and inversely to its frequency in all documents. Each document keeps only the **k** words of greatest weights.

Users can also be represented by a vector of words corresponding to their interests. The advantage of this representation is that the scoring affinity **s** of a user represented by its vector profile **u** to a document **d** represented by the vector of its words' weights can be given by the scalar product $s = u \cdot d$, or the cosine or other similarity function of this type. In Fab (Balabanovic, 1997, Balabanovic and Shoham, 1997) for instance, the function which gives the relevance of **d** for **u** is the scalar product $p(d.u)=d.u$, and the update rule of the user's profile is simply $u_{t+1}=u_t+f.d$ where $u_t$ is the current profile, $u_{t+1}$ the updated profile, **f** a real-valued score of the feedback of the user for the new document **d**.

(Pazzani and Billsus, 2007) presented several methods of content representation of items and user profiles as well as methods for learning user models. The preferences indicate the relations between some users and some characteristics of the items. We can distinguish simple preferences, called Monadic, like "I love Jackie Chan as an actor" and preferences called Dyadic comparing 2 characteristics, such as "I prefer comedy to drama". These preferences can be used to improve the recommendation, in what is commonly called post-filtering. Their main drawback is to have a coverage limited to the attribute on which they apply. The attributes with a relatively high coverage (typically an action genre in a movie catalog) will allow efficient preferences while an attribute with a quite low coverage (Jackie Chan as an actor for example) will lead to preferences with a limited impact. Recommender systems based solely on preferences, will risk asking a great deal of information before being effective to give a score or to sort all items in their catalog.

However, it is possible to extend the coverage of the preferences using a notion of similarity between attributes. For example, if Jackie Chan and Bruce Lee are considered very similar, the preference "I like Jackie Chan as an actor" could be extended to "I like Bruce Lee as an Actor". This extension, assuming it does not reduce the predictive performance, increases the coverage of the preferences.

Another classical approach to content-based filtering methods is to use a classifier that simulates the user, such as a Naïve Bayes classifier, for example. This classifier is trained with data from the user profile, with descriptions of items as input and the user's rating on these items as output. The classifier is then able to predict whether a new item, according to its description, will please or not the user (Adomavicius and Tuzhilin, 2005).

### 2.5.2 Pros and cons of content-based filtering systems

The content-based methods allow overcoming some limitations of the collaborative filtering:
- They can provide recommendations for a new item even if no rating is available for it,
- They can manage situations where different users do not share identical items, but only similar items according to their intrinsic characteristics (metadata).

However, the content-based filtering methods require rich descriptions of items and well built and well informed user profiles. These ideal cases are rare in real applications. This dependence on the quality and structure of data is the main weakness of methods based on content. Since it is difficult in many areas, to obtain structured and complete descriptions of the items, content-based methods have mainly been applied to catalogs of textual information, such as documents, pages, websites, or message forums (Pazzani and Billsus, 1997; Mooney and Roy, 1999).

Another weak point of content-based methods is their tendency to recommend items very similar to items already seen and rated by users, a phenomenon called overspecialization (Zhang et al., 2002). The "help to discover" aspect, and the originality of the recommendations, are strongly reduced as compared to collaborative filtering. On the other hand, the fact that users can get recommendations without sharing their profile ensures their privacy (Lam et al., 2006).

Content-based approaches appear symmetrical and complementary methods to collaborative filtering. Where collaborative methods involve centralized data on thousands of users, content-based approaches generally require usages of only one user. Where collaborative filtering are content-agnostic (i.e. independent to the items' metadata), content-based approaches will be effective only for catalogs of items represented by rich metadata. The complementary of content-base methods to collaborative methods made them good candidates to hybridization techniques, described below.

## 2.6 Hybrid methods

### 2.6.1 Introduction

We call hybrid filtering recommendation system a system based on different data sources of different natures. We will discuss differently the approach of managing multiple pure collaborative filtering models.

In the case of hybrid filtering, the two types of information, collaborative and thematic, are exploited. These technologies can be combined in different ways which will be detailed in this section. Other user-specific information such as demographic information, or the use of social networks, may also be included in hybrid systems.

The expected benefits of the hybridization are:
- global improvement of the accuracy, as generally achieve in data-mining using several predictive models,
- using the best of each approach to deal with single-methods' weaknesses: for instance, using collaborative filtering to help content-based filtering avoiding overspecialization, using content-based filtering to help collaborative filtering dealing with cold-start.

On the other hand, the possible drawbacks of the hybridization could be:
- increasing the complexity of the recommender system
- lowering the speed as more models are used at the same time.

### 2.6.2 Burke's classification

Burke (Burke, 2002) and (Burke, 2007) presents a classification of hybridization in 7 types we will discuss.

**Weighted**

The first natural approach for combining different data sources, or different models, is to use a voting system. We join in that the idea of mixture of models, or model assemblies, common in data-mining and machine learning (Breiman, 1996), (Dietterich, 2000), (Polikar, 2006). The Weighted approach uses N recommendation systems. Every recommender system is trained on its own data source. One can for example use a collaborative filtering system on usage logs, and a content-based filtering system using reported preferences on a catalog of items.

The generation of recommended items is done by asking candidate items to each system, and then performing a union or an intersection. Then a scoring and a sort select the recommended items. The scoring of an item **i** is done by taking the weighted average scores of each system for the item in question. In (Mobasher, 2004), for example 2 systems, one content-based, the other collaborative, are combined by a weighted average of their predictive scores, respectively 60 and 40.

The Weigthed approach can be extended to any parameterized score function. This technique has been widely used, on purely collaborative models, during the Netflix challenge, generally called "Blend". The optimization of the weights of the combination function used techniques of linear regression models, neural networks or boosting (Jaher et al., 2010).

**Switching**

The hybridization by "switch" does not propose to combine different results of recommendation systems, but to select one system dynamically at the time of the recommendation, according to a criterion.

In (Billsus and Pazzani, 2000) for example two recommendation systems are ordered and connected in cascade: a thematic engine based on a KNN algorithm is used to model the short-term profile of the user, and a naive Bayesian model is used for the long-term profile modeling of the user. The first system, KNN, is the first to be called. According to the confidence index associated with the recommendation, the system automatically switches to the second or to a last default recommender giving a static default score.

The Switch is based on a criterion and a threshold related to a particular recommender. This criterion may be difficult to define and the threshold to set (Burke, 2002). One can choose the confidence index returned by each recommender if they provide one.

**Mixed**

The Mixed approach only deals with the push of recommended items, not with the other aspects of the recommendation (rating prediction, scoring,...). Candidate selection is done by requiring each system to issue its candidates with an associated score, a predicted rating and/or a confidence index. Then a specific module performs a mix of these recommendations with a sorting and a selection based on the scores associated with the candidate items. This method of hybridization is only quoted by (Burke, 2007) and is not evaluated.

**Feature combination**

In this mode, one uses data normally used for a type of recommendation system in another context. For example, one can use data from user ratings on the items normally handled by a collaborative system. These user ratings can be added to a catalog of items, and then the catalog can be used by a purely thematic recommendation system. Burke said in fact that the principle of "Feature combination" is not a hybrid method in the strict sense as there is no use of several recommendation systems.

**Feature augmentation**

In the "Feature augmentation" principle, a recommender is used to add features to users or items before the use of these data by another recommender system. (Sagwar et al., 1998), in the "Grouplens" system, used this technique to enhance a collaborative system: software robots automatically rate items according to some of their characteristics, the triples (robotID, itemID, rating) then enriches a collaborative database. Then a collaborative system uses all of the obtained ratings, those of real users, and those of robots to build a model.

**Cascade**

The cascade hybridization method is a hierarchical method where each next recommender only refines the recommendation obtained by the previous one. For example the EntreeC system, which provided too many items with identical scores, was improved by adding a post-ranking based on a collaborative recommender (Burke, 2002).

**Meta Level**

The Meta-Level hybridization method uses as input a model made by another recommender. Compared to the Feature Augmentation principle, this method requires a total replacement of the input model by the output of the previous recommender, such as via a change of the basis of representation. A typical example, called « collaboration through content » is that of (Pazzani, 1999): first a recommender based on a linear model produced a model of user profiles as feature vectors. These vectors are then used by a collaborative engine to provide the collaborative recommendations.

### 2.6.3 Discussion

The weighted approach is the most used in the literature and has proven to be effective for accuracy (Su and Khoshgoftaar, 2009), (Bell and Koren 2007c).

The switching approach can use different models according to a context. A dynamic weighting scheme could be more general both for the weighted approach (assumed static) and the switching strategy. In one case, the weight of each model is learnt and is static. In the second case, the weight of each model is determined by another model taking as input the confidence index of each model of the first layer, or for instance a profile size.

The mixed approach is in fact equivalent to the weighted approach but limited to the push of recommended items.

Feature augmentation seems more related to data preparation when it consists in automatically adding logs from a catalog in the example of (Sagwar et al. 1998). In another hand, if the output of the recommender A is an enhanced user profile and then if this output is sent to another recommender B, the result seems to belong both to feature augmentation and meta level.

We miss another category where one model directly uses as input data from different sources, for instance a collaborative one and a content-based one. For example, the hybrid music recommender of (Yoshii et al., 2008) uses a probabilistic model called Probabilistic Latent Semantic Analysis to directly integrate a source of ratings and a source of content-based timbre analysis of the pieces of music. In our experimentation of hybridization in the chapter 3 we use as input of a KNN model a concatenation of 2 vectors as input: a collaborative rating vector, and a content based descriptor vector.

We could then consider 4 families of hybrid methods:
1. direct data integration into one model of different sources of data,
2. ensemble methods using several models at the same time with a combination function that can be static (weighed approach) or dynamic (switching approach). These approaches are well adapted for rating prediction tasks. We could assume that dynamic combinations could lead to better accuracy since they can be adapted to different users' contexts: short profile versus long profile for instance. Their main drawback, when using many models, is that they can face scalability issues or maintainability problems,
3. cascade models using several models in cascade, to refine the results of the previous step; these methods seem more adapted to the push of recommendations, where pre-filtering techniques plus post-filtering techniques are usually applied during the process of refining relevant items,
4. ensemble methods using a model as an input to another model: these methods are generally more complex. They include profile-generation methods such as (Pazzani, 1999) and data dimension reduction techniques.

## 2.7 Recommendation evaluation metrics

The evaluation of recommender systems is often done in terms of Accuracy, that is to say, of predictive performance for the task of rating prediction. This implies a dataset of user ratings on items, usually in the form of user's rating logs. As in a classical machine learning test protocol, the predicted scores on the items and the actual notes on the items are compared. In this approach, several measures are employed, the most famous being the Mean Absolute Error (MAE) and the Root Mean Squared Error (RMSE). From the Information Retrieval field, Precision and Recall can be used when the relevant recommended items are given.

### 2.7.1 Rating-based metrics

The Mean Average Error, MAE, is given by:

$$MAE = \frac{1}{|R|}\sum_{(u,i,r)\in R}|\hat{r}_{u,i} - r_{u,i}| \qquad (2\text{-}23)$$

with **R**: the set of the evaluated ratings, normally a Test Set

The Root Mean Squared Error, RMSE, is given by :

$$RMSE = \sqrt{\frac{1}{|R|}\sum_{(u,i,r)\in R}(\hat{r}_{u,i} - r_{u,i})^2} \qquad (2\text{-}24)$$

with **R**: the set of the evaluated ratings, normally a Test Set

Both measures are widely used for algorithm evaluations. Moreover, the RMSE has gained popularity since the Netflix Challenge (Netflix Prize, 2007). It is very often assumed that minimizing the RMSE (or the MAE) is equivalent to increasing the quality of the recommendations even if other metrics are also proposed (Su and Khoshgogtaar, 2009).

### 2.7.2 Classification based metrics for relevancy

The Precision is the ratio of the number of relevant recommended items to the number of recommended items

$$\text{Precision} = \frac{\text{number of relevant recommended items}}{\text{number of recommended items}} \qquad (2\text{-}25)$$

The Recall is the ratio of number of relevant recommended items to the total number of relevant items available.

$$\text{Recall} = \frac{\text{number of relevant recommended items}}{\text{total number of relevant available items}} \qquad (2\text{-}26)$$

Recall is the ratio of number of relevant recommended items to the total number of relevant available items. Recall measure is generally impractical as the global view of all relevant items for each user is not manageable for large catalogs. Precision and Recall need a binary indicator for each item to be classified into relevant/irrelevant class. Similar to Precision and Recall, Receiver Operating Characteristic (ROC) curves rely on a binary classification of the relevance. Precision and recall are often combined in the well-known F-Measure, also $F_1$-score or F-score. The standard F-Measure is the harmonic mean of precision and recall:

$$\text{F-Measure} = 2 \cdot \frac{precision \cdot recall}{precision + recall} \qquad (2\text{-}27)$$

### 2.7.3 Rank accuracy metrics

Rank accuracy metrics are more appropriate to evaluate ranked lists of items. The Normalized Distance-based Performance Measure (NDPM) was used in Fab (Balabanov and Shoham, 1997) .

$$NDPM = \frac{2C^- + C^u}{2C^i} \qquad (2\text{-}28)$$

with :

$C^-$: number of contradictory preference relations between the system ranking and the user ranking. A contradiction happens when the system says that the item **i** will be preferred to the item **j** whereas the user ranking says the opposite.

$C^u$: number of compatible preference relations, where the user rates the item **i** higher than the item **j**, but the system ranks at the same level **i** and **j**

$C^i$: number of preferred relationships of the user: the number of pairs of rated items **(i, j)** for which the user gives a higher rating for an item than for the other.

"NDPM does not penalize the system for system ordering when the user ranks are tied" (Herlocker and al, 2004): the NDPM measure can be used for binary ratings.

### 2.7.4 General protocols

**Offline evaluation**

The general protocol to evaluate recommender systems offline is the traditional Machine Learning's Train/Test dataset. This requires an initial dataset of logs of usages, generally logs of ratings from one of the available public datasets. For instance, MovieLens' datatset, and Netflix dataset are well known and widely used datasets. This dataset of logs can be represented as a table of 4 entries:

| User ID | Item ID | Rating | Date |
|---|---|---|---|
| 1 | 3 | 5 | 2005-06-01 |
| 1 | 19 | 3 | 2005-06-01 |
| 2 | 983 | 4 | 2005-06-03 |
| 2 | 252 | 1 | 2005-06-03 |
| ... | ... | ... | ... |

**Figure 2-4: Classical logs entries**

By random sampling, a proportion of the logs (in general, 80%, or 90%°) are extracted to build the Train Set. The remaining logs (20% or 10%) are kept as the Test Set. The model of the recommender system is built using only the logs of the Train Set. Then the model is evaluated on a predictive task.

If the predictive task is a rating prediction task, the protocol is well normalized in the literature. The Test Set is used to ask to the model the predicted rating of each couple (user ID, item ID) of the Test Set. Then the predicted ratings and the right ratings are compared.

If the predictive task is a recommendation task or a ranking task, the protocols are not normalized and several approaches have been carried on using generally precision/recall (Zanadi and Capra, 2008), or recall exclusively (Pizzato et al., 2010), (Cremonisi, 2010).

The offline protocol is adapted to collaborative filtering but can be used to evaluate strictly content-based recommender, under several conditions:
- the catalog of the item is joinable with the item of the logs: this can be done only with to a second table giving at least some information for each item ID, at least a title or a textual description. Fortunately in the case of MovieLens and Netflix dataset, the title and the year of release of the items are provided,
- the content-based model must be trained on the catalog's data,
- only the information of one user at one time must be used by the content-based model during the learning process, to avoid collaborative effect

**Online evaluation**

Online evaluation rely on some global indicators giving the value of the recommendations. The most used indicators in e-commerce applications are the click rates and the transformation rates, with is the rate of recommendations leading to a consumption action (purchasing, watching...): see for instance (Davidson et al., 2010), (Cremonisi, 2010), (Das et al., 2007).

The difficulties with the on-line evaluation is to be able to measure the impact of the recommender and only this impact. To neutralize other effect, an A/B testing protocol is necessary. An A/B testing protocol is a method comparing a baseline control sample A used as reference with another test sample B which differs on only one parameter. A classic way to analyze a recommender service on a website by A/B testing is to design one page containing a placebo recommendation box for a sample A of users, and one page containing the real recommendation box for a sample B of users, and to measure the performance impact between the 2 samples. Unfortunately A/B testing protocol is expensive to design and seldom used for recommender evaluation. See (Davidson et al., 2010), (Das et al., 2007) for true A/B testing protocol and (Chen et al., 2009) for a similar approach.

### 2.7.5 Discussion

(Herlocker et al., 2004) have noted that beyond the importance of the ability to predict scores, other criteria are crucial to capture the usefulness and quality of a recommendation system. The scalability is a very important point. Coverage which is the proportion of items that can actually be recommended must also be taken into account. Finally, the system's ability to give a confidence index (Basu et al., 1998) of its recommendation and to explain the recommendation will have a strong impact on perceived quality of the system (Herlocker et al. 2000; Bilgic, 2004).

The offline methods of evaluation have an important limitation: they cannot validate an innovative recommendation, not corresponding to a recorded log in the Test Set.

Another problem with off-line test protocols commonly used is that they assume that the log information is "missing at random" (Marlin and Zemel, 2009). That is, the logs used in learning and in tests were randomly drawn from all possible items and users. This assumption is obviously not tenable as there are items popular heavy users and they weigh more in the other logs. Assessments in RMSE, accuracy, etc. will therefore take more account of the heavy users and popular items. On the other hand the requests made to a system of recommendation are not random: it is not unreasonable to assume that heavy users are big users of the recommendation. But one would like the recommendation system able to value items that are not popular...

A last problem with offline methods of evaluation is that there is no standardized protocol: some people use Precision with a specific definition of the relevancy, some other people use Precision with another definition of relevancy, and many people use only RMSE assuming that decreasing RMSE always lead to better recommendations... a fact never clearly proved and that we will study the reality in Chapter 4.

Of course on-line assessment is always preferable. Industrial actors prefer operational metrics such as audience and purchases: for example Google scans the Click Through Rate (CTR), with an A/B testing framework, for its applications.

On the other hand off-line methods are very useful to design and calibrate recommender algorithms. They are cheaper than A/B testing campaigns, reproducible, flexible and not market dependant. Offline methods have still an important interest before launching a new recommender systems. Moreover we believe that there is a need of standardization and of improvements of a off-line protocol clearly defined and we will address this issue in the last Chapter of this thesis.

# 2.8 Overview of representative recommender systems

This section presents some representative recommender systems, collaborative, thematic and hybrid. The dates in parentheses refer to the years of publication of reference articles, not necessarily the date of implementation in a case of an operational system.

## 2.8.1 Tapestry: the precursor (1992)

Tapestry, is often presented as the first recommendation system to be called collaborative. It is based on the work of Goldberg (Goldberg, 1992). Tapestry allowed a flow of electronic documents to be recommended, to a user group at Xerox PARC. It was actually a hybrid system in the sense that access to documents may be made by the content of the documents but also on the basis of annotations made by other users on these same documents.

The system consisted of several modules including: a database of documents and annotations on these documents, a system to read, navigate and annotate documents, the filtering system based on a query language similar to SQL. The requests and the associated language were a major limitation of the system which made it difficult to use. Users had to create queries to retrieve documents. These queries were based on documents or annotations. The requests were then stored and periodically re-executed. It was possible to formulate queries like "give me messages talking about New York and annotated at least 20 times in the past 2 weeks".

More than the collaborative aspect, it is the hybrid aspect of Tapestry (access to content data and usage data on their contents) which was the most significant. For the collaborative feature, in fact, the system did not offer an automatic process to analyze the usages performed: users had to explicitly query the uses of other users. It is more a precursor of the automatic recommenders than an automatic recommender...

## 2.8.2 GroupLens (1994): the collaborative filtering approach

GroupLens made recommendations of newsgroup messages and was developed at the University of Minnesota (Resnick, 1994). GroupLens worked on a principle similar to Tapestry: a group of users annotate documents, in a simplified way:
First, annotations have been reduced to a rating of interest for each document.
Second, the query language, the main difficulty of Tapestry, was replaced by an automatic score prediction feature, calculated from the correlation of ratings from users.

GroupLens was evaluated for 7 weeks with 250 users, with only a qualitative report, in (Miller, 1997).

GroupLens validated the principle of collaborative filtering on a large scale, but also confirmed its weaknesses such as the cold start problem (Miller, 1997): the system was not capable of making good predictions before a certain critical mass of users. This could deter early users to become involved in the system.

GroupLens was the first article introducing the term "collaborative filtering". The GroupLens system was based on a method of K-nearest neighbors on users. The Pearson coefficient was used as the measure of similarity between the users.

GroupLens is often considered as the founder of the collaborative filtering approach.

## 2.8.3 MovieLens (1997): a reference website and a reference database

MovieLens and MovieLens Unplugged are recommender systems from GroupLens, a research team from the Department of Computer Science of the University of Minnesota, currently still very active in the field of recommendation systems. The team has applied its collaborative filtering techniques to movie recommendation. MovieLens is a Web service of movie recommendation for scientific research. It is based on the oldest works of GroupLens (since 1994). We can register on this site via a pseudonym. We are then invited to rate a number of films from a broad catalog. MovieLens asked, for example to rate at least 20 movies before making recommendations.

MovieLens' logs of ratings are publicly available on Internet. Long before the 2006-2009 Netflix's challenge, the MovieLens database was the main source of experimental data to study the algorithms for collaborative filtering.

For users of the website, an embedded version on a mobile terminal (PDA) was created (Miller et al., 2003). The version was called "unplugged" though in fact it connects from time to time to a central server. Data synchronization was occasionally made, and the system remained a central collaborative filtering system. Today the advantages seem anecdotal, and it essentially allowed a user to choose a film on the move, at the entrance of a movie theater for instance (in 2003 the Internet 3G was not yet deployed).

MovieLens is still available at: http://movielens.umn.edu/.

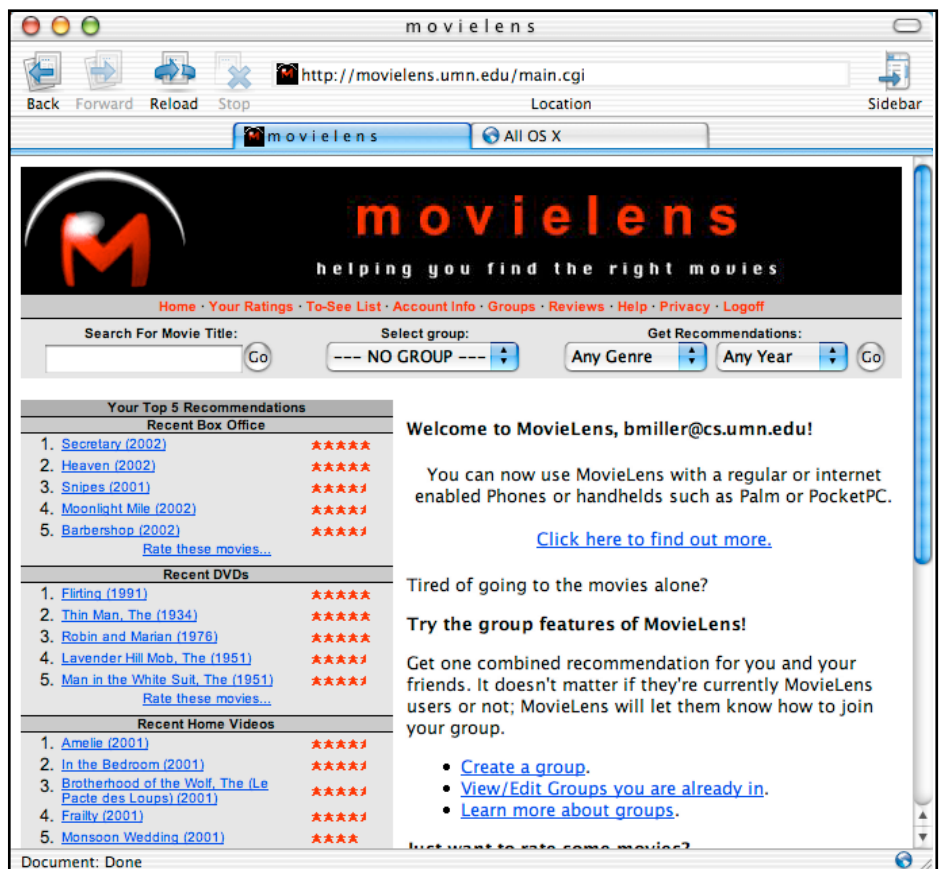

**Figure 2-5 : MovieLens' website interface**

## 2.8.4 Fab (1997): an hybrid architecture

Fab is a hybrid recommendation system for web content. Fab was created at Stanford University (Balanalovic et al., 1997a, 1997b, 1997c). It was based on an original architecture. Fab authors have defined precisely what they meant by collaborative filtering on the one hand and by content-based filtering on the other hand: in Fab, a document is recommended to a user **u** either because it corresponds to the profile of user **u** (content-based filtering) or because it has been appreciated by a user who has a similar profile to **u** (collaborative filtering).

Fab was composed of a set of software agents and a central router. Each agent had a profile based on the words in the web pages that have been rated. Two types of software agents works in Fab: collection agents, and selection agents. The collection agents search for web pages on a specific topic, for many users. The selection agents search for web pages on many topics, for a specific user. Each agent maintains a profile, based on the words contained in the rated web pages. Users are asked to rate the pages received on a 7-point scale. The ratings are used as feedback to adapt the selection agents (personal) and the collection agents (collaborative). The collection agents send the documents they find to the central router, which routes the document to the users with profiles that match, according to a certain threshold. The selection agents send documents well-rated by their own user, to similar users.

The collaborative aspect is managed at two levels: a content-based recommendation for user groups is carried out by the collection agents, and a purely collaborative recommendation is made by the selection agents. The population of collection agents is dynamic and adaptive: collection agents who bring unseen pages are deleted whereas the best collection agents are duplicated, according to a principle similar to genetic algorithms.

Fab uses the vector space model (Salton and McGill, 1983) as the representation of items. The interest of a user profile **u** for a document **d** is given by the scalar product **u.d**. The adaptation of the user profile **u** according to its rating for a web page **w** is given by **u'=u+r.w**, the rating **r** being centered on 0. This process is close to the classical Rocchio's algorithm in Information Retrieval (Rocchio, 1971). Finally, the modeling of the evolving nature of each user is performed by a simple weighting of the profile function of time: each night, each user profile is multiplied by 0.97.

The system was evaluated by 11 users. The system seems to behave as specified: the collection agents specialize in 1 or 2 topics for 1 or a few users with this topic in common. The experiment compared different sources of web pages: random, hand-selected pages, hand-selected pages in a top N of the day, and web pages recommended by Fab. The measure was a ranking of each user for the proposed web pages. The experiment demonstrated that Fab gave the best source.

The multi-agent hybrid architecture is the main originality of Fab.

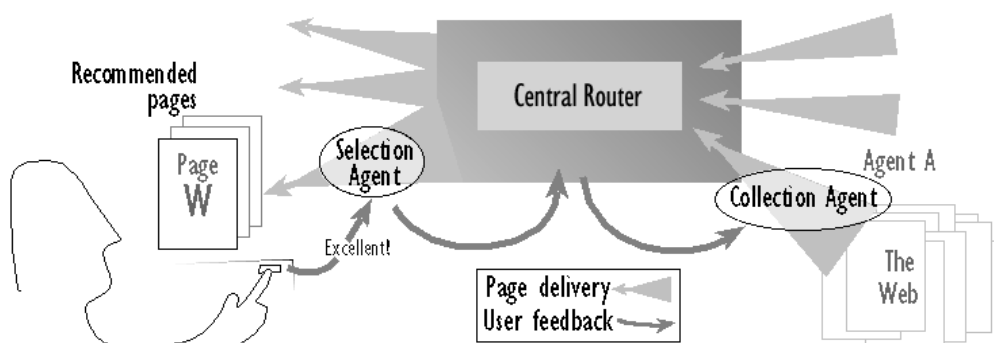

**Figure 2-6: Fab's general principle**

## 2.8.5 DailyLearner (2000): news recommendations

DailyLearner (Billsus and Pazzani, 2000) is a news recommender service. It is based on a centralized content-based recommender system called Adaptive Information Server (AIS), also developed by the authors. DailyLearner was notably deployed and tested on a PC via a Web interface, and on a mobile terminal (in 2000, the Palm VII$^{TM}$).

Web users can, after reading a news article, send feedback to the system using 3 classes: "interesting, "not interesting" and "known". Moreover, if a user requests more information on news, this news is classified as "interesting" automatically. For users of the mobile terminal, only implicit feedback is taken into account, giving a score of 80% to selected news. The news is often written on many pages and reading a news item entirely (i.e. to request all its pages) increases the implicit rating given to this news.

Internally AIS uses a score between 0 and 1 to represent the interests of users for the documents. Documents classified as already known are stored in order not to recommend them.

AIS uses two models to represent a user, a short-term model and a long-term model. The short-term model is based on a document Vector-Space representation with weights of keywords calculated using a TF-IDF principle. A K-nearest neighbor algorithm between items determines if a new document is likely to interest the user. A threshold can also determine if two documents are identical, thus classifying some documents referring to the same news as "already known" and therefore not to recommend.

If the new current document does not have sufficiently similar neighbors, the system switches to a long-term model of the user. On the long-term model, a selection of informative words (feature selection) is performed, still using a TF-IDF technique but on a much larger history of news. This selection is then used for all users. A naive Bayesian classifier is used to predict the interest of a user for a document, based on its long-term profile.

A third model, a hybrid model, uses a cascade of short-term and long-term models. If the short-term model cannot classify a document, then the document is classified by the long term model. If none of the models can classify a document, the document receives a median default score: it will not appear at the top of the list of recommendation, but will still be recommended.

The system was evaluated on real users for a web version for PC terminals (150 users) and a mobile version for mobile terminals (185 users), with the recall and precision measures. The system on the web (with an explicit feedback) proved to be more accurate than the mobile system.

The Precision (number of recommended documents actually considered interesting) for the top 4 of the recommended list was:
- 73% for the hybrid model on the web (recall 55%),
- Only 32% for the hybrid model on the mobile terminal (recall 29%).

Overall, the short-term model (KNN-items) works slightly better than the long-term model (Bayesian) and the hybrid model works slightly better than the two models separately.

The sort-term / long term user modeling, and the deployment of a recommender service on a mobile terminal, with some ergonomic constraints, are the two main innovation of the DailyLearner recommender.

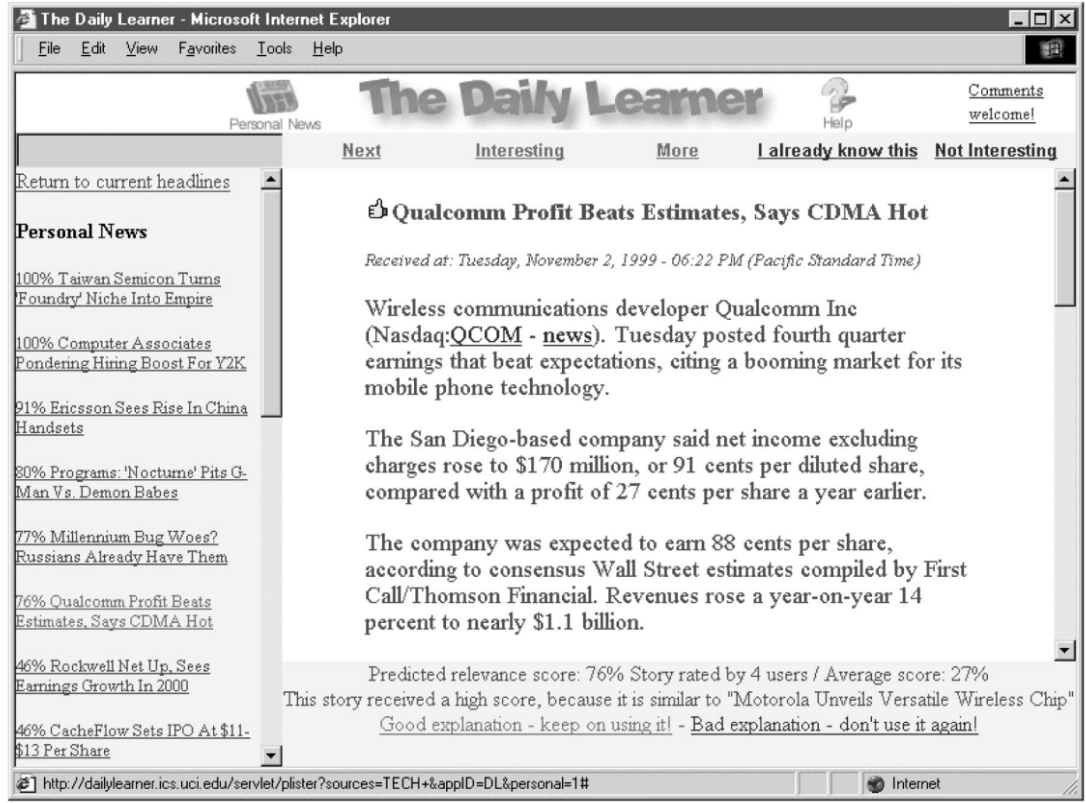

Figure 2-7: Web user interface of the daily learner

## 2.8.6 Amazon.com (2003): scalable item-to-item collaborative filtering recommendation

Amazon.com is a popular e-commerce web site that offers many cultural items (books, DVDs, CDs) and technology items (computers, cameras...). The site is available in several languages including French (www.Amazon.fr).

Amazon is a typical example of successful recommendation engine's technology. This technology is the basis of the marketing strategy of the website. The main function used is based collaborative item-to-item contextual recommendations, this was introduced very early on the site (late 90's). It is based on the logs of purchases and corresponds to the calculation of a similarity matrix of items with an optimized algorithm to scale with the imposing volumes handled by Amazon (Linden, 2003). Amazon popularized the famous feature "people who bought this item also bought these items".

An example of the interface provided by Amazon for item-to-item contextual recommendation is given Figure 2-8.

Amazon may now be classified as a hybrid system as a recommendation features based on some metadata (genre, author) also works in push mode with personalized web page and e-mailing.

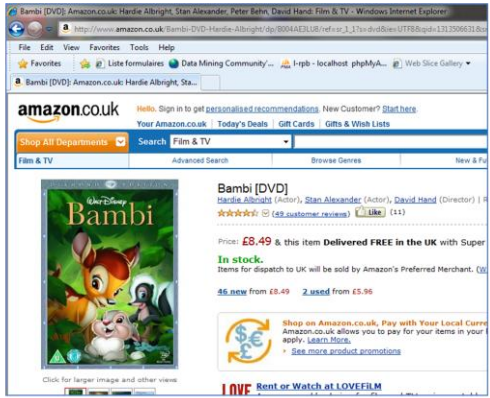
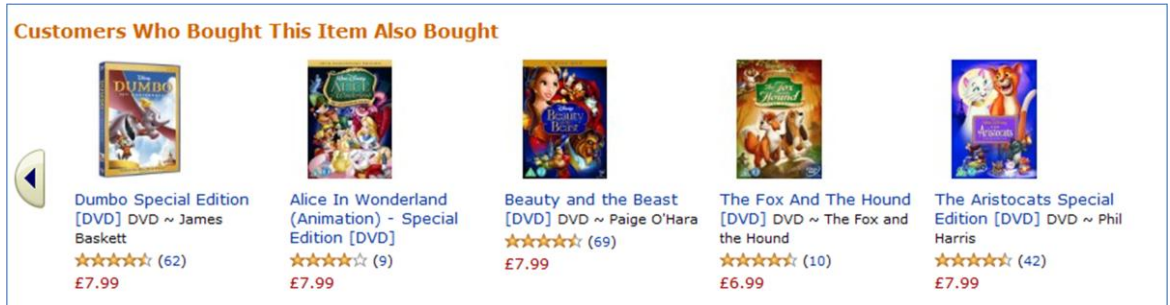

**Figure 2-8 : Amazon.com's item-to-item contextual recommendation to anonymous user**

Amazon's main innovations are:
- to have designed a high scalability recommender system,
- and to have built its entire marketing strategy on a recommender engine, intelligently integrating the recommendation functions into its website.

### 2.8.7 MORE (2006): hybrid recommendation by switch

MORE (MOvie REcommender: Caravelas and Lekakos, 2006) is a movie recommendation system prototype, using a switch-based hybridization. The switch mode between a collaborative and thematic system uses the size of the user's profile. MORE was only evaluated by simulation. It was evaluated on the MovieLens database (1 million ratings, ~6000 users, ~3700 movies).

An acquisition of thematic information was made on the IMDd database (International Movie Database, www.imdb.com): the information of genre, actors, director and keywords were joined to the titles of movies.

MORE uses a K-nearest neighbor model, applied on users in the collaborative system, or applied on items in the content-based system. In the case of the collaborative system, the chosen similarity function is the Pearson coefficient applied to the vector of ratings of users, and users can be considered neighbors of a target user **u** if and only if their similarity with **u** is positive. In the case of the thematic system, the similarity between items is the cosine of the vectors of characteristics of the items.

Two types of hybridization were tested. The system works in a collaborative mode by default because this method is considered the most efficient. In the first type of hybridization, called the "substitute" hybrid method, the system switches to the thematic filtering when a user has less than **K=5** reliable neighbors (according to a similarity threshold). In the second type of hybridization, called " switching" hybrid method, the system switches to the thematic filtering if a user has less than **S=40** ratings (threshold chosen empirically).

The authors have not done extensive testing with different parameters **K** and **S**, but noted that with the tested values the substitute hybrid method increased both the predictive performance (MAE) and overall coverage. Coverage is the percentage of items in the catalog that can be analyzed to return a score.

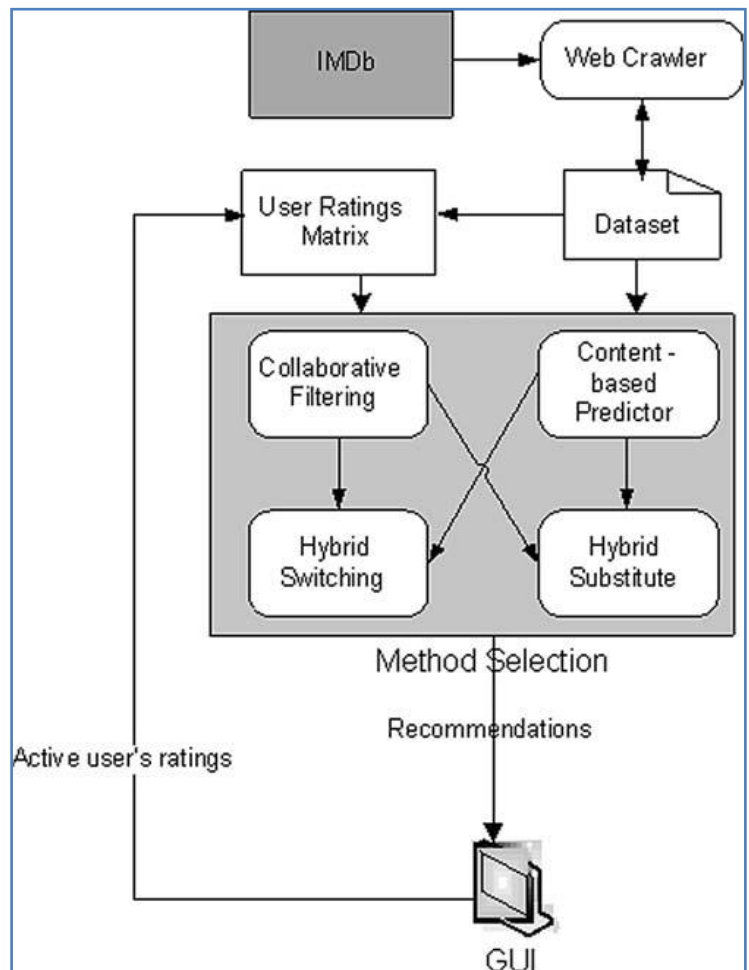
**Figure 2-9: architecture of MORE**

The substitute and switching hybridization modes are the main particularities of MORE.

## 2.8.8 CinemaScreen (2006): Hybrid recommender in cascade-mode

CinemaScreen (Salter, 2006) is a movie recommendation system, using a cascade-type hybridization. In CinemaScreen, a first collaborative process preselected a list **P** of movies potentially useful for the user. This list is built from users similar to the current user, and each item is rated by the collaborative algorithm. Then a second process, content-based, performs a ranking of the list **P**. The content-based algorithm ranks taking into account types, actors... It uses a loop approach to re-score each movie: firstly, the current rating of the movie is used to score its characteristics, such as Director, Genre, Actors. Performed on the entire list **P** of available movies, this process allows CinemaScreen to assign an average score for each actor, genre,... occurring several times in the list for the user. Then the score of each feature is used to refine the score for each movie for the user.

CinemaScreen has a software agent collecting cinema listings. The system has a pre-filter function allowing the user to restrict the recommendations of movies to those released near his home.

CinemaScreen was evaluated on the MovieLens database (1 million ratings version), mainly on the criteria of accuracy and coverage. Coverage is an important criterion because the system must be able to recommend or at least to score films present in short lists of cinemas near users.

The authors showed that hybridization could greatly increase the overall coverage of the recommendation, but without significantly increasing the accuracy of the system. CinemaScreen offers a complementary approach to that of More, using cascade hybridization instead of switching.

### 2.8.9 AIMED (2007)

AIMED is a hybrid recommendation model of TV programs proposed by (Hsu et al., 2007). Recommendations are made based on user settings as Activities, Interests, Moods, Experiences and Demographics information, hence the acronym Aimed.

The authors suggest the addition of a remote control allowing users to define their mood, to a recommender system.

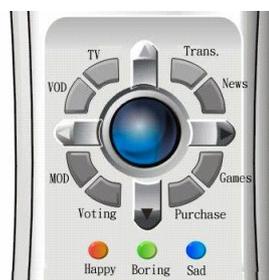

**Figure 2-10: the remote control proposed in the AIMED system**

The system remains at the proposal stage: the AIMED's model has been evaluated only by an experiment based on a simulation.

The experimentation has 109 users aged from 15 to 65. Questionnaires allow each user to define their social styles compounds of their activities, interests, age, education level and salary. This information is stored on the Set Up Box (STB TV installed in every home, here simulated) and can be sent to the central server. The central server performs a hierarchical clustering to group users. Finally the metadata of programs viewed by each user are also stored. These data are analyzed by neural networks that perform the recommendations.

During the experiment, participants were asked to rate on an agenda the TV programs they watch, their ratings for these programs, and their mood. Then a neural network model is trained to predict the rating given by users to the programs. The authors then compared the model AIMED with a classic model that did not include information on mood and lifestyle.

Authors claimed that the mood and style of life are important parameters in selecting programs, although the test protocol is not clearly explained. The basic idea of AIMED, analyzing the impact of mood and style of life on the usages is interesting, but still to be done.

### 2.8.10 Google<sup>TM</sup> News (2007)

Google News is a news aggregator that gathers news items from thousands of sources (500 in French) and depicts them on a personalized web page.

The recommendation service uses an opt-in feature to allow recording of browsing history on Google's web sites for user profiling.

The recommendation system of Google News (Das et al., 2007) was developed in 2007. It concerns the implementation of a collaborative recommendation service of news articles in an area where items change very quickly, with high scalability constraints: millions of users, millions of items. Compared to Amazon, which also has strong volume constraints, this system is characterized by very high churn rate of its catalog of items: items are news, with a constant flow. The latest news is often more interesting, so the system must regenerate the recommendation model often, which increases the problem of scalability.

Google News uses only implicit feedback to build a profile of a user: clicks on subjects. This cannot provide any information of negative feedback, and the positive feedback is noisy: one does not only click on interesting information...

The system is purely collaborative and handles only news IDs and users'IDs, without worrying about semantics. A user is represented by the news list on which she clicked.

The principle used by Google is a mix of similarity-based method and model-based method.

The similarity-based method uses only the counts of co-visitation: a co-visitation is a joint event where 2 items are clicked by the same user. The model-based method uses two clustering techniques, one Probalistic: Latent Semantic Indexing (PLSI: Hofmann, 2004) and the other MinHash (Cohen, 1997) (Cohen et al., 2001). PLSI is a probabilistic method of factorial representation of documents and their terms: here the terms are replaced by the userIDs. MinHash is a probabilistic clustering method based on the concatenation of **p** random signatures extracted from objects to group: objects here are users, random signatures are extracted from their profile. As MinHash is probabilistic, it is performed **q** times to denoise the final result of the clustering.

A parallelization framework of MapReduce type (Dean and Ghemawat, 2004) is used to accelerate the calculation of both PLSI and MinHash.

The system provides an overall score by a weighted hybrid method (Burke, 2007). The overall score of a news item **i** is given by:
$\sum_a w_a r_i^a$ , with $w_a$ the weight of the algorithm **a** and $r_i^a$ the score given by the algorithm **a** to the news **i**. The weights are learned using a predefined discrete parameter space exploration.

The system is used as a ranker: for a list of news to be recommended, the system returns a list scored and sorted.

The evaluation of the recommendation system of Google News has been done both offline and online.

The offline evaluation used a MovieLens database (1 million rating) and 2 databases from the logs of Google news, one small and one large (500 000 users, 190 000 items). The protocol was a 80% / 20% split between the training set and the test set, a model building phase on the training set, and the measures of recall and precision on the test set. PLSI always gave the best results, followed by MinHash then by the co-visitation method.

The online assessment was conducted over 6 months. A list of news items containing the recommendations of the different methods, and mixed randomly to avoid bias was used as the protocol. The performance measure was the number of clicks. A reference method such as "popular stories" was also integrated into the device. Personalized methods (combining co-visitation MinHash and PLSI algorithms) had 38% higher performance than the simple presentation of the "Top 10 items".

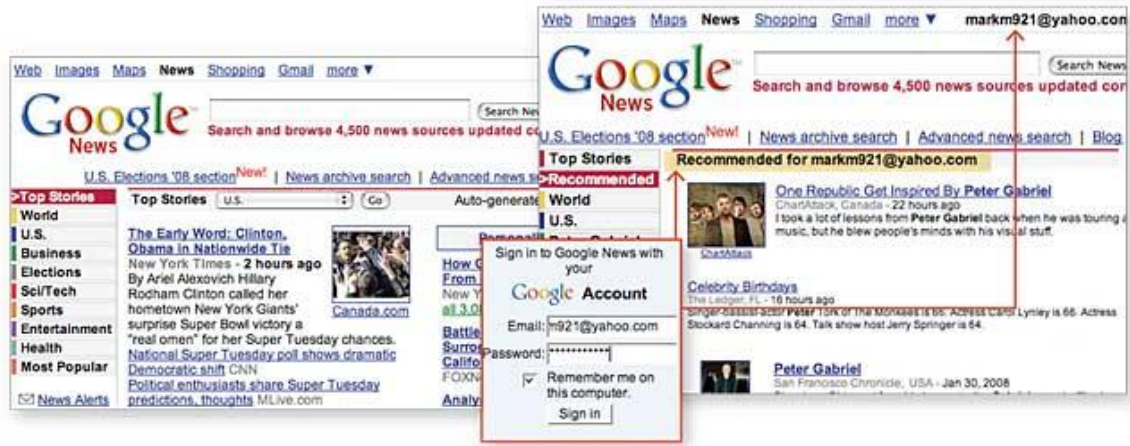

Figure 2-11: the Google™ news' recommendation interface

The high scalability of the technical solution and its adaptation to a catalog with a very high update rate are the main particularities of this recommender system.

## 2.8.11 Hybrid Music Recommender using polyphonic timbre analysis (2008)

The Hybrid Music Recommender of (Yoshii et al., 2008) is a collaborative and content-based recommender system dedicated to rank musical pieces.

The system uses a Bayesian model, referred to a 3-way-aspect model in the paper. This model is equivalent to a Probabilistic Latent Semantic Analysis (pLSA) (Hofmann, 1999). This hybrid music recommender uses 2 matrices of data:
- A collaborative matrix of user × ratings **R(u,m)** where **u** is a user, **m** a piece music a $r_{u,m}$ the rating of the user u for the piece of music m
- A content-based matrix of music × polyphonic timbres, **C(m,t)** where **m** is a piece of music, **t** a polyphonic timbre and C(m,t) the weight of polyphonic timbre in piece **m**.

A user profile u is modeled by a set of probabilities **p(u,$z_i$)** where $z_i$ is a conceptual genre. A set of conceptual genres is given by **{$z1,…,z_k$}** where **k** is the number of conceptual genres; **k** is a parameter of the model. The system learns the probabilities **p(u/z)**, **p(z/m)**... using the pLSA framework (not more detailed here). This model integrates the content-based part and the collaborative part of the system, so the hybridization is a model-integrated hybridization according to our previously proposed classification.

The goals of the authors in designing their system were: high accuracy, rich diversity of recommended artists, ability of recommending no rated pieces of music (dealing with the cold-start / new item problem), and prompt responses. The main recommender task is to rank all the nonrated data for each user to be able to recommend the musical pieces.

The Content-based part uses a bag-of-timbres which is analog to a bag-of words if we compare with text-document applications. A bag-of-timbres is a vector of weights where each component corresponds to polyphonic timbres instead of only one. The authors used T=64 polyphonic timbres to represent any music piece. Comparing with mono-timbre set decomposition this approach originally due to (Aucouturier et al., 2005 ) has many advantages:
- it does not require separating the instrument parts, which is a difficult task,
- it extracts the texture of musical pieces,
- it is based on a widely used and efficient signal processing method, the Mel-Frequency Cepstral Coefficients (MFCCs).

The system was evaluated offline, on ratings logs from the Amazon's Japanese website, Amazon.co.jp. The rating scale is 1 to 4. The values 3 and 4 are assumed good ratings. The authors used a Learning set / Test set protocol. The task was ranking and the measure was the accuracy defined as the percentage of good recommendations in the top-N ranked pieces of music of each user in the Test Set. A good recommendation is defined as a piece rated 3 or 4 in the Test Set. This is more a kind of precision measure:
- For N=1, that is to say 1 recommended item per user, the precision is 93.5%.
- For N=10, that is to say 10 recommended items per user, the precision is 80.7%.

The authors also evaluated an incremental training method which makes it possible to partially update the model, losing a maximum 5% in accuracy. More test would be necessary as the number of ratings used was low: 2.19% of a matrix of 326 users × 358 music pieces, and 10% of that in Test, so less than 256.

The main innovations of this hybrid recommender system are:
- the music-specific content based analysis of polyphonic timbres,
- the use of the pLSA framework to combine the two data sources.

## 2.8.12 Social Ranking: user's tags to customize the query results (2008)

The recommendation method of (Zanadi and Capra, 2008) is based on users' tags to improve search results. The database used for the experiments is CiteULike, a site of bookmark recommendation between researchers. Like other shareable web catalogs such as Del.icious (web pages), Flickr (photographs) and LastFM (music tracks), CiteUlike allows users to organize their shared documents by giving them tags.

Tags pose several special problems compared to conventional keywords managed centrally: their free employment causes an increase in the number of synonyms, homonyms and heterogeneous terms, which suggests that they decrease the quality of the search, rather than increasing it. On the contrary, the authors show that user's tags search, using recommendation, is possible and improves the coverage of queries.

A collaborative database using tags has 3 dimensions: users, items, and tags. The social ranking uses 2 matrices derived from this 3-entry-matrix: a [user × tag] matrix, and a [tag × document] matrix.

These two matrices can be used to build similarity matrices, one user-user, the other tag-tag. In the 2 cases, the similarity measure used is the cosine.

The user query is modeled by the list of tags she uses for her request.
The personalized answer is built in 2 steps:
- in the first step, one performs a query expansion by adding, for each tag from the query, **k** neighboring tags. For this the tag-tag similarity matrix is used.
- In the second step, a ranking of the query results is made: all items containing at least one tag of the query are returned to the user.

For a query **q = {t$_1$,..., t$_n$}**, of a target user **u** with **n** tags the ranking formula used for each document **d** returned was:

$$R(d) = \sum_{v_i \text{ who tagged } d} \left( \sum_{\{t_x | v_i \text{ tagged } p \text{ with } t_x\}, t_j \epsilon q} sim(t_x, t_j) \right) \times (sim(u, v_i) + 1)$$

For each document **d**, the formula combines the similarities of each tag of **d** with the tags of the target request. These tag-based similarities are weighted by the similarities of the users (who have given these tags) with **u**.

The authors used an extraction of CiteUlike's database of 2007, where the tags and the documents referenced only once have been deleted. The database obtained was about 100,000 documents, 55,000 tags and 28,000 users.

Experimentation of this personalized ranking method is done offline by analyzing the effects of the levels of similarity between uses and the levels of similarities between tags. The study divides also the cases where the user's tagging rate is high (users who tagged more than 50 papers), medium (between 10 and 50 papers), low (less than 10 papers), and if the document is popular (more than 5 tags) or not (less than 5 tags), leading to 3×2=6 scenarios.

The test protocol uses a document **d**, tagged by a user **u** with tags $\{t_1, t_2, ... t_n\}$. The user **u** is interested in the document **d** since she tagged it. The tags given by **u** to **d** are removed from **d** and are used to construct a query of **u** to the system: the goal is to find **d** in the answer produced by the system with a good rank index.

In the first step the query expansion is deactivated. The authors evaluated only the effect of the similarity between users who tagged the target document **d** with the same tags than those of **u**. This first experiment shows that except for the case of well-known documents and users with many ratings, the ranking is improved. The accuracy is improved for all the long-tail cases.

In the second step the whole process with query expansion is performed: for each document, the top-**k** nearest tags of each tag of the document **d** are added to the query. Adding tags increases the coverage. With **k=10**, so 10 neighbor tags added, the percentage of documents not found in the documents in the long tail decreases from 40% (no query expansion) to 23% and from 17% to 6 % for well-known documents and users giving a lot of tags.

Among the documents previously not found and now selected by the query, 40% are found in the top 100 answers. The authors show that the accuracy does not decrease up to 20 tags added, while increasing the overall coverage of the 6 scenarios analyzed.

The authors show that using folksonomy tags to perform a custom query ranking on a search engine improves coverage without sacrificing accuracy.

## 2.8.13 Beehive (2009): recommending people on social networking websites

Beehive's system (Chen et al., 2009) deals with the recommendation of people in a social networking website. Four user-to-user recommendation algorithms were evaluated, distinguishing the case of recommending people already known, and recommending unknown people.

The recommendation system evaluated was deployed on the Beehive website at IBM on 3,000 people, with a survey of 500 people. Beehive is an online social networking website for business, with over 38,000 users. Like most social networking websites, Beehive has an individual profile page, a mechanism to invite new friends, a photo sharing system, events, etc. The concept of friend in Beehive is not necessarily bi-directional as in Facebook. In Beehive, if the user **u** is friend of the user **v** that means that **u** is connected to **v**'s page: there is no request for mutual acceptance: **v** is just informed that **u** is connected and **v** has an option to connect back to **u**.

The first algorithm tested, **Content Matching**, makes the user-to-user recommendation with a content-based principle: users are linked if they publish similar contents. Each user is represented by a "bag-of-words" using a similar approach to Information Retrieval. All contents posted by the user are used by Beehive: profile page, text associated with images, tags, messages. The words are "stemmatized" (put in a canonical form) and very common words are removed. The user is then represented by a vector of the words associated with weights using a weighting principle of TF-IDF. The similarity between two users is calculated by the cosine or their respective vector-based profile.

The second algorithm, **Content-Plus-Link** is a hybrid algorithm. It hybrids content information and social networking information. Content-Plus-Link algorithm is based on the previous Content matching algorithm. But instead of simply recommending users with content-based maximum similarity scores, it boosts the similarities obtained by the link information. The similarity between the user **u** and the user **v** is increased by 50% if there is a "valid social link" between **u** and **v**. A "valid social link" is defined as:
- A maximum length sequence to connect **u** to **v** in the social network. The maximum length is 4.
- The type of valid link between two vertices **a** and **b** of the social network graph: the selected valid links are the 3 following interactions: **a** is connected to **b**, **b** is connected to **a**, **a** had commented on content of **b**.

This type of algorithm reinforces close relationships in the social network. Individuals close by their published content but not connected through their relatives are less recommended.

The third algorithm, **Friend-of-Friend**, is purely a social network algorithm. It is based on the principle: if many of my friends consider **v** as a friend then **v** is likely to be a friend of mine. Facebook's feature "People You May Know" is based on this principle. Formally, a user **v** may be a candidate to be recommended to **u** if there exists a user **a**, called mutual friend, as **Friend (v, a)** and **Friend (a, u)**. The relationship **Friend(x, y)** means that **x** has declared **y** as a friend. All mutual friends between **u** and **v** are calculated. The number of mutual friends is considered as the score associated with **v** to recommend her to **u**.

The fourth algorithm, **Sonar**, is an algorithm that aggregates information from social bases internal to IBM: publication databases, blog websites, ... The details of this algorithm are given in (Guy et al., 2008).

The four algorithms were evaluated in two experiments.

The first experiment involved 500 users of whom 230 were actually usable. Users were selected on Beehive and needed to have enough information on their profile so one can make recommendations to them.

For each user three recommendations were proposed from each algorithm, so 12 recommendations in total (in a random order). For each recommendation, a series of questions were asked:
 - Do you already know this person? Is it a good recommendation? ...

Sonar and Friend-of-Friend algorithms have a level of "good perceived recommendations " greater than those of Content-Plus-Link and Content-Matching algorithms. However Sonar cannot easily find some new friends (previously unknown), while the Content-matching algorithm is the best for this task.

| Recommended users | **Content** | **CPlusL** | **Friends of Friends** | **Sonar** |
|---|---|---|---|---|
| Unknown user | 30.1% | 24.9% | 23.8% | 6.6% |
| Already known user | 19.5% | 31.8% | 55.4% | 75.9% |

**Table 2-1: Ratio of well perceived recommendations**

The second experiment involved 2400 users of Beehive, chosen with sufficient information in their profile. A recommendation display device on the desktop was deployed. This device showed one recommended person at a time, and provided information about that person and a button to add her to the friends list. Adding a recommended person as a friend was the criterion of the good recommendation classification.

The algorithms based on social networks have an acceptance rate of the recommendation more important than those based on content data, ranging from 59% for Sonar, 48% for Friend of Friend, 40% for Content-Plus-Link and 31% for content matching. However, the rate of acceptance of the recommendation for new friends, those who were not previously known, is greater with the algorithms based on the published content.

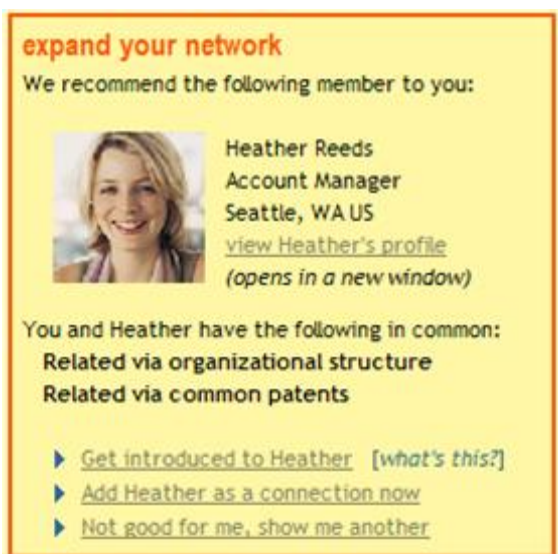

**Figure 2-12: The Beehive's widget**

For the task of recommending people to people, it is surprising that we lead to a conclusion contrary to the task of recommending item to people: the content-based recommenders using people's published information or public information provide better new recommendations than the social recommenders, that could be qualified "collaborative". This is because one often already knows some "friends of one's friends" whereas one often doesn't know many people who shares the same interests but who are not in one's first social circle.

## 2.8.14 Social Matching: Recon: a reciprocal recommender for online dating (2010)

Recon (Pizzato et al., 2010) is a reciprocal recommender system of people for an online dating website. Recommending people as a part of an online dating website shares some limitations with other systems such as job search services, or expertise in social networking search services:
- The success of the recommendation is determined by the two objects of the recommendation (on the contrary, by only one user in the case of user-item recommendation).
- Users do not stay long on this type of service, which makes the problems of cold-start more difficult to handle.
- However, users are generally willing to provide a detailed profile of who they are and what they are looking for to increase their chance of success.
- People have a very limited availability compared to the items of a catalog: each person can be recommended only a limited number of times and cannot accept too many recommendations.

- People can be proactive as in the case of conventional recommendation systems (they look for) or simply reactive (they meet).
- It is important, especially if the service is not free, that any person may be recommended.

Recon uses two components of profiling:
- The profile of the user characteristics **x** is given by a list of couples (attribute: value). Numeric values, such as age, are discretized, eg the 18-22 year old, etc. For example, a user's profile could be Alice (age:18-22, gender: Female, body: Slim,...).
- The profile of the preferences of **x** is given by the analysis of recipient users of the sent messages. Preferences of user **x** are represented by the counts, for each attribute **A**, and each of its possible value $A_v$, of the messages sent by **x** to the users with the $A_v$ characteristic. For instance if a user **x** has sent a lot of emails to users with a age: 30-35, age:35-40 and gender: Female then **x** will have an important weighting for the two segments of age (30-35 and 35-40) and for the gender: Female, in his profile of preferences.

Recon uses two phases to complete the recommendation between users:
- a compatibility index phase between users, the computed index being not symmetric: the compatibility of a user **y** to a user **x**'s preferences uses the comparison of frequency histograms of each attribute-value in the profile of **y** and each attribute-value of the preference's profile of **x**
- a phase of recommendation itself: a mutual compatibility score is computed: it is the harmonic mean of the scores of compatibility of **x** to **y** and **y** to **x**. Then these scores are ordered and for each user **x** the Top-N scores are presented.

For new users, a low but not zero compatibility function always returns 0.001. This makes it possible to recommend any user.

The implementation of Recon is performed using a list of recommendations via a subset of the users. Indeed calculating the compatibility scores of **N** users has a time complexity of **O(N²)**. The subset users' selection uses people who have the most frequent gender, an age in one standard deviation of the average age and a location in one standard deviation of the average location. The impact of these optimizations on the quality of the recommendation is not specified.

Recon was evaluated in batch on a Training Set of 90,000 users and 1.4 million messages, representing four weeks of usage on the dating website. The data from the following 2 weeks were used as Test Set. Two measurements are made during the 2 weeks of testing: the success rate, which is the number of positive messages related to recommended people sent back on the number of recommendations (the success rate is equivalent to the precision rate). Then the recall is the rate of the number of positive messages sent by a user **x** where the recipient was contained in the list of recommendations for **x**.

The authors then compared Recon with a system where no reciprocal score is calculated so with asymmetric recommendations. Pushing 5 recommendations per user, Recon gets 45% of success and about 4% of recall. The version without reciprocal score gets only 25% of success rate and less than 1% of recall. Measures made up to 100 recommendations per user shows a constant mutual domination of the reciprocal score. The average success rate of sending a message (invitation to a user **x** for a user **y**) is 17%.

Considering reciprocity in the recommendation of people increases the chances of success of a meeting from 26% to 45%. The recall is also improved. The performance of this system of linking people on website has to be confirmed on-line.

## 2.8.15 Recommender system for IP-TV : Fastweb's recommender (2010)

Fastweb (Bambini, 2010) is an internet TV service provider, the first service of this type to be deployed in Italy in 2001. The recommendation system was called ContentWise and was put into production in October 2008, on a Video On Demand (VOD) catalog, generating an average of 30,000 recommendations / day.

Fastweb provides VOD and Live TV recommendations to hundreds of thousands of users. Home users have a Set Top Box (STB), a decoder that allows authentication and interactivity. Fastweb recommends two types of items, from VOD and live TV programs.

Users are homes, as the STB does not identify a unique person. A principle based on the time slot, however, allows the system to make assumptions about the type of user: children during the afternoon, family early evening, and adults late at night.

Fastweb uses implicit feedback to acquire user profiles: a TV program watched partially generates a score proportional to the duration of viewing, and a purchased VOD generates an implicit rating of 4/5.

The algorithms included in the recommendation system are of three types: for TV, and VOD, the system uses a content-based algorithm with a program representation using Bag Of Words. Words are stemmatized, i.e. converted to their radical form (for example, "singer", "singing", "sing", "song", are converted into "song"). Then a Latent Semantic Analysis algorithm (Deerwester et al., 1990) is used. Exclusively for VOD, the system uses two collaborative algorithms, one based on Singular Value Decomposition, the other based on an item-item similarity matrix approach.

The system generates recommendations scoring the entire catalog and selecting items with the highest scores.

The system is evaluated both off-line and online with the VOD catalog on which it is deployed.

In offline, a Test set is created per user, based on actual logs of usages on the VOD catalog. For each user, a leave-one-out approach is used: a rated item is removed from the Test, then one asks the recommendation system to recommend 5 items (5 is the number of recommendations made in the normal service): If one of these 5 items matches the removed item, it is a success. The success rate on all items and all users of the file Test is a total measure of the recall.

The recall has a maximum of 19% for the item-item algorithm, 15.1% for the SVD (collaborative filtering) algorithm, 2.5% for the LSA algorithm (the content-based filtering algorithm). A simple Top10 method gives 12.2% of recall. All these measurements are made by removing 10% of most common items.

For the online evaluation, the notion of successful recommendation corresponds to the items that were actually seen after being recommended. The period of time after the recommendation is the key parameter of this evaluation. After 24 hours, the recommendations lead to almost 20% of the recommended item sold and 32% for long-tail. The protocol and the significance of this test are questionable, however: there was no A / B testing protocol for this test, so the VOD sales directly attributable to the recommendation system are not really measurable.

## 2.8.16 Youtube (2010)

Youtube (www.youtube.com) is a famous online video website community.

The recommendation system of YouTube (Davidson et al., 2010) provides contextual and personalized recommendations. The system makes recommendations but does not score the items:

this is a Top-N-Recommendation, if one chose a typology based on functionality. We restrict ourselves to the personalized Top-N recommendation.

The YouTube's data typically represents a high scalability issue: it has more than one billion videos, most without metadata, and users are multi-million daily.

YouTube's recommender system is based on association rules for short periods of navigation of each user (usually 24 hours). During this time, the system records co-visitation $c_{i,j}$ of each pair of videos ($v_i$, $v_j$) a user **u** (authenticated) watched. Note that the use of time windows makes it possible to scale with the data, as the co-visitations for longer periods would quickly become too numerous.

A score of "likeness", not necessarily symmetrical, and called "relatedness", between a reference video $v_i$ and another video $v_j$ is given by the formula:

$$r(v_i, v_j) = \frac{c_{i,j}}{f(v_i, v_j)}$$

where $c_{i,j}$ is the count of co-visitation between $v_i$ and $v_j$ and **f** a normalization function depending on each video. One can use $f(v_i,v_j)=c_i.c_j$ but other normalization functions are possible.

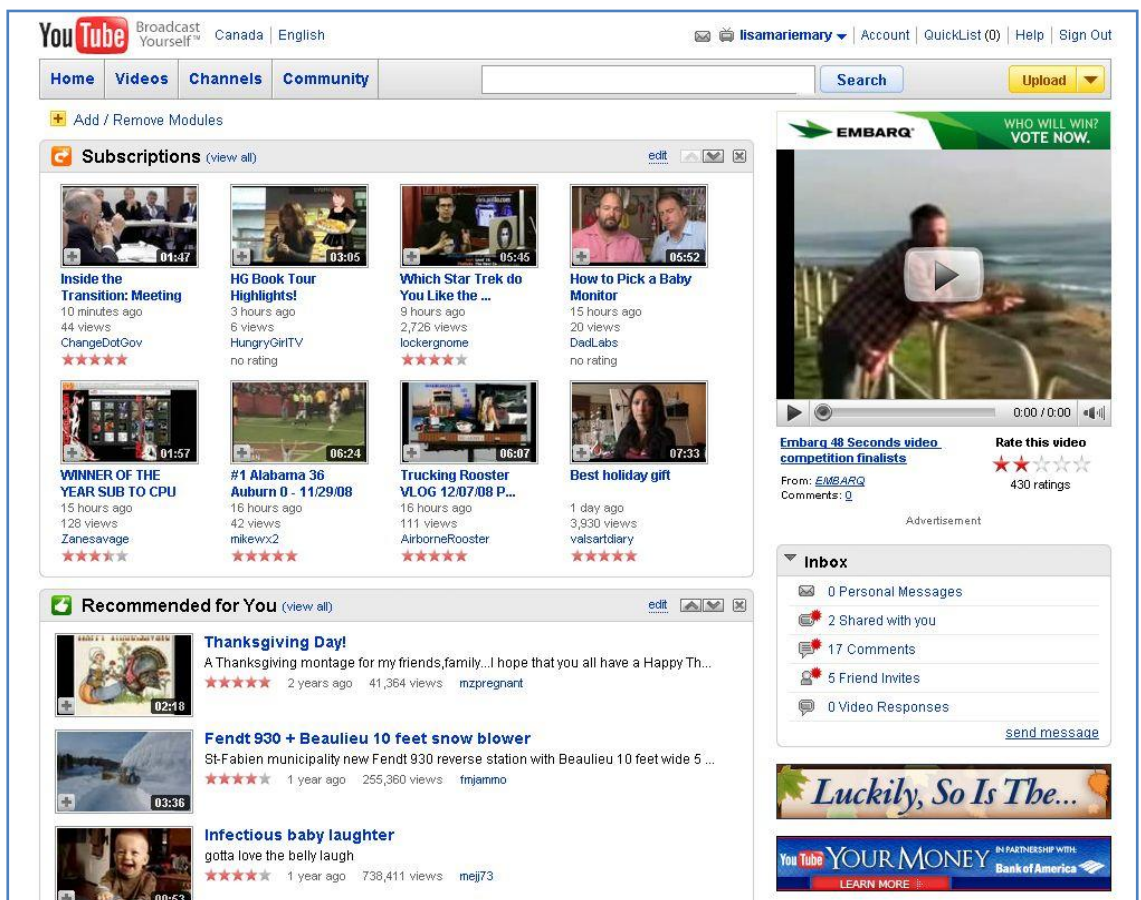

Figure 2-13: Personalized recommendation with Youtube

The personalized recommendations are made on the basis of a list **L** of seeds that are videos present in the user profile. The system then searches, taking as input the videos in **L**, and from the graph of association defined by $r(v_i, v_j)$, candidate items, ie video $v_j$ with good similarity to $v_i$. A post-filtering system avoids recommending items already known to the user. Then a post-ranking system takes into account three parameters: video quality (popularity), specificity of the user (preferences) and diversity (trying to add value to the long tail of the catalog) to order the final list of recommended items.

The system was evaluated by A/B testing. Automatic recommendation at YouTube is responsible for 60% of clicks of the home page. Within the existing modules of recommendation (most viewed videos, favourite videos, top rated videos), the personalized area gets 2 times more clicks than the 3 other non personalized modules. The personalized recommendation, therefore, generates about 24% of the audience of YouTube.

The very high scalabity on very sparse data, and the optimization techniques to address these issues, are the main particularities of YouTube's recommender.

## 2.8.17 Recommendation on Twitter: Twittomender (2010)

Twitter is a recent real time service on the Internet, enabling a community of users to post and follow short text messages based on SMS and limited to 140 characters. A message issued by a user on Twitter is called a tweet. Twitter is considered as a real time micro-blogging social network using personal textual content generation.

Users can explicitly designate the other users they wish to listen to tweets. If a user **u** decides to follow the tweets of a user **v**, **u** is said to be a follower and **v** is called a followee (being followed). The overwhelming simplicity of Twitter is its main strength but also its weakness: it is very difficult for users to find interesting tweets and followees.

The proposed recommendation system (Hannon et al., 2010) is called Twittomender and uses the Twitter API to access to user data. Twittomender recommends to users of Twitter other users to follow, therefore followees, based on their tweets and the social graph in Twitter. The system uses both content analysis of the tweets and collaborative analysis. Twittomender performs a profiling of each user taking into account his tweets, its followees and its followers.

The recommendation system is based on an open source search engine, Lucene, that indexes the tweets. Tweets, as users replace traditional documents in this framework. Tweets, as the profiles are represented by vectors of weight on terms via a classical TF-IDF principle. The terms can be words in the tweets (content-based analysis) or user IDs (collaborative analysis).

The principle of the recommender is: From a user profile, Twittomender selects the **N** most frequent words. These **N** terms are used in Lucene as a query which provides a list of users to follow.

7 profiling strategies were analyzed:
1. represent users by their tweets
2. represent users by their followees' tweets
3. represent users by their followers' tweets
4. represent users by their tweets and both their followers and followees
5. represent users by their followees' IDs
6. represent users by their followers' IDs
7. represent users by their followers'ID and their followees' ID

The first 4 strategies are content-based, the last 3 are collaborative.
Two methods of scoring, one using a combination of scores representations of strategy 1 and strategy 6, and the other based on an average rank on the 7 stragegies, were also evaluated.

A first evaluation was performed offline, based on accuracy. The accuracy was defined by the proportion of overlap between the number of recommended followees and the real followees of each individual user. The best strategy seems to be the representation 7 (collaborative profiling using both followers'ID and followees'ID). The collaborative-based representations work better than content-based ones. The maximum precision reached is 24%. A real assessment on 34 users, was done, where the magnitude of the maximum precision was confirmed, and therefore the relevance of the system.

The main innovations of Twittomender are:
- carrying out the recommendation in the context of social networks for instant communication (micro-blogging),
- using a search engine as support to the recommendation, for thematic and collaborative profiling, using document terms, itemIDs and userIDs at the same time.

## 2.8.18 Cobot (2011): the future of recommender systems?

Cobot (Sahay and Ram, 2011) is an agent-based domain-specific approach that uses socio-semantic information to provide recommendations. Cobot is connected to a conversational web interface where a community of users can talk and ask questions about health issues. Cobot dynamically recommends 3 types of information on a social website related to health issues: users, web resources such as snippets (short extracts from pages) and web links and live conversations.

Cobot has 3 components: a language understanding unit, a user modeling unit, a case based reasoning unit.

The language understanding unit has an intent classifier which analyzes the mesages of the users and selects an appropriate strategy among pre-defined conversational strategies such as: ask for more details, thank for suggestion/solution, suggest solution… Then Cobot analyzes conversations to extract concepts. It uses an ontology extractor on a medical corpus. These concepts are used to query search engines for related contents. The semantic tagging of Cobot uses Unified Medical Language System[3] (UMLS$^{TM}$) and Wordnet[4] (a lexical database for English at Princeton University) as language representation. The aim of the language understanding unit is to tag discussions and to generate queries associated to the context of the conversations. The user describes her need about a medical health issue in natural language on a discussion forum. The need is modeled via the analysis of the conversation generated by other users.

The user modeling unit uses the concepts extracted from the language understanding unit. Cobot learns users' profile by observing their participation to different conversations; different events are taken into account with the topics of the conversation: users can click and rate conversations, documents and users. A short-term profile and a long-term profile of the users are maintained. The window size of the short term profile is of few days. They are based on the frequency of users' interactions and users' topics.

Cobot proactively invites pertinent users to participate to a conversation: if they are connected, Cobot sends them an instant notification. Users developing a conversation also help Cobot to better model the topics to find good queries for relevant recommendations.

The case-based reasoning unit carries out the recommendations of web information using an external web search engine. The system stores the previous results for each previous query to the search engine. The system also stores the users' feedback to each result returned as a recommendation. For each new query the similar past results are ranked: those which have been previously selected more have a better rank index. Each case is represented by:
- A set a query terms,
- Web recommendations and hit counts,
- Conversation recommendations and hit counts,

---

[3] http://www.nlm.nih.gov/research/umls/
[4] http://wordnet.princeton.edu/

- Each new target problem, corresponding to a new query Q.

Cobot is still a new concept. It has some limitations: language understanding and modeling is still a complex task and the concept tagging is not perfect, so recommendations are not always effective, according to the authors. There is no quantitative evaluation of Cobot but the prototype has been accessible for a while online at www.cobothealth.com.

The main innovations of Cobot are:
- Recommending both users, conversations and web related information in an integrated online discussion forum in natural language,
- Using conversation graph as a part of the model of the users,
- Recommending by a query with case-base reasoning to help reuse previous queries with good social feedback.

## 2.9 Conclusion

This overview[5], although very small (sorry to many other interesting systems not listed) shows numerous approaches. We see that the main functions of recommender systems are not limited to the prediction of rating and its evaluation by RMSE. The rating prediction is rather the less used feature.

The recommendation is often coupled with a service of Search. The main functions covered by the recommender systems are:
- Help to Discover,
- Help to Explore.

For these functions, we see often the crucial importance of the notion of similarity between objects. This notion of similarity is well integrated in the methods KNN, which explains their popularity.

On the other hand type SVD methods are recognized faster and more accurate in rating prediction, useful for:
- Help to Decide,
- Help to Compare.

We will first study the performance of KNN models on classical standards, with a focus on the most powerful similarity measures, and an analysis of the cold-start problem.

Then we will study a new protocol to evaluate precisely the KNN models, and their main alternative, the fast matrix factorization techniques. This protocol will be designed on more industrial standard, detailed by evaluated tasks and objectives.

---

[5] The 3 next tables summarize this overview.

| System | Year and localization | Recommended objects | Particularities |
|---|---|---|---|
| 1. Tapestry | 1992<br>Xerox Parc<br>Palo Alto, USA | Messages / documents | 1st collaborative filtering system, still manual for querying |
| 2. GroupLens | 1994<br>GroupLens' research team, University of Minnesota, USA | NewsGroups | 1st collaborative filtering system, fully automatic ; 1st explicit reference to the term "collaborative filtering". |
| 3. MovieLens | 1997<br>GroupLens' research team, University of Minnesota, USA | Movies | Famous web recommender, log databases freely available for research purposes and widely used for benchmarks. |
| 4. Fab | 1997<br>University of Stanford, California, USA | Web Pages | Hybrid system, innovative multi-agent architecture. |
| 5. DailyLearner | 2000<br>University of California, USA | News articles | Short term and long term profiling adapted to news following. Embedded mode on a smart phone |
| 6. Amazon | 2003 (article), end of the 90's for the system. USA + worldwide localizations. | Cultural products: books, music CD, DVD, computers, cameras...<br><br>Very large Catalog. | Commercial recommender system which popularized the item-to-item recommendation feature « people who have bought this item also have bought these items ».<br>Very high scalability. |
| 7. More | 2006<br>University of Athens, Greece. | Movies | Switch Hybridization between a thematic engine and a collaborative engine, based on the size of the user's profile. |
| 8. CinemaScreen | 2006<br>Université of Surey, UK | Movies | Cascade hybridization collaborative and thematic. Recommendation of movie theaters close to the user. |
| 9. AIMED | 2007<br>University of Chia Tung, Taïwan | TV programs | Use of socio demographic profile and use of the user's humor to recommend TV programs (proof of concept). |
| 10. Google News | 2007<br>Google,<br>California, USA | Aggregated news articles (In France, from about 500 web sources) | Very high scalability.<br><br>Items with a high frequency of churn and update. |
| 11. Hybrid Music Recommender | 2008<br>Kyoto University, Japan | Music pieces | Hybrid music recommender with polyphonic timbre analysis. |
| 12. Social Ranking on CiteULike | 2008<br>College University, London, UK | Research communities' bookmarks with users' tags | Use of tag-tag similarities and user-user similarities to enhance the results of the request to a search engine. |

| System | Year and localization | Recommended objects | Particularities |
|---|---|---|---|
| 13. BeeHive | 2009<br>IBM Research Labs + University of Minnesota.<br>USA | People in a private social network (IBM) | User-to-user recommendations in a professional context. |
| 14. Recon | 2010.<br>University of Sidney.<br>Australia | People in an online dating website | User-to-user recommendations taking into account the asymmetry of user's preferences. |
| 15. Fastweb : ContentWise | 2010 (2008 for the deployment)<br>FastWeb + University of Milan.<br>Italia. | TV Programs and Video On Demand (VOD) | System deployed on an IPTV architecture, for VOD and TV programs. |
| 16. Youtube | 2010.<br>Google.<br>California, USA. | Videos uploaded by the users | Very high scalability on very sparse data. |
| 17. Twittomender | 2010<br>University of Dublin<br>Ireland. | Real-time messages on a micro-blogging social network | Use of a search engine as indexing framework to index both documents and IDs of users and items. |
| 18. Cobot | 2011<br>Georgia Institute of Technology<br>Atlanta, Georgia, USA | Web pages from web search engine, users, conversations in a domain-specific community (health). | Real time recommendation integrated within a conversational web interface, recommendation of conversation, recommendation of expert user. |

**Table 2-2. Summary of the overview**

| System | Recommended objects | Type of technique | Modeling |
|---|---|---|---|
| 1. Tapestry | Messages / documents | Language of request on an hybrid database containing both the items' descriptions and users' annotation on the items | No model used. Database of items and annotations + request language close to SQL. |
| 2. GroupLens | NewsGroups | Collaborative | K-Nearest-Neighbors user-user |
| 3. MovieLens | Movies | Collaborative | K-Nearest-Neighbors user-user (1st version). |
| 4. Fab | Web Pages | Framework with collaborative (mutual) agents and personal (individual) agents. Principle of learning close to genetic algorithms | TF-IDF representation of the viewed documents for user-profiling. K-Nearest-Neighbors user-user. |
| 5. DailyLearner | News articles | Thematic | One short-term model: TF-IDF representation of the viewed documents for user-profiling. One long-term model using naive Bayesian technique |
| 6. Amazon | Cultural products: books, music CD, DVD, computers, cameras... Very large Catalog. | Collaborative (on-line) and Thematic (mainly by e-mail) | Collaborative: analysis of co-occurrences of events (purchases, item browsing...) equivalent to a KNN item-item. Thematic: unknown, but seems simple (same author...). |
| 7. More | Movies | Hybridization by switch, using 2 models | Collaborative: KNN user-user. Thematic: KNN item-item on items' metadata. |
| 8. CinemaScreen | Movies | Hybridization of 2 models, with cascade-based triggering | Collaborative: KNN User-user Thematic: linear propagation of ratings to the characteristics of the items and vice-versa. The thematic engine re-ranks the result of the collaborative engine. |
| 9. AIMED | TV programs | Hybrid, based on items' metadata crossed with socio-demographics data and user's humor. | Neural networks: Multi-Layer Perceptron |
| 10. Google News | Aggregated news articles (In France, from about 500 web sources) | Hybridization of 3 collaborative models, weighting scheme. | Model 1: Counts of de co-visitations, Model 2: Clustering PLSI-type Model 3: MinHash clustering |

| System | Recommended objects | Type of technique | Modeling |
|---|---|---|---|
| 11. Hybrid Music Recommender | Music pieces | Hybridization using probabilistic Latent Semantic Analysis (pLSA) as integrated model. | pLSA applied on 2 matrices, a collaborative items × users matrix and a content-based items × polyphonic timbres analysis (specific to music). |
| 12. Social Ranking on CiteULike | Web pages (bookmarks) | Hybridization using user-user similarity matrix and tag-tag similarity matrix. | Query expansion using tag-tag similarity matrix, ranking model using both user-user similarity matrix and tag-tag similarity matrix |
| 13. BeeHive | People in a private social network (IBM) | 4 separated techniques user-user methods, the user being represented by her contents and/or her social network | 1. Based on the contents produced by the users<br>2. Based on 1, plus information of the users' social network<br>3. Pure social-network analysis, friends of my friends principle<br>4. Analysis of the co-occurrences of the users in different aggregated databases (publications, patents...) |
| 14. Recon | People in a online dating website | Sociodemographic data + analysis of implicit preferences (type of users generating messages accepting). | KNN user user, with a reciprocal similarity index.<br><br>Profiling using declared sociodemographic data and analysis of implit preferences |
| 15. Fastweb : ContentWise | TV Programs and Video On Demand (VOD) | Separate models<br><br>For TV programs and VOD: declared preferences.<br><br>For VOD only : 2 collaborative methods | TV and VOD :<br><br>Content-based: bag of word representation after putting words in their canonical form (stemming). Then Latent Semantic Analysis.<br>VOD only :<br><br>Collaborative: an item-item similarity model, and a SVD model. |
| 16. Youtube | Videos uploaded by the users | Collaborative. | Co-visitations (like association rules) with a time-slide window to insure scalability. |
| 17. Twittomender | real-time messages on a micro-blogging social network | Hybrid, feature combination according to the classification of (Burke, 2007). | Use of a search engine as recommendation framework.<br><br>Message indexation using bag of word representation and managing the IDs of the users (followees and followers) as terms - as if they were words. |
| 18. Cobot | Web pages from web search engine, users, conversations in a domain-specific community (health). | Hybrid content-based + social.<br>Content-based profiling is done by natural language processing and concept extractions. | Natural language processing + domain specific medical ontology to extract concepts from the conversations.<br>Short-term profiling and long-term profiling of the users.<br>Search engine (external) + case base reasoning for web page recommendations. |

**Table 2-3. Techniques used**

| System | Recommended objects | Data and users' et feed-back | offline performance evaluation | online performance evaluation or real use case |
|---|---|---|---|---|
| 1. Tapestry | Messages / documents | Documents' keywords and users' annotations | | |
| 2. GroupLens | NewsGroups | Logs of explicit ratings of users | Qualitative, 250 users | |
| 3. MovieLens | Movies | Logs of explicit ratings of users | The Movielens Database is available. | |
| 4. Fab | Web Pages | Explicit ratings of the users + TF IDF representation of the web pages | | On 11 users. Qualitative results. |
| 5. DailyLearner | News articles | On an embedded system. Implicit profiling + TF IDF representation of the news articles. | | On 185 users on the mobile terminal. |
| 6. Amazon | Cultural products: books, music CD, DVD, computers, cameras... Very large Catalog. | Logs of purchases and browsing of the users | | Conversion rate unknown. Impact estimated between 30 and 35% of the turnover (but not confirmed by Amazon). |
| 7. More | Movies | Catalog + logs of ratings | Measure of MAE | |
| 8. CinemaScreen | Movies | Catalog + logs of ratings | Measures of Precision and Coverage | |
| 9. AIMED | TV programs | Users' diaries of their usages on TV + declared mood + declared sociodemographic profiles | Simulation. Data Analysis with neural networks. | |
| 10. Google News | Aggregated news articles (In France, from about 500 web sources) | Implicit profiling: click of the users on the articles | On MovieLens' logs after binarization and on data extracted from Google News. Percentage of items correctly predicted interesting. | By A/B testing during 6 months, rate of usage increasing. |

| System | Recommended objects | Data and users' et feed-back | offline performance evaluation | online performance evaluation or real use case |
|---|---|---|---|---|
| 11. Hybrid Music Recommender | Music pieces | Content-based: Polyphonic timbres of pieces of music. Collaborative: users' ratings | Precision, artist diversity | |
| 12. Social Ranking on CiteULike | Research communities' bookmarks with users' tags | Documents' IDs, users' IDs, users' tags. | Precision/Recall | |
| 13. BeeHive | People in a private social network (IBM) | Company's social network, data from users' generated contents. | | A/B Testing Rate of good recommendation, rate of discovery. |
| 14. Recon | People in a online dating website | Declared sociodemographic data + implicit profiling using recipients of messages of invitation | Measure of Recall: rate of correctly predicted meetings | |
| 15. Fastweb : ContentWise | TV Programs and Video On Demand (VOD) | TV and VOD: bag of word representation Logs of users: VOD purchased, TV program watched including duration of viewing | Measure of Recall, on VOD only | Conversion Rate, estimated on VOD only |
| 16. Youtube | Videos uploaded by the users | Implicit Profiling: click of the users on videos | | A/B Testing, rate of increase in use |
| 17. Twittomender | Real-time messages on a micro-blogging social network | TF-IDF representation of the messages + IDs of "followers" and "followees" | Measure of Precision | |
| 18. Cobot | Web pages from web search engine, users, conversations in a domain-specific community (health). | Concept and topic building extracted from conversations via a medical ontology. Social network of the conversation. Explicit feedback by "I like"-type buttons. | | Available on-line for a while, at www.cobothealth.com. |

**Table 2-4. Data and performances**

# 3 Industrial KNN based system: the Reperio C/E Engine

*"Better a near neighbor than a distant cousin."*

Italian Proverb.

The purposes of this chapter are:
- to define a K-Nearest Neighbor (KNN) based recommender system, item oriented though generalized to other object to recommend: users, metadata's...
- to study, in an industrial context, the choice of implementation of KNN that lead to the best performances,
- to check that the KNN system are compatible with Requirements defined in Chapter 1,
- to validate in an industrial environment, system performance, especially when starting a service with the phenomenon of cold-start,
- to study the performance of the K-NN system in limit cases.

During this thesis, the presented system has become a real operational recommender engine called Reperio E/C, used on several prototypes or real services in the Orange Group.

## 3.1 Presentation of our KNN implementation: Reperio

All the analysis of the KNN-based methods' performances are done with the Reperio engine, whose developments began few months before this thesis at Orange Labs.

The Reperio engine is a prototype of a hybrid recommender engine. There are 2 branches of the engine: Reperio-C (Centralized) has a web service and works exclusively on centralized servers. Reperio-E (Embedded) does not have web service and works as a development library on any terminal with a Java Virtual Machine (JVM): this includes PCs and mobile java-based frameworks such as Google Android's smart phones. Reperio-E is adapted to limited resource systems.

The Reperio engine has allowed us to test prototypes of services in several fields: DVD recommendation, Video On Demand recommendation, TV program recommendation (embedded engine on Android smart phone), and music recommendation. Reperio is currently tested as a deployed system on the following recommendation services: books and radio stations.

Reperio is composed of
- a database containing the data sources and the models of recommendation,
- program modules: Java packages.

Reperio can handle 4 data sources:
1. Logs of users: shopping, browsing, rating on items,
2. Catalog of items, with metadata describing the items,
3. User preferences on item attributes,
4. Data of social networks: declared friends.

The models of Reperio recommendation are based on similarity matrices between objects. Generally, we build similarity matrices between items, from logs of usages (collaborative filtering) or from a catalog (thematic filtering, also known as contend-based filtering). It is also possible to build matrices of similarities between users or between item's characteristics (called descriptors).

## 3.2 KNN based recommendation models

Once a source and a representation space of objects are chosen, a k-Nearest Neighbor based model needs to specify:
1. a similarity measure to compare the objects,
2. a scoring function to predict de values associated with the objects,
3. a neighborhood size.

Reperio uses recommendation templates based on similarity matrices between data which may come from different sources. In general, the similarity matrix is a matrix known as the item-item similarity matrix (it compares pairs of items) whose sources are:
- a log table in the case of a collaborative mode,
- a catalog in the case of a thematic mode.

Other similarity matrices can be computed from tables of user preferences, or tables of declared friends, to provide personalized recommendations of descriptors (features of items) or other users.

In what follows, we restrict ourselves to the classical case of the recommended items, but it is also appropriate for the objects *descriptors* and *users*.

### 3.2.1 Choice of similarity functions in Reperio

Once the object source and space representation is defined, the first crucial choice when designing a k-NN model is those of the similarity measure.

All the similarity measures are defined below for non-null values. Non-null values are those defined in the logs. Null values are ignored. Ignoring non existing value is a key for scalability when managing huge matrices.

We remind the use the following notations: **u**, **v** denote users, **i**, **j** items, $T_i$ the set of all users who rated item **i**, $S_u$ the set of all items rated by **u**, $r_{u,i}$ the rating of user **u** for item **i**, $\bar{r}_i$ the mean rating of item **i** on all the logs of ratings, $\hat{r}_{u,i}$ the predicted rating by the scoring system of a user **u** for an item **i**. We define the following similarity measures:

$$Pearson(i,j) = \frac{\sum_{u \in T_i \cap T_j}(r_{u,i}-\bar{r}_i)(r_{u,j}-\bar{r}_j)}{\sqrt{\sum_{u \in T_i \cap T_j}(r_{u,i}-\bar{r}_i)^2 \sum_{u \in T_i \cap T_j}(r_{u,j}-\bar{r}_j)^2}} \qquad (3\text{-}1)$$

It is the similarity measure referenced by many publications (Resnick et al., 1994) (Sarwar et al., 2001) (Adomavicius and Tuzhilin, 2005) (Rao and Talwar, 2008). This formula suffers from a major bias: 2 items which have just one common user, who gave to the two items the same rating, will have a maximum similarity (Deshpande and Karypis, 2004) and (Breese et al., 1998). To correct this bias we introduce the following similarity:

$$ExtendedPearson(i,j) = \frac{\sum_{u \in T_i \cap T_j}(r_{u,i}-\bar{r}_i)(r_{u,j}-\bar{r}_j)}{\sqrt{\sum_{u \in T_i \cup T_j}(r_{u,i}-\bar{r}_i)^2 \sum_{u \in T_i \cup T_j}(r_{u,j}-\bar{r}_j)^2}} \qquad (3\text{-}2)$$

The effect of the Extended Pearson is to take into account, in the denominator, all the rating of both items.

Cosine is defined by:

$$Cosine(i,j) \frac{\sum_{u \in T_i \cap T_j} r_{u,i} \cdot r_{u,j}}{\sqrt{\sum_{u \in T_i \cap T_j} r_{u,i}^2 \sum_{u \in T_i \cap T_j} r_{u,j}^2}} \qquad (3\text{-}3)$$

See (Adomavicius, 2005) for example. Following the same principle as above, we can formulate an extended cosine:

$$ExtendedCosine(i,j) \frac{\sum_{u \in T_i \cap T_j} r_{u,i} \cdot r_{u,j}}{\sqrt{\sum_{u \in T_i \cup T_j} r_{u,i}^2 \sum_{u \in T_i \cup T_j} r_{u,j}^2}} \qquad (3\text{-}4)$$

where only the denominator changes.

In the case where the service only supports binary events such as purchases, the ratings are managed as constants. This makes inoperative a Pearson similarity: it returns always zero. To address both rating logs and purchase/browsing logs, an hybridization similarity of Jaccard and Pearson has been proposed in (Candillier et al. 2008), the « Mix » similarity:

The Jaccard similarity is defined by:
$$jaccard(i,j) = \frac{T_i \cap T_j}{T_i \cup T_j} \qquad (3\text{-}5)$$

and the Mix similarity is defined by:
$$mix(i,j) = jaccard(i,j) \times (1 + pearson(i,j))/2 \qquad (3\text{-}6)$$

Finally, mixing extended Pearson and Jaccard we can get for instance the ExtendedMix and the Weigthed Pearson similarity measures:

$$ExtendedMix(i,j) = jaccard(i,j) \times (ExtendedPearson(i,j) + 1)/2 \quad (3\text{-}7)$$

$$Wpearson(i,j) = jaccard(i,j) \times Pearson(i,j) \quad (3\text{-}8)$$

The ExtendedMix similarity measure is useful when deploying a recommender system without knowing in advance if the log data will be full binary events or ratings. This measure is robust against static ratings or pure binary logs such as browsing logs or purchase logs, as the mix measure. The Weighted Pearson (Candillier et al., 2008) is another way to take into account the size of the overlap of the two vectors: this measure is close to the ExtendedPearson similarity, and gives very good results. See (Candillier et al., 2007) and (Candillier et al, 2008) for details and results on tests of several similarity measures.

Reperio mainly uses:
- Jaccard's similarity on catalog data or on logs of sales or browsing,
- Weighted Pearson similarity in the case of logs of ratings,
- Mix similarity if the target data may change or contain both binary and real values.

### 3.2.2 Rating predictions

Reperio uses two formulas to predict scores based on the elements introduced earlier. First, the scoring formula called "Multi-users" or "mean-based" scoring function. It is adapted only to collaborative data environments.

$$\hat{r}_{u,i} = \bar{r}_i + \frac{\sum_{j \epsilon S_u} sim(i,j) \times (r_{u,j} - \bar{r}_j)}{\sum_{j \epsilon S_u} |sim(i,j)|} \quad (3\text{-}9)$$

where 
- $\hat{r}_{u,i}$ is the predicted rating of the user **u** for the item **i**
- $r_{u,i}$ is the rating of the user **u** for the item **i**
- $S_u$ is the set of the items rated by **u**
- $sim(i,j)$ is one of the previously defined similarity measures; note that as it can be negative, we use absolute values to sum the similarities in the denominator
- and $\bar{r}_j$ (resp. $\bar{r}_i$) is the mean of all the users' ratings on item **j** (resp. on item **i**)

This predictor assumes knowledge of average ratings on items of other users, which is not always the case. For example, in the case of a personal recommendation service embedded on mobile terminal, Reperio not have access to item means calculated on other users.

In the case of an autonomous embedded system, Reperio uses the formula called "Mono user"[6] (i.e; for a single user):

$$\hat{r}_{u,i} = \frac{\sum_{j \epsilon S_u} sim(i,j) \times (r_{u,j})}{\sum_{j \epsilon S_u} |sim(i,j)|} \quad (3\text{-}10)$$

### 3.2.3 Choice of the neighborhood size

In models based on k-NN item-item matrix, the number of neighbors generally increases the system performance (Bell et al., 2007a, 2007b). However, the quadratic complexity of the search and storage of the k-NN, whether in space or time, is a problem. So there's always a trade-off to decide for an industrial system (Deshpande and Karypis, 2004) (Koren, 2010).

---
[6] Reperio always uses mean-centered encoding of the ratings ; it is necessary to use the mono-user formula

Many recent studies are limited to a few tens of items for their neighborhood model (Bell et al, 2007c) (Takacs et al., 2007).

Our experiments showed that beyond tens of items of neighborhood, the performance of item-item models changes very little in reference databases such as MovieLens or Netflix: (Candillier et al., 2007) and (Candillier et al , .2008).

However, to make recommendations based on similarities for all items including those rarely rated (items in the "long tail") one must keep a neighborhood large enough. We therefore generally use a number of nearest neighbors of 200 for all tests, to have good security of effectiveness.

The recommendation model of Reperio includes calculating the similarity matrix and statistics on items and users. This process occurs cyclically in batch (e.g., 2 times a day) in centralized collaborative mode (multi-users), or in real-time without pre-computed matrix in embedded thematic mode (mono-user).

## 3.3 Implementation of the core functions

In what follows, the active similarity matrix can be, depending on use case:
- the item-item similarity matrix computed on the logs (collaborative recommendation)
- the item-item similarity matrix computed on the catalog of items (thematic recommendation)
- a hybrid of two previous matrices (hybrid recommendation)
- logs or external catalogs whose itemIDs were intersected with the main catalog (hybridization with external data).

### 3.3.1 Rating prediction

**The default predictors**
The default predictors are triggered in Reperio in the following cases:
- when the user has not rated any item
- when the item **i** to be scored does not have a shared neighborhood with the user's profile: the similarity matrix did not keep enough neighbors for **i**.

**Default predictor using user's preferences**
There is a simple formula for rating prediction in Reperio based solely on user preferences, and thus using the rating on the descriptors.

Each descriptor is associated with an internal weight, from the catalog: if the catalog is from a reliable source, all descriptors have an associated internal weight to 1. If the catalog has been generated by methods of aggregation and automatic text-mining, the descriptors can have weights between 0 and 1 equivalent to a confidence index.

Each descriptor is associated with a weight from its attribute. The dictionary of attributes is defined at deployment of the system: the weights of each attribute are determined at this time by an expert.

The general formula for predicting ratings based on preferences, in thematic mode, is:

$$\hat{r}_{u,i}^{pref} = \frac{\sum_{d \in D(i), d \in P(u)} p_{u,d} \cdot w(d) \cdot W(A(d))}{\sum_{d \in D(i), d \in P(u)} w(d) \cdot W(A(d))} \qquad (3\text{-}11)$$

where $D(i)$ is the set of descriptors for item $i$, $P(u)$ the set of descriptors recorded in the preferences of $u$, $w(d)$ the weight of the descriptor $d$ (usually 1), $p_{u,d}$ the rating of u for the descriptor **d** and $W(A(d))$ the weight of the attribute $A$, defined in the dictionary of attributes.

Following user testing on our prototype of embedded agent for a recommendation of TV programs, we found that preferences with strong negative values meant "suppress elements with these characteristics", while very positive preferences meant "I may love the items with these characteristics". Negative preferences are more important carriers of information as they generally mean complete suppression whereas positives preferences generally just mean possible selection.

Reperio uses a configurable preference scale. The preference rating scale has an external representation and an internal representation. The external representation depends on the man-machine interface and is for instance from 1 (I dislike) to 5 (I like very much).

The internal representation for preferences in Reperio ranges from a $min$ value to $max$ value so that $min = -100 \times max$. This preference scale is asymmetric and emphasizes negatives ratings of items' characteristics.

**Default Predictor using average ratings**

Predictors by means of ratings are default predictors used in thematic mode, collaborative mode and hybrid mode. They simply return a constant, or the mean rating of the item, or the mean rating of the user, or the mean of the user's means and the item's mean, based on available information. A constant value is returned in the worse case, when no mean information is available about the means.

**Case of thematic mode, or case of embedded system**

Reperio-E is an engine designed to operate also in embedded environment in single-user mode with thematic data only. In this configuration, it is not possible to access logs of usages on other users. It is therefore not possible to obtain information on the mean ratings of items in a population of users or information on average each user.

The only possible scheme of predicting a rating for an item **i** for user **u** exploiting the item-item similarity matrix in this case is to perform a direct computation of a weighted average of the similarities of the target item **i** with the items rated by **u** that are close to **i**. This is shown in the Figure 3-1, using the rating prediction formula "mono user".

**Collaborative case**

Where there is access to logs of usages on the items, various complementary solutions to improve the "mono user" rating prediction formula exist. Our choice was prioritized after several tests to deviations of the users' mean ratings for each item on the total population, but other options are also possible, see (Candillier et al., 2007). This formula is known as "multi-users" (see Figure 3-2) in opposition to the first formula.

### 3.3.2 Ranking

Using Reperio the ranking is directly provided by the rating prediction rating. A list of items **L={i₁,…iₙ}** is then simply sorted in decreasing order of predicted score. Options allow:
- use of preference-based rating prediction: this allows for example to filter in a query result, items that have characteristics not desired by a user,
- zeroing of items being too similar to existing items in the user profile.

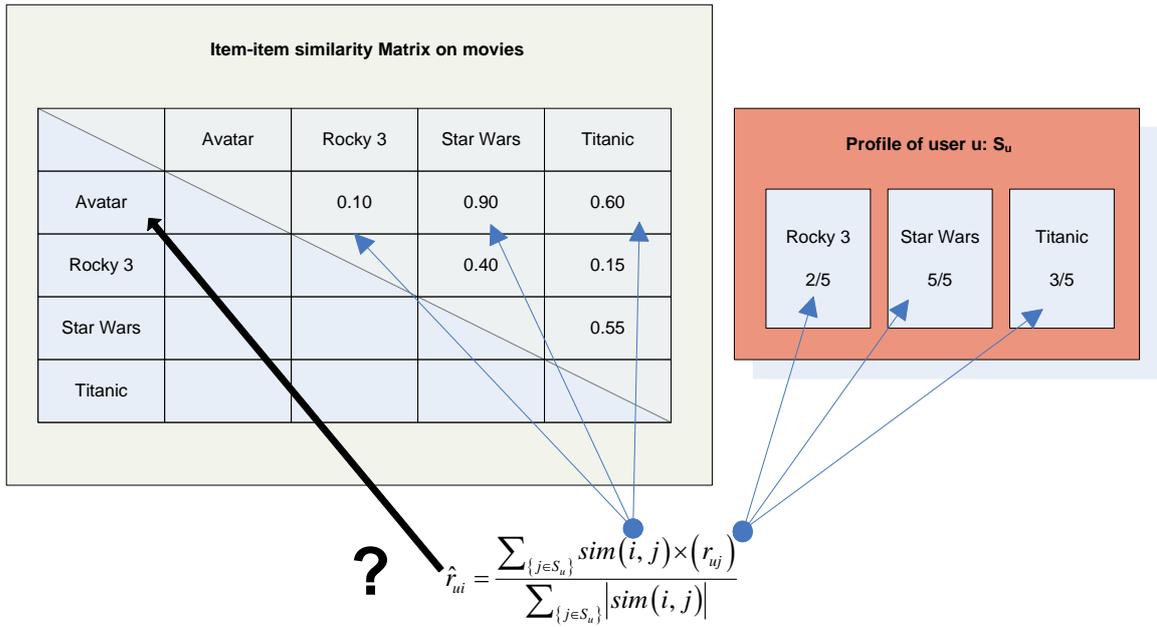

**Figure 3-1: Rating prediction in the case where only one profile is available**

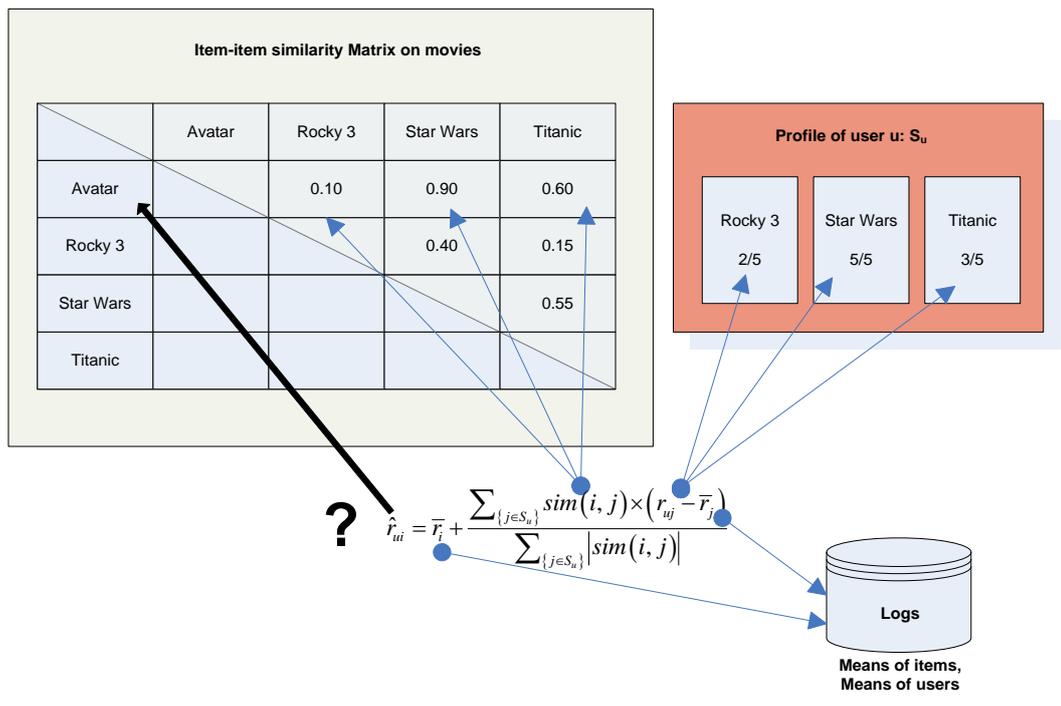

**Figure 3-2: Rating prediction: collaborative case**

### 3.3.3 Anonymous item-to-item recommendation

Item-to-item recommendation is straightforward as it is shown is the following figure. The similarity matrix is directly accessed to find the most similar items of the item used in context.

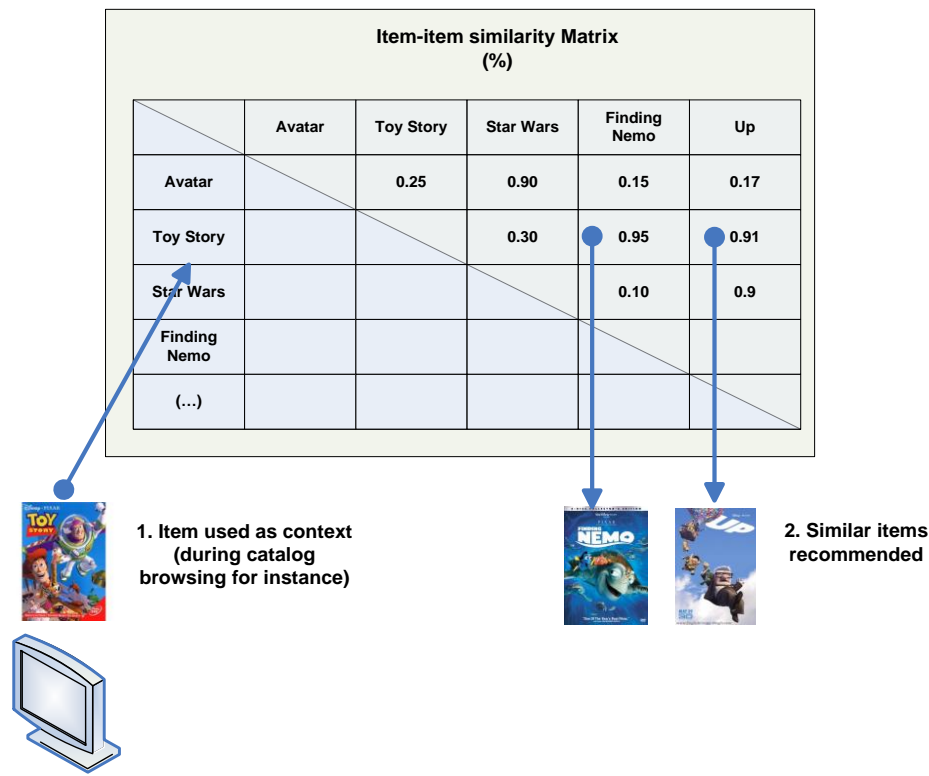

**Figure 3-3 : Principle of item-to-item recommendation to anonymous users**

### 3.3.4 Seed-based personalized recommendation of items

The seed-based personalized item recommendation is possible if the user profile contains rated items. To manage shopping or browsing logs, ratings equivalent to each user action may be generated.

In Reperio, the seed-based personalized recommendation of items consists of 5 steps:

**1. Determination of seed items (possible short / long term profile management)**

During this step, Reperio chooses up to $G$ reference items called seeds. Normally these items are highly rated from the user profile. If we do not have highly rated items, we take the rated items, which are usually an indication of the interests of the user. Options allow Reperio to select items more or less recent: it makes it possible to work with the short-term or long-term user's profile. By default, Reperio searches up to 10 highly rated items by the user (items above the average rating of the user), with randomization to insure diversity.

**2. Determination of candidates (possible management of the long tail)**

Once the $G$ seeds selected, Reperio searches $C$ candidate items by seed ($C$ is configurable and set to 100 by default). For this, it uses the item-item similarity matrix and searches for similar items. A parameter specifies whether one only searches in the most often rated items, or rarely rated items. Separating long tail / short head is performed under a 80/20 rule. The 20% of the most rated items are considered in the short-head, the remaining 80% are considered in the long-tail. The choice to search in the long tail or in the short-head, and the short-head / long tail threshold, are made by the service using Reperio (according to some business rules). The candidates are stored in a hash-table insuring that each candidate is unique.

**3. Deletion of items already known by the user**

Reperio removes items from the list of candidates already in the user profile.

**4. Ranking of candidate items**

Reperio ranks the remaining candidates to sort items in order of decreasing predicted rating with respect to the targeted user.

**5. Selection of a randomized Top-N item list with a diversity parameter *D***

Reperio then selects at random the *N* requested items, among the $N \times D$ first items in the ordered list of candidates. The parameter *D* is called diversity factor: set to 1, diversity is null and Reperio takes the Top-N candidates. Set to 10 Reperio take random items among the **10 *N*** first candidate items available.

These steps ensure the recommendations to be:
- close to the user profile, encouraging a confident collection of recommendations,
- with items from well-known catalog list or rare item list (short head / long tail),
- non trivial (not present in the user profile),
- ordered by relevance,
- non repetitive.

The Figure 3-4 gives an overview of this process in the illustrative case of a movie recommender. In 2010, an experiment was conducted with Reperio-E on a TV-program recommender embedded on mobile smartphone. The system recommended TV programs to users based on their personal feedback. The recommendation strategy used was the seed-based strategy applied on 3 use cases: instant recommendation, daily recommendations, weekly recommendations. The diversity parameter was set to 0: as the catalog was changing every week there was no need to use a principle of diversity. After 2 months of tests, the 10 users involved in the experiments were interviewed on various aspects of the service. Concerning the quality of the recommendations, 9 users on 10 declared to be very satisfied with the system.

## 3.4 Checking the industrial requirements

This section reviews the requirements previously defined in chapter 1 to illustrate how Reperio manages its compliance with them.

### 3.4.1 Ability to manage multiple data sources

Multiple source data management in Reperio includes 3 areas: suitable engine architecture, standardized format and generic data sources, adapted similarity functions.

**Engine's adapted architecture**

The cycle for updating data table consists in adding new data from back office (this is especially the case for re-delivery of catalogs), then trying to delete old or outdated information. The removal of old information is not trivial because we must be careful, for example, not to delete items, or descriptors of the items, which are still listed in user profiles: the "rated item descriptors" table is used to memorize items deleted from the Catalog but still referenced in at least one user profile. In Contend-Based mode, the item-item similarity matrix is computed using the items of the Catalog and the the item in the "rated item descriptor" table. A special process regularly monitors the volume of the database and carries out the pruning of the oldest information when it becomes necessary (oldest rated items).

**Standardized and generic data source format**

The similarity matrix computation process takes as input a table which is a multi-indexed sparse matrix with 3 fields: row, column, value. Other fields, such as date, comments ... are available for specific selections. The 4 main data sources of Reperio all share this format. It is then easy to generate item-item similarity matrices of each type of objects, user-user or descriptor-descriptor, as shown in the Figure 3-5. Reperio is able to provide up to 8 different types of recommendations.

**Adapted similarity function**

A similarity function adapted to the real data or binary data has been proposed and tested (Candillier et al., 2008). This function, called "Mix" makes it possible to deploy Reperio on services based on rating logs or on logs of purchases where user's actions are encoded with constant pseudo-ratings. The use of constant ratings is not consistent with a Pearson's similarity (the similarities of vectors with constant features are all zero when using a Pearson similarity).

### 3.4.2 Management of the cold start

Management of Cold-Start is provided by the possibility of using models based on thematic or hybrid item-item similarity matrix in mono-user mode if necessary, and the possibility of using external data sources. The cold start is studied in a specific section later in this chapter.

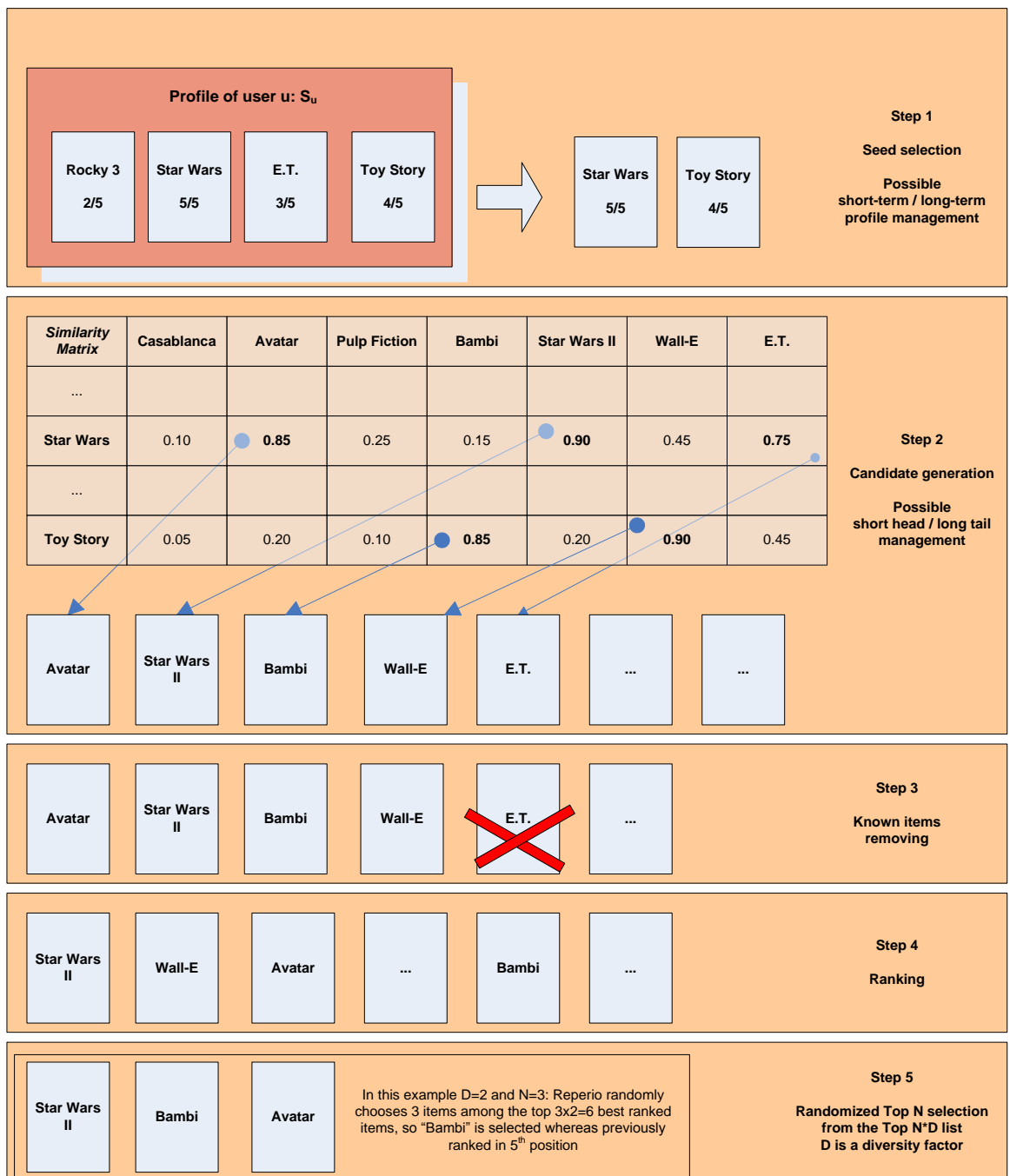

**Figure 3-4 : Reperio's general process to generate personalized recommendations of items**

Page 108

### 3.4.3 Management of noisy or corrupted data

We distinguish here two cases:
- noisy data of a catalog of items, where some attributes are very informative and some not at all,
- corrupted data, for example by fake users.

**The noisy data from a catalog of items**

On the catalogs of items, Reperio allows to select or to specify the weight of each attribute in a dictionary of attributes. For example, for a movie recommendation service, we can specify the following pairs (attribute weight): (Title, 0) (type, 3) (Actor, 4) (Director, 3), (Supervisor, 3), (Country, 0), (Producer, 0) ... The weights are determined empirically by cross-validation during service initialization. A feature weighting scheme, such as Relief-F (Robnik-Sikonja, 2003) is being studied to automate this process. Descriptors may also have a specific weight. This function is used in an embedded prototype of TV program recommender (on Smartphone Android$^{TM}$): for example, the different actors of the films or guests of emissions have a weight inversely proportional to the order they appear in the item's description of the Electronic Program Guide (EPG, the electronic TV catalog).

**Data corrupted by fake users**

Mitigating the impact of corrupted data is managed by the possibilities of hybridization. Reperio can merge different similarity matrices by linear combination of their item-item similarities. For example, to neutralize the impact of fake users, it is possible to hybridize a collaborative item-item matrix (with fake users) and a thematic item-item matrix computed from a catalog of data contained by the service provider.

### 3.4.4 Scalability

Reperio E/C includes 3 systems to insure scalability for medium to large datasets:
1. A-built-in optimized indexed data structure in the kernel of Reperio allowing to load in RAM up to 100 millions of logs in 2Gb of RAM.
2. A Ram-Cache mechanism to reduce the number of requests to the database: in general, coupled this the optimized indexed data structure, the database is just read once.
3. A parallelization framework making it possible to parallelize the item-item k-NN matrix building on **M** machines, leading to a process time nearly divided by **M**.

Reperio E/C has been tested up to 500,000 "items" using a transposed of the Netflix's logs' matrix. This configuration takes several hours to perform the KNN search (See user-user k-NN models in Candillier et al., 2007). Then Reperio has been tested up to 1,000,000,000 of logs by duplicating users' rating with new user IDs. Increasing the items' dimensionality (number of ratings or number of descriptors) increases linearly the time of the KNN search. Beyond these figures Reperio E/C needs new mechanisms to deal with scalability issues.

Examples of speed performances of Reperio are given later in this chapter.

### 3.4.5 Reactivity

In Reperio all personalized recommendation function are based on an on-line user profile. This profile is always updated in real-time and any change in the profile, adding or changing ratings on the items, has an immediate impact on the personalized recommendations. The functions of recommendation are simple and of linear complexity with respect to the size of the user profile. It is not necessary to re-compute the recommendation model for each profile modification: the item-item similarity matrices change slowly and can be updated cyclically every day or every week.

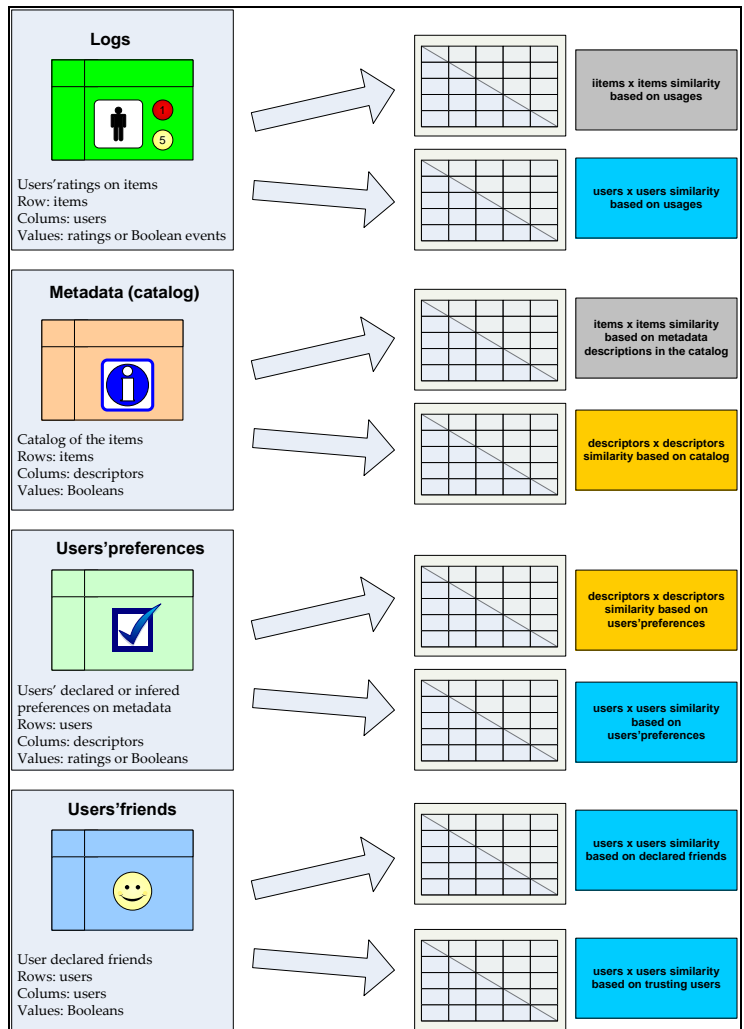

**Figure 3-5: List of data sources and possible similarity matrix generations in Reperio**

### 3.4.6 Trust-based relationship management

There are three simple functions in Reperio related to the trust relationship, transparency of the system and its explanation: the traceability of any recommendation, the possibility of generating a confidence index, and the linking of any item presented with preferences of the user.

**Traceability of the recommendations**

Each personalized recommendation can be traced to find in the user profile, which most similar item seed was used to generate the recommendation or the predicted rating.

**Confidence index**

An index of confidence may be associated with any recommendation or predicted rating. This is done using the similarity index between recommended item i and the item in the user profile the most similar to **i**.

**Highlighting preferences related to items recommended**

For any user **u** and item **i**, Reperio can provide a list of descriptors of i referenced in the preferences of u. When viewing an item **i** recommended to **u**, we can highlight descriptors of **i** that were rated positively by **u**, and conversely highlight in another way (other colors…) the list of descriptors of i that were rated negatively by **u**.

### 3.4.7 Management of the long-tail

Management of the long tail is directly integrated into the personalized recommendation function of Reperio. A specific mode can be used to search for particular candidate items in the short-head or in the long-tail. When selecting the final list of items to recommend, a diversity parameter is used to define a random selection more or less diverse in the sorted list of the candidate items.

## 3.5 Expected accuracy in industrial contexts

In this section, we study, on a reference logs database (Netflix), the performance of the KNN recommender systems on collaborative thematic and hybrid data sources. We first show that models based on item-item similarity matrices give good performances if they use appropriate formulae and calculation of similarity scores. We study first the case of a system already operational with many data. Then we simulate a cold start process. We show that if the collaborative systems seem insurmountable in terms of performance when users have long profiles, however they are not very competitive, if user profiles are not rich, which is a common case in practice. To manage the cold-start in a recommender system, we show that "cascade" approaches, thematic then hybrid and then collaborative, would be more appropriate.

### 3.5.1 Presentation

We have seen in this chapter that an effective technology in the field of collaborative filtering is based on calculating the item-item similarity matrix (Karypis, 2001) (Sarwar et al. , 2001), (Linden et al., 2003) (Deshpande and Karypis, 2004). When building an item-item similarity matrix based on usage logs, when the service starts, we find ourselves confronted with the well known problem of the cold-start (Adomavicius and Tuzhilin, 2005). Therefore, using content-based approaches, that is to say based on items' descriptive metadata, in combination with collaborative methods, is a recurring issue. Hybrid recommendation methods, which use collaborative and thematic data were studied in (Balabanovic and Shoham, 1997) (Salter et al. 2006). But they have been little studied, to our knowledge, on the reference data widely accessible as Netflix (Netflix Prize 2007).

In this section we restrict ourselves to models that are based on an item-item matrix, this matrix is calculated on logs data or items' metadata. We study the quality of item-item similarity matrices that can be produced:
- from rich thematic data,
- from both collaborative data and thematic data (hybridization).

according to several contexts of cold start, with short profiles and long profiles.

### 3.5.2 Experimental protocol

**Similarity Measures**

We will use the following similarity measures, previously presented:

$$\text{Pearson}(i, j) = \frac{\sum_{u \in T_i \cap T_j}(r_{ui} - \overline{r_i})(r_{uj} - \overline{r_j})}{\sqrt{\sum_{u \in T_i \cap T_j}(r_{ui} - \overline{r_i})^2 \sum_{u \in T_i \cap T_j}(r_{uj} - \overline{r_j})^2}}$$

$$\text{ExtendedPearson}(i, j) = \frac{\sum_{u \in T_i \cap T_j}(r_{ui} - \overline{r_i})(r_{uj} - \overline{r_j})}{\sqrt{\sum_{u \in T_i \cup T_j}(r_{ui} - \overline{r_i})^2 \sum_{u \in T_i \cup T_j}(r_{uj} - \overline{r_j})^2}}$$

$$\text{Cosine}(i,j) = \frac{\sum_{u \in T_i \cap T_j} r_{ui} \cdot r_{uj}}{\sqrt{\sum_{u \in T_i \cap T_j} r_{ui}^2 \sum_{u \in T_i \cap T_j} r_{uj}^2}}$$

$$\text{ExtendedCosine}(i,j) = \frac{\sum_{u \in T_i \cap T_j} r_{ui} \cdot r_{uj}}{\sqrt{\sum_{u \in T_i \cup T_j} r_{ui}^2 \sum_{u \in T_i \cup T_j} r_{uj}^2}}$$

$$mix(i,j) = jaccard(i,j) \times (1 + pearson(i,j))/2$$

$$jaccard(i,j) = \frac{|\{T_i \cap T_j\}|}{|\{T_i \cup T_j\}|}$$

$$\text{ExtendedMix}(i,j) = jaccard(i,j) \times (1 + \text{ExtendedPearson}(i,j))/2$$

**Function for rating predictions (scoring methods)**

Firstly, the formula called **Mean-based**, or **Multi-users** to insist on the fact that it is usable only in collaborative contexts:

$$\hat{r}_{ui} = \bar{r}_i + \frac{\sum_{\{j \in S_u\}} sim(i,j) \times (r_{uj} - \bar{r}_j)}{\sum_{\{j \in S_u\}} |sim(i,j)|}$$

In the case of a single-user system, one uses the formula known as **Not-Mean-Based**, or **Mono-User** to insist on the fact that it is compliant with mono-user recommender systems, where items are represented by vectors of metadata instead of vector of many users' ratings.

$$\hat{r}_{ui} = \frac{\sum_{\{j \in S_u\}} sim(i,j) \times (r_{uj})}{\sum_{\{j \in S_u\}} |sim(i,j)|}$$

**Parameters of the KNN models used**

As previously mentioned (cf 3.2.3), in k-NN item-item models, the number of neighbors generally increases system performance (Bell et al., 2007a, 2007b). However, the quadratic complexity of the search for k-NN, whether in time or space, is quadratic. So there's always a tradeoff for an industrial system (Deshpande and Karypis, 2004) (Koren, 2010) to select a reasonable value for K. Moreover, many recent studies are limited to a few tens of items for their neighborhood model (Bell et al, 2007c) (Takacs et al., 2007). So we use a number of nearest neighbors of **K=200** for all tests, to have good security effectiveness.

**Data sources used: Netflix + IMDb**

For the collaborative part we used the logs from Netflix (www.netflix.com), which are now a reference database. For our tests of relevance and cold-start we used a sample of this database, 25,000 users. This sample was split, for each user profile, in 2 parts, Learning set and Test sets of respectively 90% and 10% of the ratings.

For the Cold Start analysis, we conducted two tests :
- one using long profiles of more than 180 ratings per user on average in the Training set and 20 rating-profile in the Test set.
- another using short profiles of 20 ratings per user on average in the Training set and 180 rating-profile in the Test set (we simply reversed Training and Test).

The following table summarizes the characteristics of our tests for the collaborative part (excluding latest tests of cold-start from 10 to 100,000 users).

| | |
|---|---|
| Number of selected users | **25,000** for the hybridization test with many data<br>From **10** to **100,000** for the cold-start simulations |
| Number of selected items | **17,770** (100% items) |
| % of sampling in the Learning set (case of the long profiles) | **90%** |
| % of sampling in the Test set (case if the short profiles) | **10%** |
| Size of the Learning set for the long profiles | **4,707,540** |
| Size of the Test set for the long profiles | **516,979** |
| Average size of the long profiles (in number of ratings) | **188** rated items |
| Average size of the short profiles (in number of ratings) | **20** rated items |

**Tab. 3-1 Characteristics of the data for the tests**

For the content-based filtering part we used data from the Internet Movie Database (www.imdb.com). This database contains many information on the movies referenced internationally. We used the following attributes to enrich metadata of Netflix's movies: genre, director, actors, tags of users (these are keywords or terms associated to the movies), the country of production ... Each pair (attribute, value) of a IMDB's movie's information is considered a Boolean called descriptor that can be linked to an item. For example, the item (Star Wars I) will have the following descriptors: (Genre, science fiction) (Actor, Harrison Ford), (Director, George Lucas), etc.

IMDb data were crossed with those of Netflix using the information of the title + year. We calculated the Levenshtein's distance between the title of the movie from Netflix and IMDb for that, once freed from non-alphanumeric characters. In the same time, we calculated the distance between the date of the movie release from Netflix and from IMDb. An automatic procedure then performs the join of the Netflix's movies' title with the closest title in the IMDb database, provided that the release dates are not different by more than 1 year. The error rate was then estimated by sample, manually analyzing the associations of Netflix-IMDb's titles. The overall number of different descriptors was: 861,507. The number of items for which we could link a descriptor was 15,953 or 90.13% of the items. The error rate of the joint was estimated by sampling at 6%: it was basically due to sports or wildlife documentaries that were not listed in IMDb but were available to rent in the Netflix database.

### 3.5.3 Test protocols
**General principle of the modeling**
The test procol at each measure step is the following:

Select **N** Netflix's users randomly
Select all the logs $r_{u,i}$ of the **N** users (all items rated by these users)
Split randomly the logs obtained in 2 parts, Learning and Test (a user profile **u** will have a learning profile $u_L$ and a test profile $u_T$)
For each type of method, build a item-item similarity matrix-based model using all the items' rating known in the Learning set:

- In **Collaborative mode**: use the logs to calculate similarities (the items × users matrix), and then use the Multi-user Mean-Based scoring method to predict the ratings.
- In **Thematic mode**: use IMDB metadata to calculate similarities (the items × metadata matrix), and then use the Mono-user Not-Mean-Based scoring method to predict the ratings.
- In **Hybrid mode**: use IMDB metadata to calculate similarities (the items × metadata matrix), and then use the Mean-Based scoring method to predict the ratings.

The RMSE is then calculated on the ratings belonging to the user profiles in the Test set, comparing actual ratings and predicted ratings.

**Simulation of the Cold-Start**

We vary the number of users logarithmically. We work in two contexts:
- long profiles, where we use 90% of each user profile, stored in the Learning set, which corresponds to an average user profiles of 180 rated items (the rest of the profile being used in Test set to evaluate the RMSE).
- short profiles, where we use 10% of each user profile, stored in the Learning set, which correspond to an average of only 20 items scored (the remaining 90% being in Test).

**Measures: RMSE and its variations RMSE-In and RMSE-Out.**

The RMSE is a popular measure for evaluating a predictive model for ratings. If we consider a set of ratings to predict in a Test set, noted $R_T$, we want to compare ratings with a set of predicted ratings $R_P$, then the RMSE is given by:

$$rmse(R_T, R_P) = \sqrt{\frac{1}{|R_T|} \sum_{r_{u,i} \in R_T} (r_{u,i} - \hat{r}_{u,i})^2}$$

with

| | |
|---|---|
| $r_{u,i}$ | rating of the user **u** for the item **i**, this rating known in test |
| $\hat{r}_{u,i}$ | rating of the user **u** for the item **i**, this rating predicted by the model |
| $|R_T|$ | Size of the Test set (number of ratings to predict) |

In fact for models that could not be able to answer for all recommender requests (it is especially the case on KNN models) the RMSE can be calculated using two measures: either directly by comparing the right rating **r** with the rating predicted by the model $\hat{r}_{u,i}^{model}$ (here a item-item KNN model) if possible, or by comparing the true rating r with the default rating $\hat{r}_{u,i}^{default}$ of the default predictor if none of the items of the user profile allows the use of a k-NN model.

So we make two RMSE measures, called RMSE-in and RMSE-out. The RMSE-out corresponds to the calculation of the total error, using the main predictive model or the default predictor if the main predictive model is unavailable.

The RMSE-in is the internal error of the main predictive model. It corresponds to the calculation for cases where the calculation of the predicted rating $\hat{r}_{ui}$ is possible via the item-item K-NN model, so without using the default predictor. This indicator lets one see the real performance of a K-NN model regardless of the calls to the default predictor (which will confuse the measurement).

### 3.5.4 Results with long profiles: collaborative, thematic, and hybrid methods

We first evaluate the performance of a system already fitted in charge, according to three cases of classic uses:
- pure collaborative system,

- pure content-based system (no logs of usage available),
- hybrid system (to try to optimize performance).

### Default predictor

We first evaluate the performance of the default policies that are based on a cascade model of average ratings. The first to be used if a model of similarity cannot answer for a rating of an item **i** for user **u** is **(Mean of u + Mean of i) / 2**. If the information about **i** or **u** is not available, then a cascade to the average ratings of **u**, or the average ratings on **i** is performed. Sampling during the separation Learning / Test can indeed lead to situations where an item or a user is unknown in the Learning set. The table above summarizes the performances of the default predictors used. Note the relatively good performance of the default predictor "**(Mean u + Mean i) / 2**".

| Default predictor | RMSE-out |
|---|---|
| **Mean i** | 1.0103 |
| **Mean u** | 1.0070 |
| (Mean i + Mean u) / 2 | 0.9651 |

**Table 3-2: RMSE-out for the default predictors used in these experiments on 25,000 users**

Note that for pure thematic engine, working with only one user at once, we shall only use the "**Mean of u**" default predictor.

### Optimized collaborative methods

These measurements were made with 25,000 users randomly sampled among the (approximately) 480,000 users of Netflix with 90% of their profile in the Learning set and 10% of their profile in the Test set. We distinguish:
- performance prediction with RMSE-in applying directly the formula to predict the ratings only when the prediction is possible,
- final performance prediction with RMSE-out: application of the default predictor in the cases where the ratings could not be predicted by the K-NN model.

The results comparing rmse-in and rmse-out are shown in Table 3-3 and Table 3-4. This study shows that the k-NN methods on item-item matrices are competitive for the optimization of the RMSE if we have:
- a good similarity function, because one can have a very strong performance variation using classic Pearson similarity (called Pearson in our bench) or using Extended Pearson's,
- a good rating prediction scheme,
- a good default predictor.

Many articles use the "classic" Pearson's similarity and therefore under-estimate the performance of item-item k-NN type methods, which could affect the estimate of the real gain of content-based or hybrid methods.

|  | RMSE-in / Scoring Mode | |
|---|---|---|
| **Similarity function used** | **Mono user / not Mean-Based** | **Multi users / Mean-Based** |
| Jaccard | 0.9193 | 0.8768 |
| Extended Cosine | 0.9940 | 0.9881 |
| Cosine | 1.4729 | **1.2915** |
| Extended Pearson | 0.8938 | 0.8621 |
| Pearson | **1.4815** | 1.2583 |
| Extended Mix | 0.9131 | 0.8723 |
| Weighted Pearson | **0.8745** | **0.8355** |
| Mix | 0.9056 | 0.8678 |

**Table 3-3: RMSE-in for multi-user mean-based, or mono-user not-mean-based scoring methods**

|  | RMSE-out / Scoring Mode | |
|---|---|---|
| **Similarity function used** | **Mono user / not Mean-Based** | **Multi users / Mean-Based** |
| Jaccard | 0.9226 | 0.8795 |
| Extended Cosine | 0.9092 | 0.8720 |
| Cosine | 1.0684 | **1.0214** |
| Extended Pearson | 0.8987 | 0.8677 |
| Pearson | **1.0695** | 1.0128 |
| Extended Mix | 0.9165 | 0.8752 |
| Weighted Pearson | **0.8804** | **0.8425** |
| Mix | 0.9093 | 0.8709 |
| Reference: default predictor | 0.9651 | 0.9651 |

**Table 3-4: RMSE-out for several similarity measures (collaborative mode, Netflix logs)**

The first point to notice is that RMSE-out compress the results to that it is difficult to see the differences in performance of the different similarity measures: the default predictor has masked a large part of the impact of the different similarity measures.

If we have a look at the RMSE-in performances in the Table 3-3, using the same scale, the difference in performance are visible: the classic measures of Cosine and Pearson, when they are limited only to the shared dimensional of both the two items to compare, are not reliable. On the contrary, the extended measures of Cosine and Pearson give good similarity information.

The multi-user "mean-based" scoring scheme is always better than the mono-user "not mean-based scoring scheme.

Note that Mix similarity (RMSE-in: 0.8678) , the Extended Pearson similarity (RMSE-in: 0.8621) and Jaccard similarity (RMSE-in: 0.8768) give good performances. It is surprising for the Jaccard similarity as this measure misses the value of the ratings, taking into account only the presence or absence of the ratings.

Finally, the best RMSE performances are achieved with the Weighted Pearson similarity:

**Content-based recommendation**

We will rely first on a single source of thematic data, without any collaborative information during modeling. We use the "Not Mean-based" formula to predict the ratings, even if in fact we have multiple users in the Learning set that we could use to calculate averages on items or users. The protocol is identical to the collaborative system: instead of using a table (itemID, userID, rating) it uses a table (itemID, DescriptorID, constant). In this case we used a Jaccard similarity measure because there are no ratings for the pairs **(u, i)**. All the results are based on test profiles of 25,000 users, taking as input the learning part of their profile, but user by user (no access to global averages over all users).

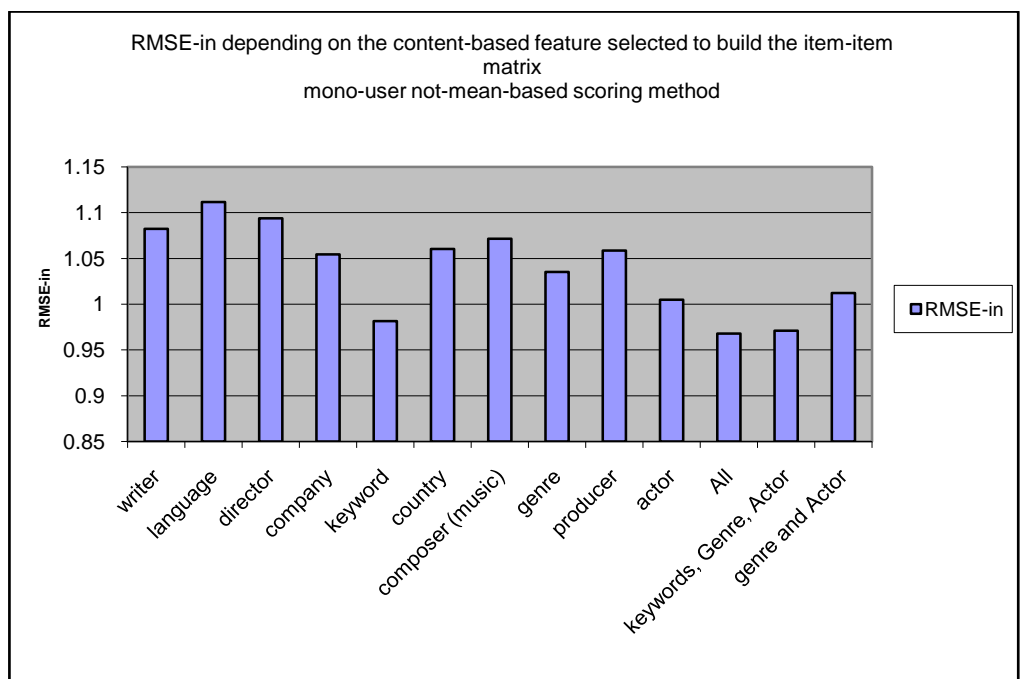

**Figure 3-6: RMSE-in according to the metadata used (content based recommendation)**

We are interested in the RMSE-in, which is not very good in general except for keywords and combinations of descriptors using the keywords. The information contained in the metadata seems to have a low predictive power, except in regard to keywords. Note that keywords can sometimes be seen as implicit ratings because in IMDb those keywords (mainly added by registered IMDb's users) are often "Cult" "Good movie" and so on.

In the field of movie recommendation, the selection of the genres, the actors and mainly the user-generated keywords if available seems the best choice.

**Collaborative - Thematic Hybridization**

We present a method of light hybridization (industrially realistic), thematic and collaborative. It arises in a case of actual use of a centralized system: we have to use logs, to estimate correctly the average ratings of the items, however we do not have enough logs to apply collaborative filtering methods. Here, only the similarity matrix is calculated using the metadata with the Jaccard similarity function. For the rest the Multi-user formula is used on the basis of collaborative learning for all of the 25,000 user profiles (as for collaborative tests). The RMSE-in is usually greatly improved. All these results are based on the profiles of 25,000 users, taking as input the learning set's part of the users' profiles and computing the RMSE on the test set's part of the users' profiles.

The Table 3-5 provides a summary of the results obtained with different approaches. We can see that for large volumes of data and detailed profiles for each user, collaborative methods are superior to purely thematic methods and hybrid methods tested.

The thematic methods has low predictive performance compared to collaborative methods even using averages users' ratings' default predictors. Thematic approaches with a light hybridization show a RMSE 6-7% higher than the RMSE obtained by a "good" collaborative approach: moving from a RMSE of about 0.84 to a RMSE from 0.91 to 0.92 with a purely thematic method, this difference being very significant on data from Netflix's. On large volumes of data usage, purely collaborative methods seem to be definitively the best technical choice.

| **Filtering mode** | **Data source** | **Similarity function** | **Scoring method** | **RMSE-in** | **RMSE-out** |
|---|---|---|---|---|---|
| Collaborative filtering | Netflix's logs | Jaccard | Multi-users mean-based | 0.8768 | 0.8795 |
| Collaborative filtering | Netflix's logs | Weighted Pearson | Multi-users mean-based | **0.8355** | **0.8425** |
| Thematic filtering | IMDB metadata, all attributes | Jaccard | Mono-user not-mean-based | 0.9679 | 0.9717 |
| Hybrid filtering light | IMDB metadata, all attributes + means of items | Jaccard | Multi-users mean-based | 0.9113 | 0.9151 |
| Hybrid filtering light | IMDB metadata, Keywords, +means of items | Jaccard | Multi-users mean-based | 0.9243 | 0.9292 |
| Hybrid filtering | Netflix's logs + IMDB metadata, All attributes. Fusion in one matrix | Jaccard | Multi-users mean-based | 0.8760 | 0.8785 |

**Table 3-5: Summary of the different approaches**

### 3.5.5 Simulation of a Cold-Start

We continue to use the principles of collaborative modeling (using all the logs of users' profiles in the Learning set), pure thematic modeling (using thematic data for modeling and then use of the profile of each user in the learning set separately just during rating prediction) and light hybrid modeling (using thematic data and average ratings of items based on users logs on the Learning set during rating prediction). We vary the number of user logarithmically. We consider two contexts:
- long user profiles, using the wealth of logs from Netflix, with 180 rated items in the learning set, on the average,
- short user profiles, where each user has rated (in the learning set) only 20 items on the average.

The aim of this analysis is to check if the thematic filtering and the hybrid filtering can actually be useful during some deployments with cold-start. The thresholds are just fixed to roughly represent 2 typical use cases: items easily evaluable and rated (music pieces, movies,...) or items long or expensive to evaluate and to rate (trips, long novels, expensive restaurants,...).

The results are shown in the two next figures Figure 3-7 and Figure 3-8.

**Global results on long profiles**

Note that in all the cases, the system sometimes uses default predictor cascading[7], whose performance depends on the size of logs. This explains the slight variation in performance even of the pure thematic model which is able to use, as default predictor, the mean of the ratings of each current user's learning profile. On long profiles in cold-start, a thematic recommender sounds interesting until about 100 different users. Beyond 100 users, the collaborative system is more advantageous. The hybrid system does not seem very interesting compared to collaborative filtering. Even during the beginning, the very low difference of the performances does not justify itself only to the choice of hybridization, much more complex to implement than a collaborative system.

**Global result on short profiles**

The Content-based Recommendation varies quite widely, with a RMSE between 1.1 and 1.2. It should be seen as above as the effect of the default predictor called every time the scoring by the item-item matrix cannot give answers, but also the effect of the variance, which is more important on short profiles than on long profiles. In the case of short profiles, thematic filtering is generally the most interesting until about 1,000 users. Beyond the light hybridization becomes more interesting, to about 100,000 users as collaborative filtering becomes competitive.

In conclusion, if collaborative methods seem the best methods on large volumes of data in case the user profiles are large (around 200), they are in difficulties when users are less well represented by small profiles, of about 20 ratings. The most realistic cold-start situations, are the ones where users have not yet completed their profile much. In this case, a thematic engine is interesting to start a service to approximately 1,000 users. In addition, from 1,000 to 100,000 users, hybridization becomes interesting.

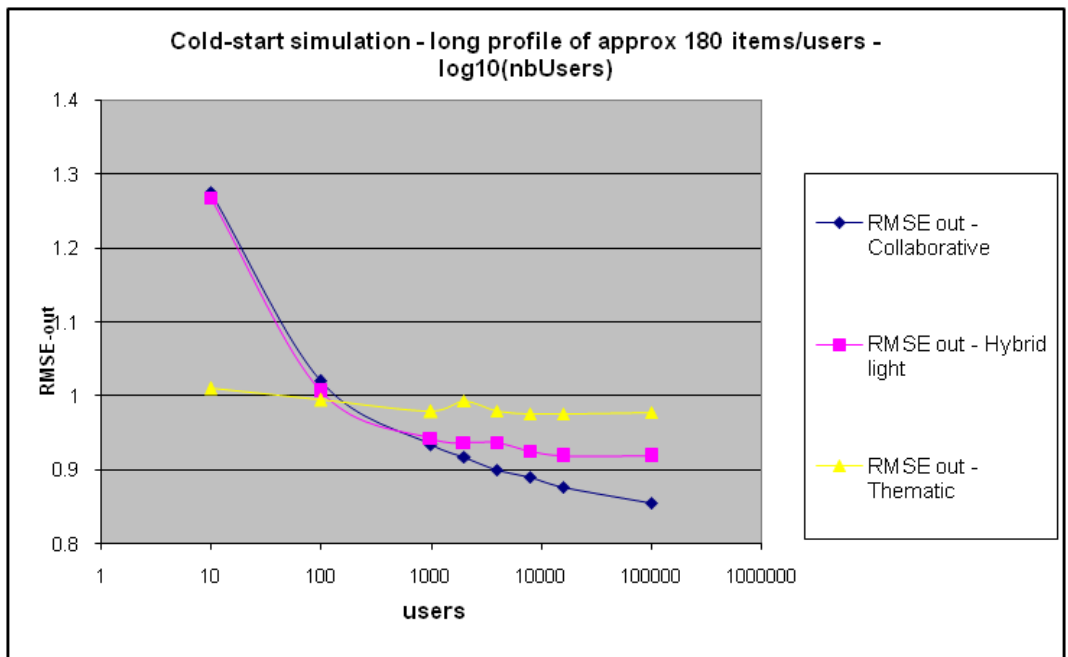

**Figure 3-7: Cold start simulation on long profiles**

---

[7] for thematic system: for about 5% of the predicted ratings, using mean of the current user as default rating

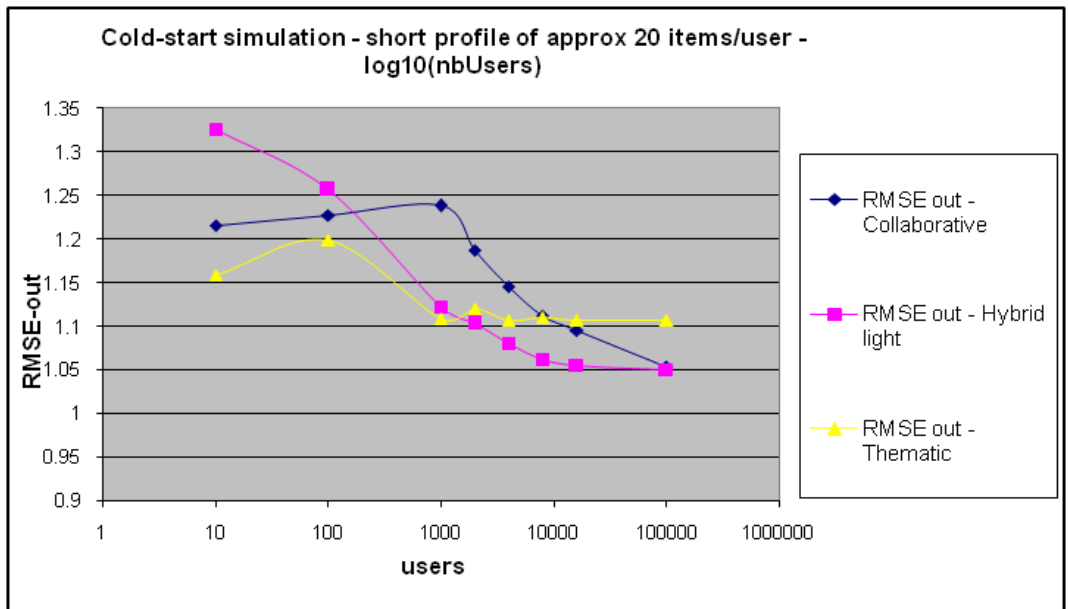

**Figure 3-8: Cold start simulation on short profiles**

### 3.5.6 Conclusion for KNN-based thematic and hybrid recommendation

We studied the performance of recommendation systems based on techniques of item-item similarity matrix of reference data (Netflix) and with different protocols: already loaded system, cold-start with long profiles, cold-start with short profiles.

Thematic-based and hybrid-based techniques with item-item matrix modelling seem to have an interest in some cases identified at least during the start of real on-line services. Thematic and hybrid methods are interesting when the data on users are limited. Thematic and hybrid methods are even more interesting when the user profiles are short. We should now work on the following steps:
- Evaluate complex methods of hybridization, for example mixing the collaborative and thematic data to construct the item-item matrix of similarity, using different attributes' weighting schemes,
- Assess more accurately the phenomena of cold-start when user profiles are short, for technical or legal rasons, for example.

Using the classification (Burke, 2007), test a hybrid system with dynamic "Switch", which would use a thematic, collaborative or hybrid system depending on the size or on other characteristics of the profiles.

## 3.6 Examples of KNN's speed performances

We present some speed performances of Reperio, for similarity and rating calculations, which are critical in industrial deployments.

The 2 use cases presented are:
- one case with a medium catalog dataset of about 11,000 items of very low dimensionality (about 15 descriptors per item on the average). This case is evaluated on Reperio-E.
- one case with a medium log dataset of 17,700 items of medium dimensionality (about 5000 non null ratings on the average). This case is evaluated on Reperio-C.

**Reperio *Embedded***

*Hardware platform used:* Android specification version 1.5, 528 MHz ARM processor Qualcomm MSM7600A, RAM: 288 MB Disk Micro SD 2 GB. Reperio uses an electronic program guide of 60 TV channels (French channels) containing 15 attributes and an average of 1600 items / day. Each item has an average of about 15 descriptors.

**Scoring of items of Electronic Program Guide (EPG)**
Scoring is done with a profile of 25 items scored, using a Jaccard similarity.

|  | Time | Speed |
|---|---|---|
| Time slice (100 items) | ~ 2.2s | ~ 45 scores /s |
| On 1 Day (about 1600 items) | ~ 35.3s | ~ 46 scores /s |

**Table 3-6: Scoring speed in thematic on-line embedded mode**

**Computation of Item-Item similarities**
The tests are carried out from the item "Tropic Thunder", it contains 17 descriptors (descriptor close to the maximum per item). We use the Jaccard similarity.

|  | Time | Speed |
|---|---|---|
| Time Slice (100 items) | ~ 0.08s | ~1 250 similarities /s |
| On 1 Day (about 1600 items) | ~1.2s | ~1 300 similarities /s |

**Table 3-7: Similarity computing speed in thematic on-line embedded mode**

**Reperio *centralized* with all Data in RAM**
For centralized collaborative applications, Reperio is generally used with a powerful server with a lot of memory.

*Hardware platform used:* PC Linux 64-bit OS, 3.40GHz 2-cores CPU, 32GB RAM, 64-bit, 300 MB Disk. Computation time and scoring are based on Netflix's dataset: ~100 million logs, 17 770 items, ~480 000 users. Each item has an average of about 5000 user ratings.

**Computation of similarity matrix and applying a model on a cross-validation**
Scoring is done using information of user profiles in the learning set (in the 90 millions logs), with the Pearson similarity and with the multi-user scoring formula.

|  | Time | Speed |
|---|---|---|
| Learning time of a item-item model (computing the similarity matrix and selecting of 70 neighbors / item) for 90 million logs | ~ 2 h 30 | ~17 500 similarities /s |
| scoring time for 10 millions logs | ~1 minute | ~166 000 scores /s |

**Table 3-8: Similarity and scoring speed on collaborative offline mode**

These first measures seem reassuring: the methods based on KNN approaches can be, if well optimized, very effective even on relatively large volumes of data. However, these measurements for small size of items, or medium number of items.
On-line K-NN search is efficient only if
- The items are represented by a catalog,
- we control the maximum number of descriptors items,
- and each item is represented by a maximum of a few dozen property.

Exact K-NN search is of quadratic complexity in computation time. The current public databases have sizes in number of items very reasonable: less than 4,000 for MovieLens, 17,770 items to Netflix. So the exact K-NN item-item matrix computation is not a problem if well implemented.

However for some industrial applications the catalog may contains several millions of items: for instance, with web sites centralizing users generated contents.

# 3.7 Critical cases: KNN-based system, centralized, with all data stored in database

The previous speed performances were measured in two specific situations: low dimensional item representation, or medium size of the catalog of items. But for huge DBMS (Databases Management Systems), if we have tens of thousands of items or more in high dimensional space such requests last several milliseconds each. On huge database, a KNN based recommender system will face 3 kinds of issues we briefly present:

**Access to vector items bottleneck**

The DBMS will constitute the first critical problem. If the set of logs (or the catalog) is huge it cannot be stored in RAM and then will be stored in a DBMS. Item's KNN exact search require to access to each item's record (users' rating in the collaborative case, item's descriptors in the thematic case). If the catalog is huge, the random access to a record of variable length inevitably takes several milliseconds with the available technologies: we face this problem on various DBMS when the number of entries go beyond one million, sometimes before.

This problem can be solved by extracting in a flat file all the logs before computing the model. This requires just one sequential browsing of the DBMS, much faster than random access. In Reperio, accesses to the DBMS to update the models are done sequentially. Another approach could be to process the logs entry per entry using for instance the Gravity algorithm to build an intermediate data representation stored in a full matrix in low-dimensional space. This will be studied in the next chapter.

**Increasing scoring time for long user profiles**

Even after the model is computed, KNN model uses on-line profile of any user **u** to compute the predicted rating of **u** for an item **i**. This is one strength of this approach, as any new user or user with an updated rating can be scored easily, without rebuilding the KNN item-item model. This is also one weakness: for on-line prediction, user's profile will still have to be accessed by random access. In this case, profile-caches in RAM are necessary.

This problem can be solved by giving a maximum size for the users' profile. A pragmatic approach is for instance to reduce the analyzed profile to the last **P** events, for instance **P**=100. In Reperio, especially in Reperio-E (the embedded version), we used such a technique. Another approach could be to use a short-time and a long-time profiles as in the DailyLearner (Billsus and Pazzani, 2000) and to use a fixed-size clustering for the long-time profile representation.

**Quadratic complexity of the KNN search**

The KNN search requires inevitably a double loop with direct access to each item **i** and direct access to item **j**. The **n×(n-1)** comparisons become very time-consuming as when each comparison costs an average of several millisecond.

This problem can be addressed with the parallelization of the similarity computation process. This is the solution taken by Reperio E/C, which is acceptable for catalogs of several tens of thousands items. For larger catalogs however, other techniques become necessary. Several approaches are currently studied: grid-based collaborative filtering such as AURA (Castagnos and Boyer, 2007) and fast approximate KNN search techniques such as LSH (Gionis et al., 1999), and MinHash (Cohen et al. 2001). We are also considering using Gravity and a classical clustering method to compute fast KNN search.

# 3.8 Conclusion

We show that the K-Nearest-Neighbors methods are very competitive as long as you use a good similarity measure to define neighborhoods. This is not trivial but very important. We show that with a wrong but widely used similarity of Pearson not taking into account all the nonmissing values of the 2 vectors being compared, but just the common values of the two vectors, the internal error of the model increases from 0.86 to 1.25 of RMSE. It is easy to show that the Pearson's similarity reduced to the common values of the two vectors is wrong and leads to poor results.

Yet it is the definition of the Pearson's similarity leading to the 45% additional error that is often used in the literature as reference: for instance: (Sawar et al., 2001) (Adomavicius and Tuzhilin, 2005) (Su and Khoshgoftaar, 2009 ). How can this go unnoticed? First, using a cascaded default predictor with a KNN model one can reduce the RMSE increasing to about 16%. Second, the KNN methods are today less fashionable, and are more used as baseline references to compare with other recent methods: it is understandable that in these cases KNN methods are not always optimized. But again it is urgent to implement the KNN methods to see what are the real gains of the proposed new methods.

It is traditionally expected that hybridization techniques leads to better result that single methods (Burke, 2002). Although it is certainly the case when we hybridize several algorithmic methods (Bell and Koren, 2007c), it might not be the case when we hybridize several data sources, except during the cold start.

Our experiment on the cold-start behavior of collaborative filtering, content-based filtering and hybrid filtering show that we should use sequentially:
1. Content-based filtering during the cold start of a service, with few ratings per user and less than 100 users,
2. then Hybrid filtering, up to 1000 users or to 10000 users if the profiles are still small,
3. then collaborative filtering, which at last provides the best performances.

Finally, we show that KNN-based methods make it possible to efficiently provide all the core functions and implement all the requirements of an industrial recommender system.

# 4 Evaluating Matrix Factorization and KNN in industrial context

"Use recommendation to drive the demand down the long tail"

Chris Anderson, 2004.

This chapter is intended to define a new methodology to analyze the performance of recommender systems. Where are the recommendation methods efficient, regarding to the four previously defined core functions? On which users' and items' segments? What is the added value of the different algorithm techniques?

This chapter should enable us to answer several questions: are matrix factorization methods such as Gravity and KNN methods complementary or does one of the methods systematically outperforms the other one? If the methods are complementary, in which cases should we use Gravity and in which cases should we use KNN? When one needs to produce a matrix of item-item similarity, is it worth using the reduced dimension space provided by Gravity?

# 4.1 A general industrial offline protocol

This section presents a general protocol used to analyze in an industrial and marketing point of view, the behavior of recommendation algorithms such as the KNN algorithm and the Gravity algorithm. We focus only on off-line evaluation protocol: it was not possible to conduct an A/B testing protocol (see 2.7.4) during the time of this work. We will use Train/Test[8] datasets and k-fold cross validation to measure the performances of the algorithms.

## 4.1.1 Data sets

We chose the 2 best known datasets MovieLens (1M) and Netflix. These two datasets have the advantage of public access and to allow performance comparisons with many other techniques. The two datasets were fully analyzed with many recommendation algorithms, but, as far as we know, not with an industrial-driven protocol. These two sources have each a main table of logs. Each log is of the form <user ID, item ID, rating, date>.

Note that the 2 datasets are collected from non-neutral services using recommender systems which could have modified the natural behavior of the users. It is likely that the users' choices for the rated items were changed. Thus, the distribution of couples (user, rated item) is probably different from a distribution that would be observed on a service with no recommendation system. However, we assume that the tastes of the users were not affected significantly: a recommendation system offers items but the user is ultimately the evaluator. The two following tables give a brief overview of the two datasets.

| | |
|---|---|
| Number of Logs | 1,000,209 |
| Number of users | 6,040 |
| Number of items | 3,952 |
| Rating's scale | [1;5] (ratings are integer) |
| % of missing values of the underlying user $\times$ item matrix | 95.6% |
| Source | Ratings collected by a website recommending movies, during 2000. Each user has at least 20 ratings, required before using the recommender service of the website. |
| Possible bias | Effects of the Grouplens' recommender system attached to the website on people choices. |
| Average Train Set's size in number of logs | ~900,000 logs - 90% |
| Average Test Set's size in number of logs | ~100, 000 logs - 10% |
| Validation Set (for Gravity) | 0.5% of the train set |
| Number of folds used (Learning/Test splits) | 2 |
| Average number of ratings per item (Train sets) | ~ 243 |
| Average number of ratings per user (Train sets) | ~ 150 |
| Mean of rating (Train sets) | ~ 3.6 |

**Table 4-1: MovieLens 1M dataset's statistics**

---

[8] We use train/test sets or learning/test sets terms interchangeably

| | |
|---|---|
| Number of Logs | 100,480,507 |
| Number of users | 480,189 |
| Number of items | 17,770 |
| Rating's scale | [1;5] (ratings are integer) |
| % of missing values of the underlying user × item matrix | 98.8% |
| Source | Ratings collected from Netflix's DVD rental service from October 1998 to December 2005. No minimum number of rating per user |
| Possible bias | Effects of Cinematch, the DVD recommender used by Netflix, on people choices. |
| Average Train Set's size in number of logs | ~90,432,000 |
| Average Test Set's size in number of logs | ~10,048,000 |
| Validation Set (for Gravity) | 0.5% of the train set |
| Number of folds used (Learning/Test splits) | 5 |
| Average number of ratings per item (Train sets) | ~5090 |
| Average number of ratings per user (Train sets) | ~190 |
| Mean of rating (Train sets) | ~3.8 |

**Table 4-2: Netflix dataset's statistics**

## 4.1.2 Agnostic thresholds for Long Tail / Short Head analysis

**The long tail controversy**

For a given distribution such as sales of a catalog of items, when the Pareto principle applies, the first 20% of the item hits accounts for 80% of sales. The definition of the long tail is the complementary aspect of those of Pareto's principle: the most frequently-occurring items represent less than 50% of occurrences of the logs (Anderson, 2004).

The existence of the phenomenon of long tail is the subject of many recent debates. Chris Anderson (2004, 2006) indicates that for many internet players the sum of infrequent item purchases exceeds the sum of frequent item purchases. For example, of its 10,000 albums, Ecast, a digital jukebox company, sold at least one track per album per quarter, representing a 98% coverage of its catalog (Anderson, 2006). (Brynjolfsson et al, 2003) found that a large proportion of the books sold by Amazon were not available in conventional stores, giving a competitive advantage to Amazon: Barnes and Noble conventional stores provide around 130,000 books whereas 1/4 of Amazon's sales are done on other books. These examples seem to prove the potential interest of the long tail. But there are also counter-examples: For instance (Tan and Netessine, 2009) found no evidence of long-tail effect in the well-known Netflix's dataset:
- customers who watched long tail items (called niches) are in fact heavy users,
- light users tend to focus only to popular items,
- over time, the demand for hits increases.

In fact we did not find, either for the Netflix's dataset or the MovieLens' dataset, a clear long tail behavior: For MovieLens, items with a number of ratings under the average number of rating per item (243) account for about 70% of the items but only to about 20% of the ratings (usages). For Netflix, items with a number of ratings under the average number of rating per item (5090) account for about 83% of the items but only to about 13% of the ratings (usages)

We did find a Pareto's principle for the Netflix's dataset: less than 17% of the most-frequent items account for more than 86% of the ratings. But for the MovieLens dataset, one needs more than 31% of the most frequent items to gather about 80% of the ratings, therefore we can infer that the 20-80 principle does not strictly apply.

We believe as (Tan and Netessine, 2009) that the definition of the long tail / short head threshold must be relative and catalog-dependent. We prefer to use agnostic thresholds to define segments for both items and users. We used a split around the average number of ratings, both for items and for users. If a long-tail phenomenon or a Pareto distribution existed they would have been noticed with this simple process, allowing refinements if necessary. On the other hand, taking 80% of the ratings directly in a long tail if the distribution is quite uniform would have been arbitrary. As we did not find any long tail effect, either in the MovieLens 1M dataset or in the Netflix dataset, we will use the terms of light/heavy user segment and of unpopular/popular item segment instead of using long tail / short head concepts. Table 4-3 and Table 4-4 give some statistics for these average thresholds for a typical Train Set of 90% of the data, for MovieLens 1M and Netflix.

| | |
|---|---|
| Number of ratings per item (average) | 243.2 |
| Number of ratings per user (average) | 149.0 |
| Number of logs in the Train set (average) | ~900,000 |
| | |
| **Threshold for the Unpopular Item Segment** | **< 240 ratings** |
| Number of items in Unpopular Item Segment | 2537 |
| Percentage of items in Unpopular Item Segment | 68.6% |
| Number of items in Popular Item Segment | 1162 |
| Percentage of items in Popular Item Segment | 31.4% |
| Number of ratings in Popular Item Segment | ~715,000 |
| Number of ratings in Unpopular Item Segment | ~185,000 |
| | |
| **Threshold for the Light User Segment** | **< 150 ratings** |
| Number of users in Light User Segment | 4,156 |
| Percentage of users in Light User Segment | 68.8% |
| Number of users in Heavy User Segment | 1884 |
| Percentage of users in Heavy User Segment | 31.2% |
| Number of ratings in Heavy User Segment | ~640,000 |
| Number of ratings in Light User Segment | ~260,000 |

**Table 4-3: Segments' statistics for the MovieLens' Train Sets (90%)**

| | |
|---|---|
| Number of ratings per item (average) | 5089.0 |
| Number of ratings per user (average) | 188.0 |
| Number of logs in the Train set (average) | ~90,430,000 |
| | |
| **Threshold for the Unpopular Item Segment** | **< 5090 ratings** |
| Number of items in Unpopular Item Segment | 14,832 |
| Percentage of items in Unpopular Item Segment | 83.5% |
| Number of items in Popular Item Segment | 2,938 |
| Percentage of items in Popular Item Segment | 16.5% |
| Number of ratings in Popular Item Segment | ~78,025,000 |
| Number of ratings in Unpopular Item Segment | ~12,405,000 |
| | |
| **Threshold for the Light User Segment** | **< 190 ratings** |
| Number of users in Light User Segment | 335,706 |
| Percentage of users in Light User Segment | 69.9% |
| Number of users in Heavy User Segment | 144,333 |
| Percentage of users in Heavy User Segment | 30.1% |
| Number of ratings in Heavy User Segment | ~68,765,000 |
| Number of ratings in Light User Segment | ~21,665,000 |

**Table 4-4: Segments' Statistics for the Netflix's Train Sets (90%)**

### 4.1.3 Review of the prerequisites of the core functions

What is a good recommender system? If we restrict to an off-line protocol, how can we define a good recommender? We will review the core functions of a recommender and we will define and justify the indicators we choose.

**Decide**

The main use case is a user watching an item description on a screen (a DVD jacket for instance) and wondering if he would enjoy it. Giving a good personalized rating prediction will help the user to choose. For this task, an accurate rating predictor is mandatory. If the rating scale is not simply binary but for instance a typical 1-5 rating scale, the error in prediction doesn't have a linear impact. The larger the error of the rating prediction, the most likely the error in decision: predicting twice a 4 instead of a 5 is less bad than predicting once a 4 instead of a 2.

The "Help to decide function" can be given by the rating prediction function and must be measured by an accuracy metric. Extreme errors must be penalized.

**Compare**

The main use case is a user getting an intermediate short list of items after having given her preferences: this user then wants to compare the items of this short list, in order to choose the one she will enjoy most. We assume in this class of use cases that they can involve pre-filtering processes, that is to say a drastic sub-selection of the catalog. The main implication is that we cannot use a ranking function optimizing only the best ranking result on the catalog for each user: this kind of ranking system will give good ranking results on very popular items but the quality of the ranking won't be guaranteed on unpopular items.

The "Help to compare" function needs a ranking function with an homogeneous quality of ranking over the catalog.

**Discover**

The main use case is a user getting recommended items in a push-based way. We can define 3 main quality properties for this kind of recommendation:
- Obviously, the recommendation must be relevant: the system must recommend items the user will like.
- The recommendation must be useful: it is generally not useful to recommend very popular items as they are generally already known by the user. Helping people to discover is not pushing trivial items, but confidently suggesting 'golden nuggets'.
- The recommendation system must build a trust relationship: this can be formalized in this way: if the system recommend popular items, even if the user does not like them the user will still understand the recommender system: generally the popular items are liked by people ; whereas if the system recommends risky items, it must be sure of its targeting: a user will less likely forgive a bad recommendation on a unknown item.

At this stage, we must distinguish the method of ranking on one hand and the generation strategy of the recommendation on the other. The method of ranking is the general algorithm producing a rank index for each item of the catalog: it can simply use a sorting based on the predicted ratings. The recommendation generation strategy is the process used to select the actual recommended items. The strategy can involve pre-filtering processes and post-filtering processes: for instance, select items from different marketing segments, select items from a white list, delete items too close to those already rated by the user, etc.

To analyze the algorithms we chose to rely on the evaluation of a global ranking of the catalog:
- this strategy is the simplest strategy to push top-N recommendation given a catalog as a list of items,

- this strategy is of course not adapted to real application (where diversity at least by randomization is needed) but gives a good view of the ranking quality of each algorithm for top-scored items,
- this strategy will exhibit the natural behavior of the algorithm: to which item segments and to which user segments they tend to give importance.

The "Help to Discover" function requires at least 3 measures:
1. a measure to determine if a recommended item is appreciated or not,
2. an accuracy measure to determine the percentage of acceptable items (items liked by the user),
3. a value measure to determine the added value of the recommendations.

The protocol for generating recommendations will be to rank all the catalog and then to select the Top-N items.

### Explore (Navigate)

The use case is the anonymous user who is watching an item description on a screen, or the user who is selecting an item to purchase. On pure e-commerce websites, these are crucial instants to add a product to a watch list, or to add another item to the basket. More generally, the item-to-item contextual recommendations can be viewed as a kind of neutral related search. In a way it also helps people to Discover, but is specific enough to be analyzed separately:
- it is the only not trivial automatic recommendation method appropriate for anonymous user,
- it is the most widely used recommendation feature in the industry and the most requested feature by marketing departments for instance at Orange.

Here the generated recommendations will depend directly on a contextual item. The recommendations are not required to be always good: if a user is currently watching an item she knows she doesn't like, the related items are not required to be well perceived by this user. On the other hand, in the case of an interesting contextual item, the related items should be interesting too.

In fact, to assess the quality of the item-item similarity matrices, we have a problem: it is impossible, without a user, to measure the quality of a semantic similarity between two items. For example, for a given user **u**, the movie E.T. may be more similar to Star Wars than to Raiders of the Lost Ark, and this may be the other way with another user **v**. There is no absolute right answer: E.T. and Star Wars are both popular science fiction movies. And on the other hand, E.T. and Raiders of the Lost Ark are all two Steven Spielberg's movies whereas Star Wars was directed by George Lucas. And a third user **w** finally could find a greater similarity between the last two films played by Harrison Ford.

If now we try to evaluate the quality of a similarity matrix by simulating the item-to-item recommendation offline, by associating with each context item **N** most similar items, and using a performance measure, we have another problem : it can be more effective to associate each context item with **N** items optimized for each user, rather than **N** items similar to the context item. Basically, instead of linking the movie White Snow to Bamby, Peter Pan, The Aristocats..., it may be more efficient, to optimize precision for instance, to associate general-and-often-high-rated blockbusters such as Titanic, Star Wars, Lord of the Rings. Now we want to assess the quality of the Help to Explore (navigate) function. We want a good semantic similarity for each associated item, not a crude optimization of accuracy. But only an experiment with real users can assess the semantic similarity. In fact the similarity measures we use, such as Pearson are just factual. They are known to give good correlation measures for our usage data.

We can assess the overall quality of the matrix of similarity between items by an indirect method: the simplest is to check that a KNN model using this matrix will give good performance for other aspects of the recommendation. This is the approach we take, by focusing on measures such as RMSE, precision, ranking.

For the KNN type algorithms, the analysis is straightforward and simple: the similarity matrix is the kernel of the model. The algorithms that are not directly based on a similarity measure need a method for extracting the similarities between the items. For Gravity, this corresponds to a method to compute similarities between the factors of the items.

The Table 4-5 gives a summary of the key points for each core function evaluation.

| Core Function | Key Points - Prerequesites |
|---|---|
| **Help to Decide** Given an item **i**, user **u** wants to know if he will appreciate **i** Rating prediction | The rating prediction must be accurate Extreme errors must be penalized because they increase the risk of wrong decision |
| **Help to Compare** Given **n** items, user **u** wants to know what item to choose Ranking prediction | The predicted ranking must be as correct as possible for every couple of items of the catalog. |
| **Help to Discover** Given a huge catalog of items, user **u** wants to find **k** new interesting items Personalized recommendation | For each user, the system must be able to select in a list of items the most preferred items. We have to identify good / bad recommendations The recommendation must be relevant The recommendation must be useful The recommendation must be trusted |
| **Help to Explore (Navigate)** Given an item **i**, user **u** wants to know what are the related **k** items Item-to-items recommendation | The item-item similarity matrix must be good. This can't be assessed for the semantic point of view without users. But we can evaluate the global quality of the matrix by using it in a KNN model: if the model performs well on the other task, then the item-item similarity matrix is good. |

**Table 4-5: Summary of the prerequisites**

# 4.2 Evaluation process and quality Metrics

## 4.2.1 What we will measure in fact?

The protocol we are going to carry out rely on a Learning/Test set methodology: the initial dataset is split into 2 datasets Learning and Test. The models are built on the Learning set whereas the evaluation is done on the Test set.

We use 2 datasets of logs of ratings of users on items. Let be **R(u,i)** the initial rating matrix, **U** the set of the users, **I** the set of the items, |**U**| the number of users, |**I**| the number of items, **L(u,i)** and **T(u,i)** respectively the Learning set and the Test set randomly extracted from **R**, |**T**| the number of logs in Test. For the MovieLens data **R** has 95.6% of missing values and for the Netflix data **R** has 98.8% of missing values. As for both datasets we cannot assume that the values of **R** are missing at random (Marlin and Zemel, 2009), we could also not assume that the values of **L** and **T** are missing at random.

We assume a loss function named **loss(x,y)** giving the cost of the error between **x** and **y**. We also assume an ideal matrix with no missing value, $\widetilde{R}(u, i)$, where all information about every rating of every user on every item are known.

The error evaluation phase has a bias if we look for the global error as defined below:

$$E = \left(\frac{1}{|U|} \times \frac{1}{|I|}\right) \times \sum_{u \in U, i \in I} loss\left(P(u, i), \widetilde{R}(u, i)\right)$$

Indeed, we are going to carry out our evaluation on the Test set **T(u,i)** so with:

$$E_{Test} = \left(\frac{1}{|T|}\right) \times \sum_{u \in U, i \in I, (u,i) \in T} loss(P(u, i), T(u, i))$$

$E_{Test}$ is not an estimator of **E** because **T** has the same distribution as **R**, and its values are not missing at random. An estimator of **E** would be:

$$E_{Estimated} = \left(\frac{1}{|U|} \times \frac{1}{|I|}\right) \times \sum_{u \in U, i \in I, (u,i) \in T} \left(loss(P(u, i), T(u, i)) \times \frac{1}{Prob((u,i) \in T)}\right),$$

with $\mathbf{Prob}\big((u, i) \in T\big)$ : the probability that the value of **(u, i)** is present in the Test set **T**.

The problem is that we do not know the probability distribution for **T** as we do not know that for **R**. We can try to approximate the probability distribution for **T**, but we have to assume that the distribution of the users and the distribution of the items are independent [9].

With this hypothesis, we have: $\mathbf{Prob}\big((u, i) \in T\big) = \frac{K(u)}{|T|} \times \frac{K(i)}{|T|}$,

with $K(u) = |\{j, (u, j) \in T\}|$, $K(i) = |\{v, (i, v) \in T\}|$. **K(u)** and **K(i)** are respectively the number of items rated by u and the number of users who rated i.

Finally an estimator of the error would be, with the hypothesis of independence distribution of the users and the items :

$$E_{Estimated} = \left(\frac{1}{|U|} \times \frac{1}{|I|}\right) \times \sum_{u \in U, i \in I, (u,i) \in T} \left(loss(P(u, i), T(u, i)) \times \frac{|T|^2}{Prob((u,i) \in T)}\right)$$

Another way is to consider that:

$$E_{Test} = \left(\frac{1}{|T|}\right) \times \sum_{u \in U, i \in I, (u,i) \in T} loss(P(u, i), T(u, i))$$

is still an interesting measure. In fact we do not have to assume that the request to the recommender system will have an uniform distribution. It is also possible to assume that this measure is closer to the real most interesting value we would like to estimate: the average error of the system in deployment. De facto, heavy users will use the service and the recommender system more often that light users and should be considered more carefully than light users. And although the recommendation systems should enhance the long tail, the requests for the "help to decide" and the "help to compare" tasks, at least, will more often be related to popular items. We will take this last hypothesis in our protocol, but keeping in mind that other hypothesis are also possible, and that our protocol is not an ideal one.

---

[9] This might not be true at least with the Netflix dataset, see (Tan and Netessine, 2009).

Finally an ideal methodological process would be:
- using a dataset of logs of usages from a service not using a recommender system nor any kind of automatic promotional system (advertising),...
- and using a dataset where the feedback of the users on the items is measured randomly, to insure that the data are missing at random.

But, except for the second constraint where (Marlin and Zemel, 2009) indicated that Yahoo could provide such a data, no dataset compliant with these constraints are today publicly available.

### 4.2.2 Evaluation protocol for the accuracy (Help to Decide)

After modelling, every log (**u, i, r$_{u,i}$**) for a user **u**, an item **i** and the rating **r** in the Test Set is used to ask the predicted rating $\hat{r}_{u,i}$ to the model, then the RMSE between all **r$_{u,i}$** and **$\hat{r}_{u,i}$** is computed

$$RMSE = \sqrt{\frac{1}{|T|} \sum_{(u,i,r) \in T} (\hat{r}_{u,i} - r_{u,i})^2} \qquad (4\text{-}1)$$

with **T**: the Test Set

*Note*: The RMSE has the same unity as the ratings; the lower the RMSE, the better the global relevancy.

### 4.2.3 Evaluation protocol for the ranking (Help to Compare)

After modelling, for each user u and for each couple of item **(i, j)** in the Test Set rated par **u** with **r$_{u,i}$>r$_{u,j}$** or **r$_{u,i}$<r$_{u,j}$**, the preference given by **u** is compared with the predicted preference given by the recommender method, using the predicted ratings $\hat{r}_{u,i}$ and $\hat{r}_{u,j}$.

The measure used is the Normalized Distance-based Performance Measure (NDPM) of (Herlocker et al.,2004) :

$$NDPM = \frac{2C^- + C^u}{2C^i} \qquad (4\text{-}2)$$

with :
$C^-$: number of contradictory preference relations between the system ranking and the user ranking. A contradiction happens when the system says that item **i** will be preferred to item **j** whereas user ranking says the opposite
$C^u$: number of compatible preference relations, where user rates item **i** higher than item **j**, but the system ranks at the same level **i** and **j**
$C^i$: number of preferred relationships of the user: the number of pairs of rated items **(i, j)** for which the user gives a higher rating for on item than for the other.

If the NDPM's scale does not separate the measures enough, the percentage of compatible preferences could be used. The percentage of compatible preferences is not dataset-independent but this is not a problem for our benchmark if we analyze MovieLens and Netflix separately.

*Note*: the NDPM is an index ; the lower this index, the better the ranking performance. However for the percentage of compatible preferences, the greater the percentage, the better the ranking performance.

## 4.2.4 Evaluation protocol for the relevancy (Help to Discover)

After modelling, every user **u** in the Train Set receives **N** recommended items.

The recommendation strategy is to score all the catalog then push the 10 items with the best predicted score, not already known in Train Set.

This approach would not be suitable for real recommendation because it is deterministic and too expensive: it always returns the same Top-N items (for a deterministic algorithm) and scoring full catalog takes time. This approach, however, to evaluate the "best ranking" behavior of the algorithm studied. This method promotes a ranking engine that focuses on the high scores. This is the general approach of the search engines for which the first results of a query are the most important because they are the only ones evaluated by users.

An item **i** recommended for the user **u**
- is considered **relevant** if **u** has rated **i** in the Test Set with a rating greater than or equal to her rating mean
- is considered **irrelevant** if **u** has rated **i** in the Test Set with a rating lower than her mean of rating
- is not evaluated if not present for **u** (not rated by **u**) in the Test Set.

The users mean are computed on the Train Set; with the random sampling between Train and Test they are assumed stable.

**H** will stand for the set of evaluable recommendations in the Test set, that is to say the set of couples **(u,i)**, **i** being the recommended item to the user **u**, which existed in the Test Set. |**H**| is the size of **H**, in number of couples **(u i)**.

$$\text{precision} = \frac{\text{number of relevant recommended items}}{|\mathbf{H}|} \quad (4\text{-}3)$$

*Note*: The precision is given in percentage of good recommendations: the greater the precision, the better the global relevancy.

## 4.2.5 Evaluation protocol for the Impact of the recommendations (Help to Discover)

We first define a Measure of Impact of a recommended item (MI) which is the normalized invert frequency of this item - equivalent to the invert of the sum of its counts multiplied by +1 if the item is finally a good recommendation and multiplied by -1 if the item is finally a bad recommendation.

First we give an absolute definition, noted aMI, of the impact of the recommendation of the item **i** to the user **u**:

$$\mathbf{aMI_u(i)} = \frac{1}{\text{count}(i)} \times \begin{cases} -1 \text{ if u doesn't like i} \\ +1 \text{ if u like i} \end{cases} \quad (4\text{-}4)$$

where **count(i)** is the number of ratings of item i on the Train Set.

As we defined, for the fields of rating-based logs of usages, a good recommendation as an item which is, in final, is rated by the user greater than her rating average, we can write :

$$\mathbf{aMI_u(i)} = \frac{1}{\text{count}(i)} \times \text{sign}(r_{u,i} - \bar{r}_u) \quad (4\text{-}5)$$

with $\bar{r}_u$, the mean of ratings of the user **u**.

The aMI measure will always give low decimal results as on the average, an item i of a catalog I of size |I| will have an impact of 1/|I|. For the Netflix's catalog for instance, this average impact is 1/17770, which is about 0.000056. We chose then to normalize the absolute Measure of impact by multiplying it by the size of the catalog. This will also allow to compare this measure across different datasets.

Finally we obtain:

$$\mathbf{MI_u(i)} = \frac{1}{\text{count}(i)} \times \text{sign}(r_{u,i} - \bar{r}_u) \times |\mathbf{I}| \tag{4-6}$$

with $\bar{r}_u$: the mean of ratings of the user **u**, and |**I**| the size of the catalog (the total number of available items).

The basic idea is that, the more frequent a recommended item is, the less impact the recommendation has. This is summarized in the next table:

|  | **Impact of the recommendation** | |
| --- | --- | --- |
|  | **Impact** if the user **likes** the item | **Impact** if the user **dislikes** the item |
| Recommending a popular item | **Low**:<br><br>The item is likely to be already known at least by name by the user. | **Low**:<br><br>Even if the user dislikes this item she can understand that as a popular item this recommendation is likely to appear... at least at the beginning. |
| Recommending an unpopular item | **High**:<br><br>The service provided by the recommender system is efficient. The rarest the item was, the less likely the user would have find it alone. | **High**:<br><br>Not only the item was unknown and did not inspire confidence, but it also was not good. |

**Table 4-6: Principle of the measure of impact of the recommendation, offline**

The global Sum of Measure of Impact of a list **Z** of recommendations is the sum of the MI only for the relevant recommended items for each user.

$$\mathbf{SMI_u(Z)} = \sum_{(u,i) \in Z, (u,i) \in H} \frac{1}{\text{count}(i)} \times \text{sign}(r_{u,i} - \bar{r}_u) \times |\mathbf{I}| \tag{4-7}$$

where **H** denotes the set of the evaluable recommendations in the Test set.
where **Z** denotes a set of couples (user, item), representing a set of recommendations.

An finally, the Average Measure of Impact of a list of recommendation is the averaged SMI:

$$\mathbf{AMI_u(Z)} = \frac{1}{|H|} \sum_{(u,i) \in Z, (u,i) \in H} \frac{1}{\text{count}(i)} \times \text{sign}(r_{u,i} - \bar{r}_u) \times |\mathbf{I}| \tag{4-8}$$

*Note*: the absolute impact of a recommendation (aMI) has a minimum of -1 and a maximum of +1. The MI could in theory range between **-|I|** and **+|I|** which corresponds to the recommendation of an one-count item for a user who enjoyed it. The AMI has the same range, but only in theory. Just remember that the greater the AMI, the better the positive impact on users.

## 4.2.6 Evaluation protocol for the quality of an item-item similarity matrix (help to Explore)

For this evaluation, we will need a item-item similarity matrix. This is straightforward in the case of a KNN model, but in the case of a Matrix Factorization model, we have to find a way to compare any couple of items. Once we have a similarity matrix given by the model, we can use it in a classic KNN model and check all the other measure of performances: we can assume that the most accurate the KNN model, the best the similarity matrix. We will use one baseline similarity matrix to have performance references: a similarity matrix build at random: a uniform random function is used to compare each pair of items and then the K most similar items are selected to each item of the catalog.

| Key Points - Prerequisites | Adapted measures |
|---|---|
| **Help to Decide** <br><br> The rating prediction must be accurate. <br><br> Extreme errors must be penalized because they increase the risk of wrong decision. | The RMSE will be well adapted for the "Help to Decide" performance measurement. This measure penalizes extremes errors whereas for instance the MAE do not. |
| **Help to Compare** <br><br> The predicted ranking must be good for every couple of items of the catalog. | The NDPM (Herlocker et al., 2004) will be well adapted for the "Help to Compare" performance measurement. This measure of ranking doesn't give more importance to the best ranked items. <br><br> The percentage of compatible ranks between any couple of rated items and its associated couple of predicted ratings of the items could also be used, if it discriminate better the performances. |
| **Help to Discover** <br><br> For each user the system must be able to select in a list of items the most preferred items. <br><br> We have to identify good / bad recommendations. <br><br> The recommendation must be precise. <br><br> The recommendation must be useful. <br><br> The recommendation must be trusted. | The average rating of the user **u** can be used as a threshold to determine if a recommended item **i** was liked by **u**, once we have the rating $r_{ui}$. <br><br> Once the good recommendations identified, the classical Precision measure can be used to evaluate the precision of the recommender. <br><br> Useful and trusted recommendations will be measured via the recommendation impact measure we defined in this section. <br><br> We will use 2 indicators, Precision and AMI. |
| **Help to Explore** <br><br> The simple item-to-item recommendation must lead to precise recommendations <br><br> We have to identify good/bad recommendations. <br><br> The recommendation must be precise. | A good similarity matrix for the task "Help to Explore" is a similarity matrix leading to good performances, in accuracy, relevancy, usefulness and trust. |

**Table 4-7: Adapted measures for each core function**

### 4.2.7 Complementary information about our test

**Working mode of our Gravity-based model**
Unless otherwise specified, we use the following parameters for the gravity-based model:
- Learning rate: 0.030
- Regularization factor: 0.008

All factors are constrained in [-1;+1] and ratings are internally normalized in [0;1] and de-normalized during RMSE computing (avoiding some rare but possible numeric divergences).

We use a learning process with early stopping to prevent overtraining:
- Validation Set: 1.5% of the Training Set.
- We will use 2 modes of early stopping.
- Limited convergence time mode: stop learning after 1 hour and an half, or if the RMSE increases successively after 3 iterations on Validation Set whereas RMSE is still decreasing on the Train Set .

The choice of the early stopping by time is guided by industrial constraints: we can't rely on a good algorithm if this algorithm reaches its optimum performance too slowly. In our experiments, all KNN models are built within 1 hour and an half, so we used this time threshold for Gravity.

We will use bias for user and bias for item (1st factor user, 2nd factor item) as in the BRISMF version of Gravity (Takacs et al., 2009). See also 2.4.3.

**Working mode of our KNN-based model**
Unless otherwise specified, we will use the following settings for the KNN-based model:
- KNN type: item-item similarity matrix
- Similarity used: weighted Pearson
- Scoring method: collaborative (multi-user mean-based, see 3.2)
- Default predictor: (mean user + mean item) / 2, or the available mean.

**Baseline Predictor 1: Uniform Random Predictor**
A uniform random model generates uniform random ratings for each rating prediction. These ratings are used both to measure the accuracy of the system (RMSE) and the simple relevancy of the system (precision for 10 recommendations) using a full catalog scoring and keeping the 10 best scores.

**Baseline Predictor 2: basic Default Predictor (mean u + robust mean i)/2**
This is the default predictor used in the Reperio engine, with a robust mean for items. The robust mean of an item **i** is calculated only if **i** has at least 10 ratings. Otherwise the predictor will return the mean of the user **u**'s ratings. If **u** has no rating (in the Training Set, with some sampling this may happen), then the Default Predictor returns the global mean of the ratings of the logs.

**Our configuration**
All our test are carried out on this configuration: Personal Computer with 12 GB Ram, processor Intel$^{TM}$ Xeon$^{TM}$ W3530 64-bit-4-core processor running at 2.8 GHz, hard disk of 350 GB. All algorithms and the benchmark process are written in Java$^{TM}$.

**Notes:**
We will use the following abbreviations for the segmentation of the performance:
**Huser**: Heavy users
**Luser**: Light users
**Pitem**: Popular items
**Uitem**: Unpopular items (the meaning of unpopular is rather "rare", "infrequent")

In order to make results easier to read, we give information about MovieLens performances only if they differ significantly from those of Netflix's.

## 4.3 Results for score-based functions

### 4.3.1 Help to decide: rating prediction, and speed performances

**Baseline predictors' performances**

We first measure the RMSE performances of our reference predictors called baseline predictors. The result, globally and for each segment, are given in the next two tables.

| Baseline predictor | dataset | measures | Global RMSE | rmse Huser Pitem | rmse Luser Pitem | rmse Huser Uitem | rmse Luser Uitem | Scoring time (s) |
|---|---|---|---|---|---|---|---|---|
| **Basic default predictor** | **MovieLens** | **mean** | **0.9573** | 0.9386 | 0.9484 | 1.0038 | 1.0552 | 0.22 |
| Basic default predictor | MovieLens | stand. Dev. | 0.0018 | 0.0022 | 0.0012 | 0.0054 | 0.0098 | 0.02 |
| **Basic default predictor** | **Netflix** | **mean** | **0.9640** | 0.9399 | 0.9986 | 1.0071 | 1.0683 | 49.63 |
| Basic default predictor | Netflix | stand. Dev. | 0 | 0.0001 | 0.0002 | 0.0006 | 0.0005 | 3.39 |

**Table 4-8: Baseline predictors's performance in RMSE: Default Predictor**

| Baseline predictor | dataset | measures | Global RMSE | rmse Huser Pitem | rmse Luser Pitem | rmse Huser Uitem | rmse Luser Uitem | Scoring time (s) |
|---|---|---|---|---|---|---|---|---|
| **Random Uniform** | **MovieLens** | **mean** | **1.7073** | 1.6955 | 1.7689 | 1.648 | 1.7197 | 0.31 |
| Random Uniform | MovieLens | stand. Dev. | 0.0030 | 0.0026 | 0.0051 | 0.0113 | 0.0169 | 0.05 |
| **Random Uniform** | **Netflix** | **mean** | **1.6962** | 1.6774 | 1.7520 | 1.6832 | 1.7498 | 81.40 |
| Random Uniform | Netflix | stand. Dev. | 0 | 0.0001 | 0.0001 | 0.0003 | 0.0012 | 37.40 |

**Table 4-9: Baseline predictors's performance in RMSE: Random Uniform Predictor**

Globally, we see that in the worse case, with the Random Uniform Predictor, the RMSE is about 1.7 for a rating scaling of 1 to 5. The basic Default Predictor has an error rate 45% lower than that of the Random Predictor. Its behavior both on MovieLens data and on Netflix data is good and stable.

We also see that heavy users (Huser) and Popular items (Pitem) are easier to model with the basic Default Predictor than the light-user-unpopular-item segment.

**KNNs' RMSE performances**

We carried out tests for different sizes of neighborhood, compliant with our tasks in an industrial context. Increasing the number of KNN generally increases the performances but it will be complicated to build matrix of **K** Nearest Neighbors and associated similarity weights with high values for **K** (>500 or >1000) as this matrix must be kept in Random Access Memory. For very large catalog applications, the size of the KNN matrix must be reasonable.

The KNN method performs well except when **K** is small and except for the light-user-unpopular item segment (Luser Uitem). This fact is confirmed on MovieLens data. The results are consistent across both datasets. We see a significant gap between the RMSE for the light-user-unpopular-item segment and the RMSE of the heavy-user-popular-item segment. Clearly, the KNN model is not adapted to the former, whereas it performs well on the later.

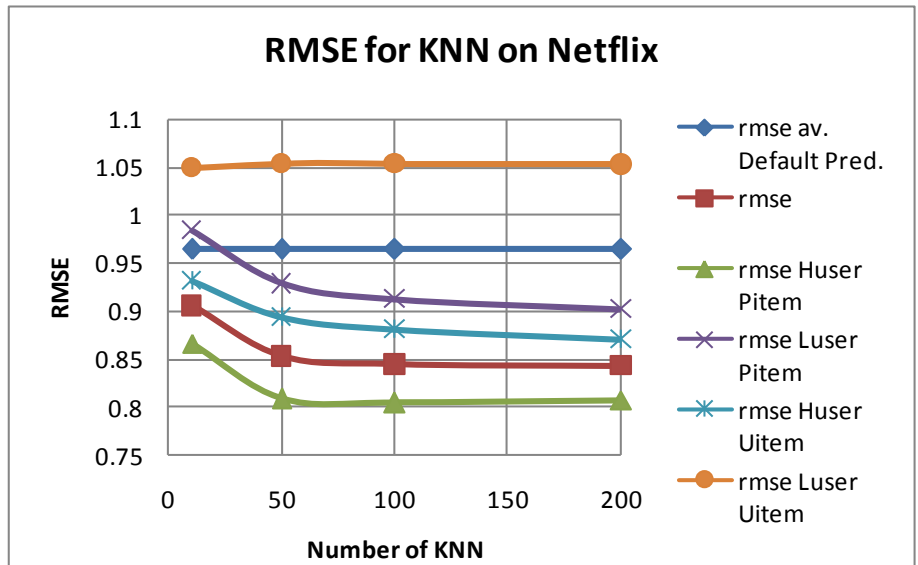

**Figure 4-1: KNN's RMSE performances on Netflix data**

### Gravity's RMSE performances

We carried out tests for different number of factors. We see that Gravity has difficulties modelling the light-user-unpopular-item segment: on this segment the RMSE never decreases under 0.96. On the contrary the RMSE for heavy-user-popular-item is close to the 0.81 value between 16 and 32 factors, and the two symmetrical segments light-user-popular item and heavy-user-unpopular-item both have also good low RMSE. The RMSE decreases when the factor increases up to around 20 factors. After, the RMSE increases. It is a consequence of our time-constrained early stopping condition: remember that we wanted Gravity to last no more than a KNN model so the algorithm is given 1 and a half hours of computation max. On Netflix this corresponds to about 140 passes on the Train dataset. The optimal number of factors seems to be between 16 and 32. We found similar results with MovieLens, with no significant RMSE improvements after 32 factors.

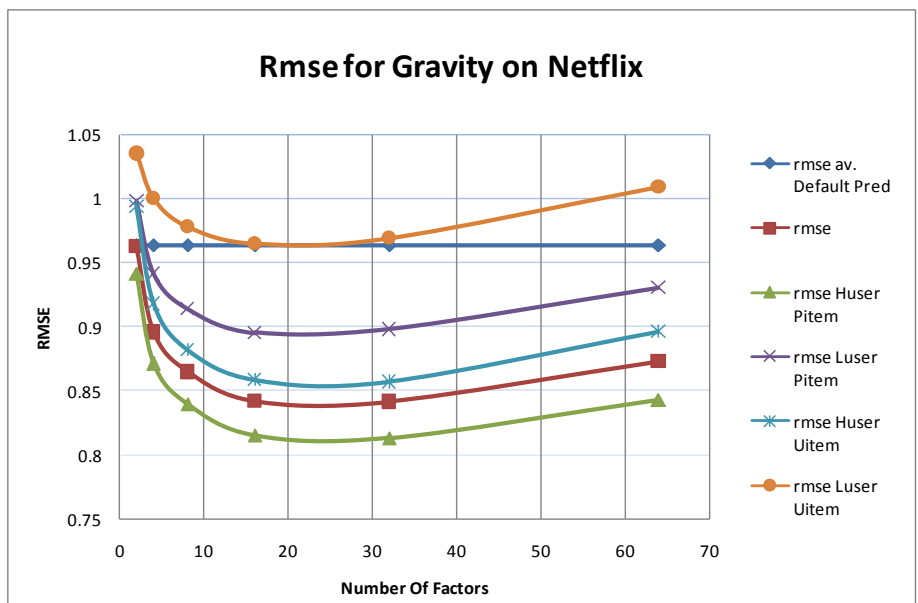

**Figure 4-2: Gravity's RMSE performance on Netflix data - limited time to converge: 1 and a half hours.**

If we restrict Gravity to have a computation time of 1 and an half hours (5400 seconds) corresponding to the time of a KNN search, the optimum results are around 24 factors. Gravity outperforms the KNN methods for the RMSE, especially for the Light-user-unpopular-item segment, even if in fact the performances for this segment are quite similar to those of the average of the Default Predictor (but the Default Predictor has a RMSE of 1.07 on this segment).

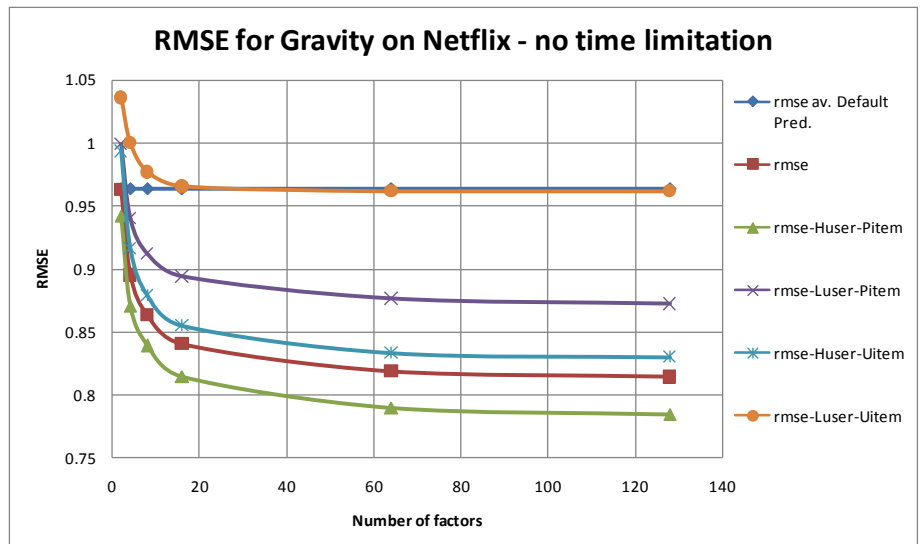

**Figure 4-3: Gravity's RMSE performance on Netflix data - no limited time to converge**

When we don't restrict the run of gravity within 1 and a half hours, the RMSE can decrease, for the 2 heavy-user segments, under 0.85, even under 0.80 for the heavy-user-popular-item segment. But for the light-user-unpopular-item segment the RMSE is always greater than 0.95, close to the Default Predictor global performance (RMSE around 0.96 on Netflix). The convergence time during the model building phase is not in favor of Gravity. Searching the K nearest neighbors of **N** items must be done by computing the similarities of all couples $(\mathbf{i}, \mathbf{j}) \in \mathbf{N} \times \mathbf{N}$. Then for each item **i** a selection of the **K** most similar items to i is carried out. So searching for 10 or 200 nearest neighbors does not change a lot the processing time, as we see in the next figures: all the KNN search are computed, on the Netflix data (17,770 items) in less than about 1 and a half hours. On the other hand, the convergence time of Gravity is very slow if we only rely on a RMSE sudden rising curve on a validation set. For instance, it takes more than 10 hours with a 64-factor model and almost 40 hours for a 128-factor model.

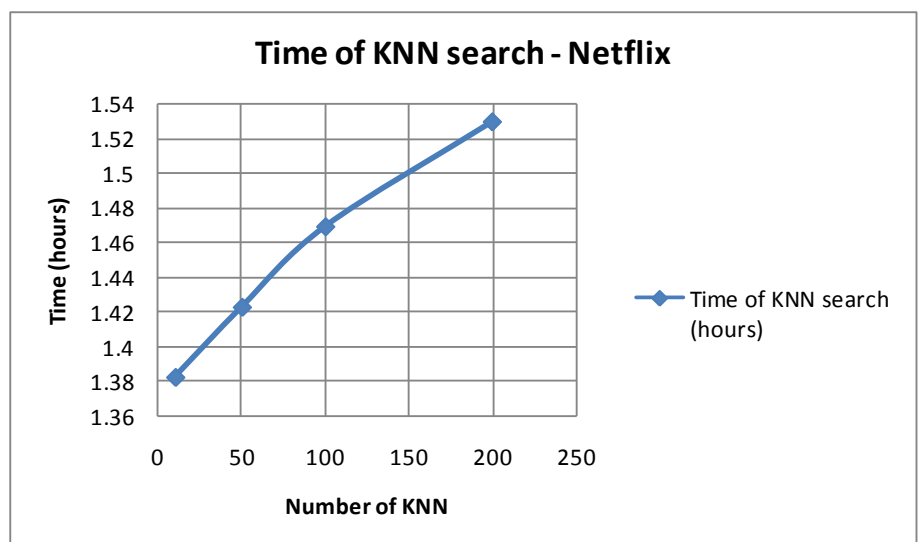

**Figure 4-4: Time for the KNN search on Netflix's data**

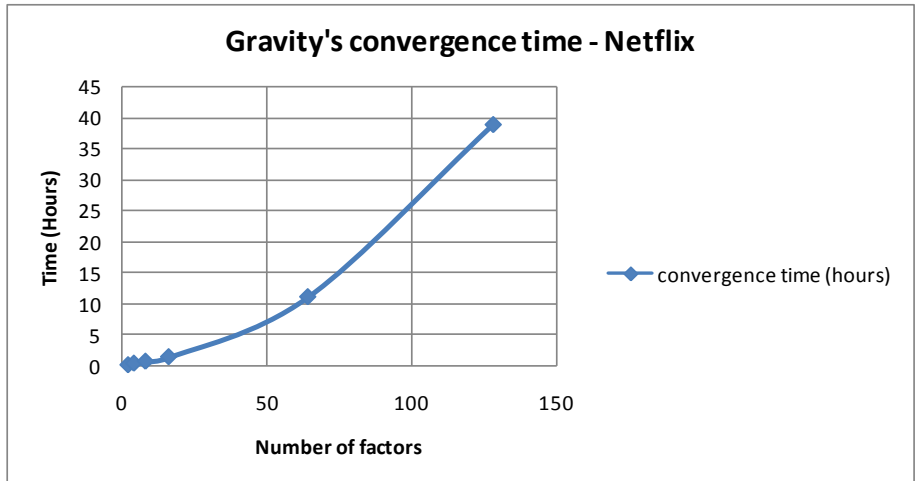

**Figure 4-5: Gravity's time to converge with early stopping without time max, on Netflix's data**

The next figure gives an example of the average scoring speed of the different algorithms (except the Random Predictor which is not relevant in this case). We take for the KNN model a neighborhood size K=100 and for Gravity a number of factors F=16 to compare the speed algorithm on similar accuracy performances. The figure gives the number of rating predictions, that is to say the number of predicted ratings for a user **u** and an item **i**, computed per second. Note that this time includes I/O. KNN is twice slower than Gravity; a KNN model has a scoring time linearly dependant to the size of the users' profile whereas Gravity, using a scalar product of vectors of fixed size, is more efficient for this task.

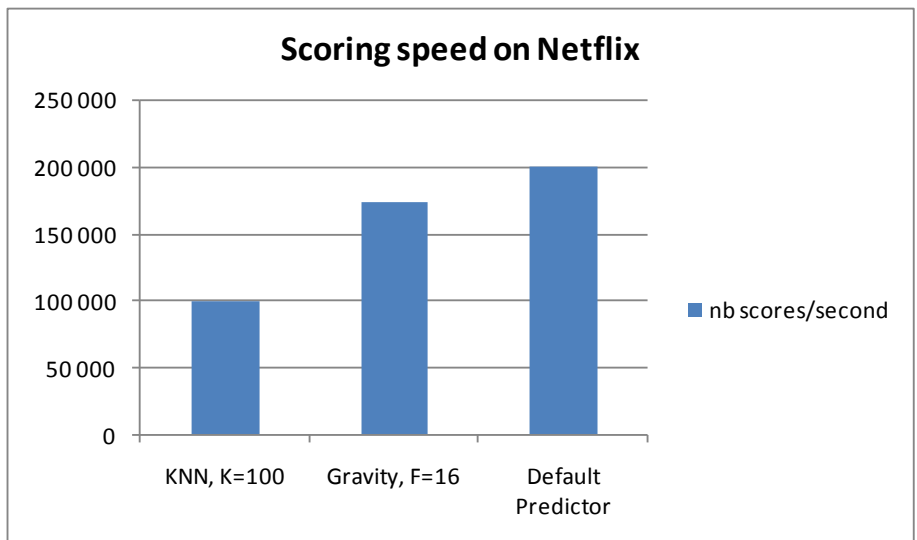

**Figure 4-6: Average number of rating predictions (scores) per second for the 3 main predictors**

**Analysis for the RMSE**
We keep for this analysis the results of Gravity with time-based early stopping.
A global comparison focusing on Netflix results shows that:
- Gravity dominates KNN for the light-user-unpopular-item-segment (the Default Predictor has a RMSE of about 1.07 on this segment, despite a global mean of 0.96.
- KNN dominates Gravity on Netflix for the heavy-user-popular-segment. This is not true when Gravity is allowed to converge without time restriction, but in this case the convergence time becomes a problem.
- Gravity dominates KNN for the light-user-popular-item segment and for the heavy-user-unpopular-item segment.

| Best accuracy (RMSE) per segment | | |
|---|---|---|
| | Popular items | Unpopular items |
| Heavy users | KNN | Gravity |
| Light users | Gravity | Gravity |

**Table 4-10: global analysis for the RMSE: best algorithm for each segment**

## 4.3.2 Help to compare: full catalog ranking

**Baseline predictors' ranking performances**

For the ranking performances, we note no real differences between the segments as shown in Figure 4-7. On the other hand, we note the relatively good performances of the default predictor. The number of preferences actually evaluated is the number of couples of item **(i,j)**, for any user **u** in test, where **u** has expressed a preference, i.e. either $r_{ui}>r_{uj}$ or $r_{ui}<r_{uj}$. This number of preferences is the same for all the analyzed algorithms, given a dataset.

| Predictor | NDPM | % compatible | nb compatible. | nb contrad. | nb pref evalued | Stand. Dev. of NDPM |
|---|---|---|---|---|---|---|
| Random Uniform | 0.7500 | 0.4999 | 688,299 | 688,504 | 1,376,802 | 0.0003 |
| Default Pred | 0.6340 | 0.7319 | 1,007,707 | 369,096 | 1,376,802 | 0.0006 |

**Table 4-11: Statistics for Baseline Predictors for the full ranking task on MovieLens**

| Predictor | NDPM | %compatible | nb compatible. | nb contrad. | nb pref evalued | Stand. Dev. of NDPM |
|---|---|---|---|---|---|---|
| random Uniform | 0.7500 | 0.5000 | 109,000,000 | 109,000,000 | 218,000,000 | 0.000001 |
| Default pred. | 0.6587 | 0.6826 | 149,000,000 | 69,250,000 | 218,000,000 | 0.000087 |

**Table 4-12: Statistics for Baseline Predictors for the full ranking task on Netflix**

With the Default Predictor, the percentage of compatible preferences is significantly better (greater) on MovieLens than on Netflix. The NPDM is also better on MovieLens (lower than that on Netflix). Remind that there is no Pareto's effect on the MovieLens' datatset whereas there is this phenomenon in Netflix's. Many items in Netflix have few ratings and no stable mean. The ranking of the Default Predictor, based on items' mean, is then less reliable.

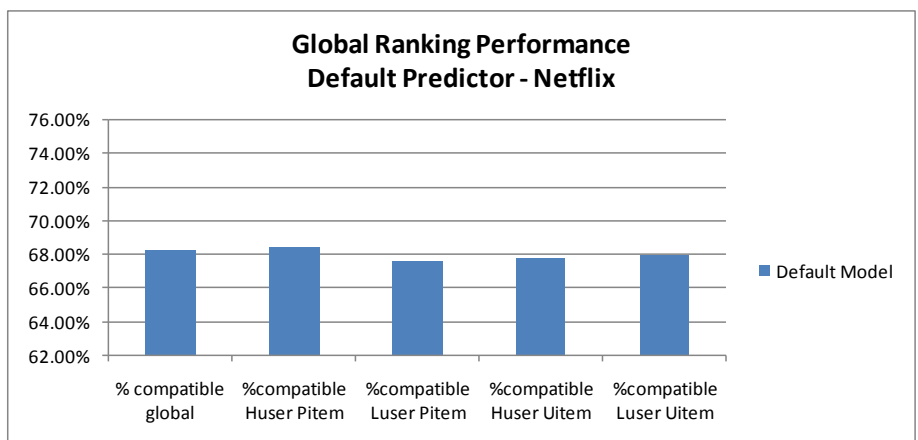

**Figure 4-7: Performances in global ranking for the Default Predictor on Netflix**

*Note that as the percentage of compatible ranking discriminates better the results (larger scale) we will now use, on the separate datasets, this measure instead of the NDPM.*

Page 142

**KNNs' ranking performances**

On Netflix, KNN performs well as shown is the next figure, except for the light user segment. Again the maximum of ranking compatibility is around 77% for heavy users' segments.

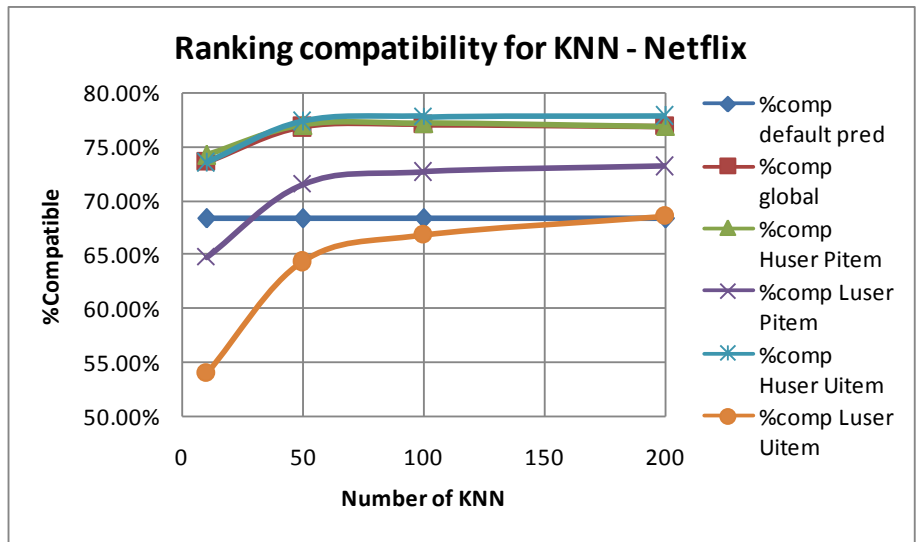

**Figure 4-8: Ranking compatibility for KNN on Netflix**

Comparing the default model with the KNN on MovieLens shows that globally the KNN models doesn't perform well for global ranking: it does not outperform significantly the Default Predictor model. It gives better performances only of the Heavy users's segments, up to 77% of preferences compatibilities.

**Gravity's ranking performances**

We give the result for the time limited version of run for Gravity: Gravity can converge only during 1 hour and a half. For the Light-user Unpopular-item segment Gravity outperforms the KNN model, quickly modeling the ranking for this segment with the same performances as for the Light-user-popular-item segment. For the rest, the performances are similar to those of KNN. The results for Gravity are consistent across both dataset, except that on MovieLens the curves do not decrease after 32 factors as Gravity has enough time to converge.

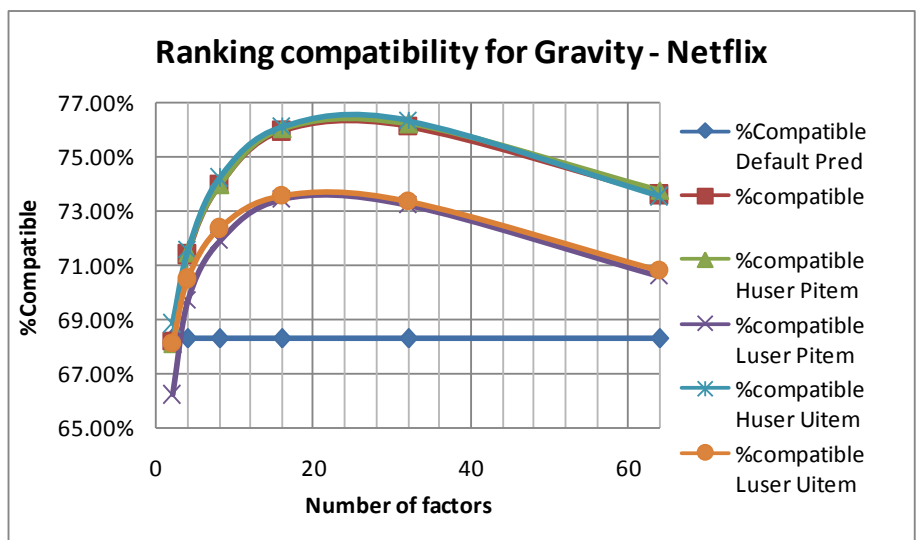

**Figure 4-9: Ranking compatibility for Gravity on Netflix.**

**Analysis of the ranking performances**

Again we focus on Netflix and we compare the result for the ranking.
- Gravity clearly dominates for the 2 segments of light users
- KNN slightly dominates for the 2 segments of heavy users even if Gravity could work almost as well as KNN for the good number of factors.

| Best ranking (%compatibility) per segment | | |
|---|---|---|
| | Popular items | Unpopular items |
| Heavy users | KNN | KNN |
| Light users | Gravity | Gravity |

**Table 4-13: Global analysis of the ranking task (help to compare)**

## 4.3.3 Help to Discover: push of useful recommendation

**Analysis using the Precision Measure**

**Baseline predictors' precision performances**

On Netflix, the precision of random predictor is around 50% whereas the precision of the default predictor is very good, about 92.8%. If we analyze the performances according to our different segments we see a good stability of the results both for the Random Predictor and the Default Predictor.

On MovieLens, the precision of the Random Predictor is around 55%, with a peak at 61% for the heavy-user-popular-item-segment and a minimum at 39.6% on the light-user-unpopular-item segment. The reason of the peak is quite simple: it is popular items, even at random, which are counted in the heavy-user-popular-item segment and these items are generally enjoyed. The Default predictor has an average precision of 86.7% which is quite good for its simplicity. Its performance decreases only on the light-user-unpopular-item segment.

| Precision of baseline predictors on Netflix | | | |
|---|---|---|---|
| Predictor | Precision | Stand. Dev. Of Precision | nb reco eval |
| Random Pred | 0.5304 | 0.0002 | 5916 |
| Default Pred | 0.9286 | 0.0000 | 38250 |

**Table 4-14: Precision of baseline predictors on Netflix**

With an average of 92.9% of good recommendations for a Top-10 best rated items (with a robust mean), and a remarkable stability of the performances on the different segments, the Default Predictor is very competitive. The Default Predictor compares very well with the literature: see the performances of precision of the overview (chapter 2).

| Precision of baseline predictors on netflix - with segmentation | | | | |
|---|---|---|---|---|
| Predictor | prec Huser Pitem | prec Luser Pitem | prec Huser Uitem | prec Luser Uitem |
| Random Pred | 0.5374 | 0.5467 | 0.4747 | 0.4868 |
| Default Pred | 0.9300 | 0.9247 | 0.9299 | 0.9249 |

**Table 4-15: Precision of baseline predictors on Netflix - with segmentation**

**KNNs' precision performances**

On Netflix the precision increases as the number of KNN increases. But the results are not significantly better than that of the Default predictor.

| Precision of a KNN model on Netflix | | |
|---|---|---|
| Nb of KNN | Precision | Stand. Dev. |
| 10 | 0.8675 | 0.000198 |
| 50 | 0.8983 | 0.001407 |
| 100 | 0.9167 | 0.002260 |
| **200** | **0.9339** | 0.001970 |

**Table 4-16: Precision of a KNN model on Netflix**

On Figure 4-10 we see that the precision is better than the default predictor for only 2 segments and only for at least K=200. Under K=100, it seems better to use a default predictor than a KNN predictor for ranking tasks.

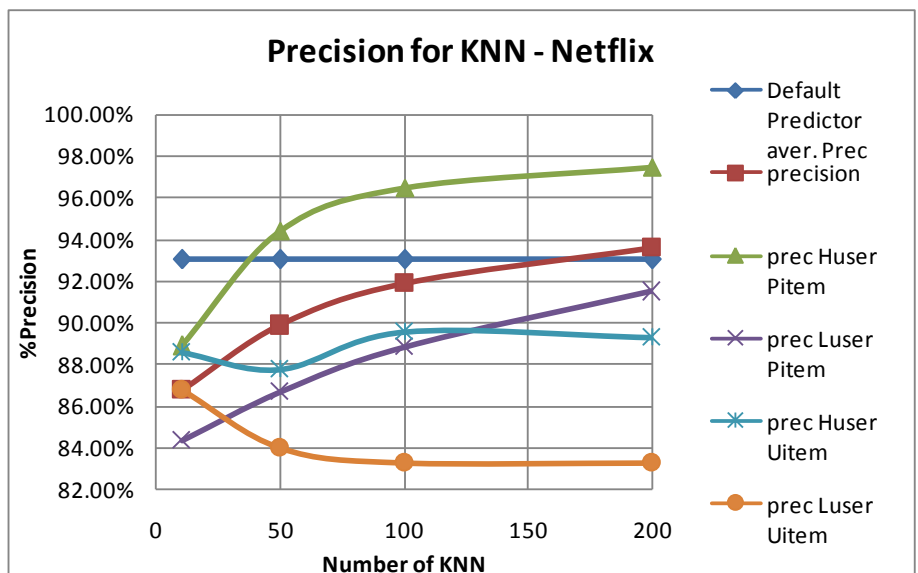

**Figure 4-10: Precision for KNN - Netflix.**

Nevertheless the heavy-user-popular-item segment is well modelled: the precision for 10 generated items for the KNN model is greater than 97% for the model with 200 neighborhood for the heavy-user-popular-item segment.

**Gravity's precision performances**

On Netflix Gravity shows a better behavior than the KNN model, especially for the light-user-unpopular-item segment as shown in **Figure 4-11**.

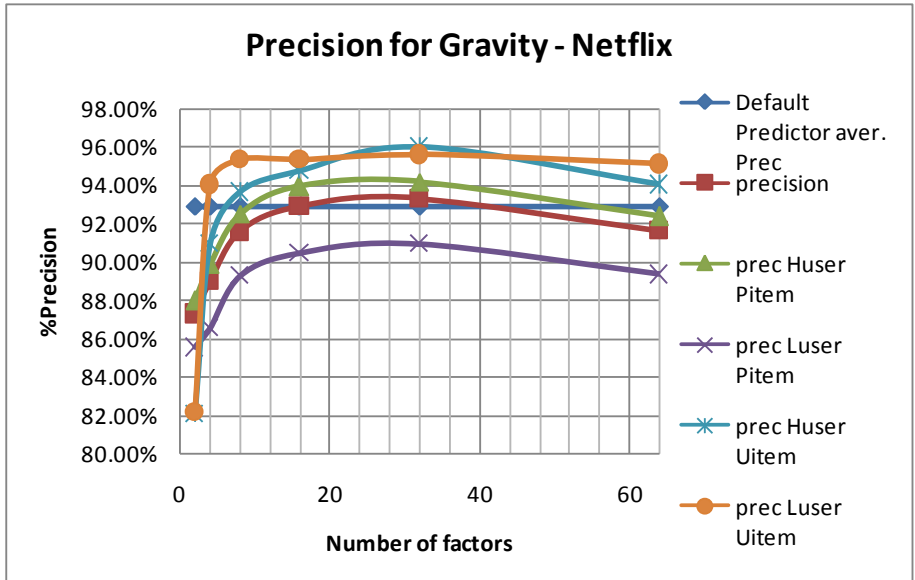

**Figure 4-11: Precision for Gravity - Netflix**

The global precision of Gravity on MovieLens increases slowly when the number of factors increases, varying from 87.3% to 91.7%, which is the same order as the performances of the KNN. The impacts of the recommendation are on the average lower than those of the KNN models.

**Analysis concerning the Precision**

The global precision of Gravity is slightly better than that of KNN but Gravity improves especially the light-user-unpopular-item segment. KNN is very efficient for the heavy-user-popular-item segment.

| Best Precision for a Top 10 recommender, per segment | | |
|---|---|---|
| | Popular items | Unpopular items |
| Heavy users | KNN | Gravity |
| Light users | Default Predictor | Gravity |

**Table 4-17: Global analysis for precision: best algorithm for each segment**

Globally, the Default Predictor is a very competitive ranking method for generating Top-N recommendations, and should be used for every new user / light user.

The precision measure, used with a global catalog scoring scheme, and a Top 10, is questionable to see the added value of a recommender system: the most surely enjoyed items are not necessary the most useful ones as no information of the rarity of the items are taken into account. The very good performance of the Default Predictor is clearly an indication of the limitation of the precision measure: The Default Predictor is quite equivalent to the "most popular items" or "best rated items" panel widely used on e-commerce websites: every user has experienced that this information is sometimes useful but not always relevant and certainly not adapted to personalized discovery. Moreover, a Top 10 recommending scheme it is just usable for a first round, except for fast changing catalog: other items, riskier to recommend, will have to be pushed after.

**Analysis using the Average Measure of Impact**

On Netflix, the AMI gives slight negative performances of the Random Predictor and a small performance to the Default predictor: The Default Predictor "wins" its impact values on well known high popular items.

| Baseline Predictors | Average 1/(items' count) | Stand. Dev. | Average Measure of Impact |
|---|---|---|---|
| Random Pred | -0.000035 | 1.1269E-06 | **-0.6219** |
| Default Pred | 0.000028 | 4.7525E-07 | **0.4976** |

**Table 4-18: Average Measure of Impact of baseline predictor on Netflix**

The KNN model behaves significantly better that the Default Predictor for the AMI.

| Nb of KNN | Average 1/(items' count) | Stand. Dev. | Average Measure of Impact |
|---|---|---|---|
| 10 | 0.000086 | 0.000000 | **1.5282** |
| 50 | 0.000108 | 0.000001 | **1.9191** |
| **100** | **0.000115** | **0.000037** | **2.1146** |
| 200 | 0.000112 | 0.000012 | **1.9724** |

**Table 4-19: Average Measure of Impact of KNN on Netflix**

For Gravity, the behavior is much worse than that a KNN model. In general, the impact of Gravity is similar to or lower than that of the Default Predictor.

| Factors | Average 1/(items' count) | Stand. Dev. | Average Measure of Impact. |
|---|---|---|---|
| 2 | 0.000017 | < 0.000001 | **0.3020** |
| 4 | 0.000019 | < 0.000001 | **0.3376** |
| 8 | 0.000025 | < 0.000001 | **0.4442** |
| 16 | 0.000031 | < 0.000001 | **0.5508** |
| **32** | **0.000033** | < 0.000001 | **0.5864** |
| 64 | 0.000022 | < 0.000001 | **0.3909** |

**Table 4-20: Average Measure of Impact of Gravity on Netflix**

For MovieLens the results are similar but not with the same magnitude: the Random Predictor gives a strong negative AMI of -4.7, the Default Predictor gives an AMI of about 2.7, Gravity is around 3.8 and the KNN model gives 6.85. The order of the algorithm's performances are identical to those on Netflix but the AMI are amplified. We believe that the relative small size of the dataset and the fact that each user has at least 20 ratings makes MovieLens easier for the task of recommending rare items.

If we now select the best Average Measure of Impact for each predictor, we obtain the next figure.

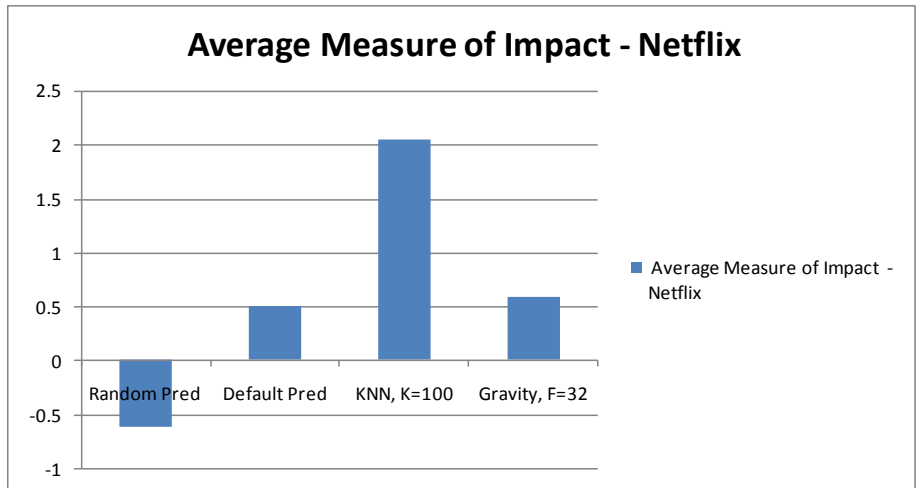

**Figure 4-12: Best Average Measure of Impact on Netflix**

On Netflix an analysis according to the segmentation gives a more detailed view of where the impacts are. Note that the support for the different evaluated segments are very different and the weights of the two popular item segments are significantly higher.

|  | Huser Pitem | Luser Pitem | Huser Uitem | Luser Uitem |
|---|---|---|---|---|
| Gravity F=32 | 0.38 | 0.26 | 8.93 | 10.61 |
| KNN K=100 | **0.71** | **0.43** | 9.59 | 8.84 |
| Default Pred | 0.29 | 0.25 | **21.22** | **12.31** |
| Random Pred | 0.00 | 0.03 | -5.13 | -0.53 |
| Best algorithm | KNN | KNN | Default Predictor | Default Predictor |

**Table 4-21: AMI according to each segment**

**Analysis concerning the Average Measure of Impact**

The results are consistent across both datasets. For the recommending task, the KNN models have better Average Measure of Impact than the Gravity models. The Random Model of recommendation has a negative Average Measure of Impact. The Default predictor has a AMI performance close to that of Gravity. Gravity, taking into account more global aspect of the data, has a behavior close to the Default Predictor.

| Best AMI for a Top 10 recommender, per segment | | |
|---|---|---|
|  | Popular items | Unpopular items |
| Heavy users | KNN | Default Predictor |
| Light users | KNN | Default Predictor |

**Table 4-22: Best models for the AMI depending on the segments**

We still must be careful with these results directly computed on simple algorithms and with a simple global catalog scoring scheme. The results just show the global natural bias of each algorithm. It is very easy to use strategies such as post-filtering techniques based on the popularity of the items to change the behavior of the recommender system, then to change the average impact of the recommender.

Nevertheless we believe that the Average Measure of Impact is an useful measure to evaluate how useful a recommender system is in the task of helping to Discover interesting items. The better an algorithm can fit the user's profile, the rarer and the riskier items it can choose. The rarest and the riskiest items are also, when relevant, the most interesting ones in term of user's satisfaction.

### 4.3.4 Summary of the test for scoring-based functions

We analyzed 4 models, a KNN model, a Gravity model, a Random model, and a Default Predictor model on 3 tasks adapted to a scorer-based recommenders: Decide, Compare, Discover and on 4 user-item segments: heavy-user-popular-item, heavy-user-unpopular-item, light-user-popular-item and light-user-unpopular item. The summary of the results are given in the Table 4-23.

If we analyze the results by segments, we see that globally, KNN is well adapted for the heavy-user segments, and Gravity, and the Default Predictor, are well adapted to light-user segments.

Globally, for the tasks "Help to Decide" and "Help to Compare", Gravity is the best-suited algorithm of our tests. For the tasks "Help to Discover" KNN is more appropriate.

A switch-based hybrid recommender, based on the items' and users' segmentation could exploit this information to improve the global performances of the system.

|  | **Heavy Users Popular items** | **Heavy Users Unpopular items** | **Light Users Popular Items** | **Light Users Unpopular Items** |
|---|---|---|---|---|
| **Decide** <br><br> RMSE | KNN | Gravity | Gravity | Gravity |
| **Compare** <br><br> % Compatible preferences | KNN | KNN | Gravity | Gravity |
| **Discover** <br><br> Precision | KNN | Gravity | Default Predictor | Gravity |
| **Discover** <br><br> Average Measure of Impact | KNN | Default Predictor | KNN | Default Predictor |

**Table 4-23: Best models depending on the tasks and the segments**

## 4.4 Results for the similarity based function

### 4.4.1 Help to Explore: item-to-item

We use the protocol we defined before in this chapter: a good similarity matrix for the task "Help to Explore" is a similarity matrix leading to global good performances, when used in a KNN model. We choose a similarity matrix with 100 neighbors for each item: this is largely enough for item-to-item tasks where generally a page displays 10 to 20 similar items.

For this kind of model, a native-data-based KNN model with K=100 and a Gravity-Emulated KNN model with 16 factors give the following results:

|  | Native KNN<br>K=100<br>Similarity WPearson | KNN computed on Gravity's factors<br>K=100, number of factors=16<br>Similarity Pearson |
|---|---|---|
| RMSE | 0.8440 | 0.8691 |
| Ranking: % compatible | 77.03% | 75.67% |
| Precision | 91.90% | 86.39% |
| AMI | 2.043 | 2.025 |
| (Global time of the modeling task) | (5290 seconds) | (3758 seconds) |

**Figure 4-13: Quality of a item-item similarity matrix according to 4 measures: results on Netflix**

This results show that for all performance measures the similarity matrix obtained with Gravity are slightly worse than that obtained on native data. We tried different similarity measures for the gravity-based representation: cosine, normalized-variants of the cosine, inverted Euclidian Distance, with no better results. Increasing the number of factors of Gravity slightly increases the performances, but those of the native KNN are still better.

The analysis by item's segment and user's segment is complex and in our opinion has no real meaning here as it would have to be done for each segment with a 4-dimensional indicator: RMSE, Percentage of compatible ranking, Precision and AMI. For the items, it is likely that popular items will globally be better modeled than unpopular items as it is always the case for both KNN and Gravity. Note that for the users, the segmentation is pointless as the item-to-item recommendation is dedicated to anonymous users;

The global results indicate that Gravity is slightly less effective for the item-to-item task ("Help to Explore").

We built also a random KNN similarity matrix to have a baseline comparison model. This is not the same model as a pure random predictor which only returns a random rating. In the case of a random KNN similarity matrix, applying the mean-based scoring formula, on the average the predicted rating can be close to the items' mean plus the user's bias. The results of the random KNN model are shown in the next section dedicated to the discussion of the Gravity-emulated model. It is always worse than the other models.

As the Default Predictor model based on items' means and users' means cannot by itself produce a similarity matrix, it is disqualified for this task.

We conclude in final that the KNN model is the best model for the "Help to Explore" task.

### 4.4.2 Using Gravity as emulated mode for KNN modeling

Another approach of using a Gravity model to emulate the native data is to deal with huge catalogs. When the number of items (or any other objects) to compare increases, the KNN's quadratic complexity becomes a problem.

The previous results show that a Gravity-Emulated KNN always performs slightly worse than a native KNN model, but the trade-off between scalability and performances seems good. We present the results for the KNN models with K=100, comparing KNN computed on Gravity's factors, native KNN and a Random KNN used as baseline. Note that the Random KNN model performs better than a Random Uniform model. The former uses a random similarity matrix in conjunction with the mean based scoring scheme and default predictor cascading scheme of the KNN model we use in our Reperio's framework (see chapter 3). The later simply returns a random number between 1 and 5.

For the RMSE, the Gravity-Emulated KNN model looses 0.025 point going from 0.844 to 0.870. Compared with other models it still performs correctly.

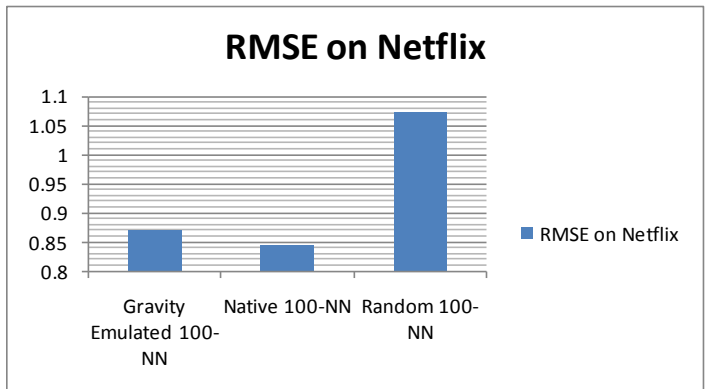

Figure 4-14: RMSE for native-KNN, Gravity-Emulated KNN and Random KNN

For the global ranking the difference between the Gravity-Emulated model and the native KNN model is still low, whereas a random KNN model perform very badly.

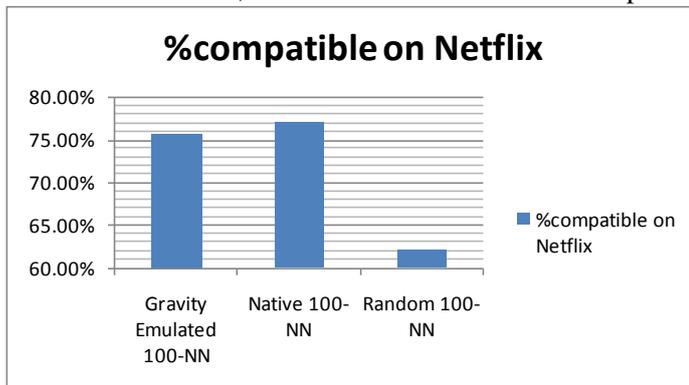

Figure 4-15: Global ranking for native-KNN, Gravity-Emulated KNN and Random KNN

For the precision for a Top-10 ranking the Gravity-Emulated KNN model performs significantly worse than a native KNN model.

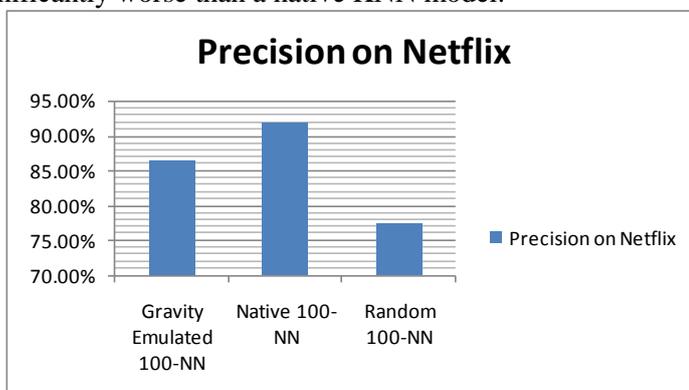

Figure 4-16: Top-10 ranking for native-KNN, Gravity-Emulated KNN and Random KNN

For the Average Measure of Impact, the Gravity-emulated KNN model and the native KNN model performs almost identically.

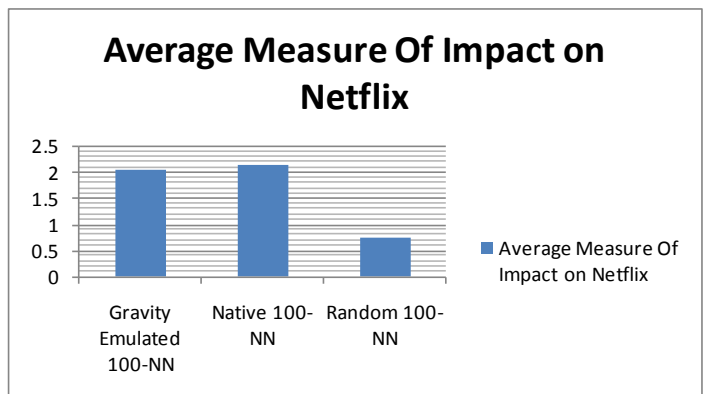

**Figure 4-17: Impact of recommendations for native-KNN, Gravity-Emulated KNN and Random KNN**

We can conclude that Gravity can be used as a dimension reduction algorithm to perform faster KNN search. A more powerful process should use:
- a dimension reduction carried out by gravity, then a parallelization of the KNN search as already implemented in Reperio.
- or a dimension reduction carried out by Gravity, then a clustering technique to group the items, then a KNN search restricted within each cluster[10].

## 4.5 Conclusion

We have presented a new protocol to evaluate precisely the quality of recommender algorithms on the core function tasks we have defined in Chapter 1. This protocol can be used to analyze algorithms or complete recommender systems. We have experimented this protocol on two state-of-the art algorithms, Item-Item KNN and Gravity, and two baseline algorithms. We have demonstrated the utility of this protocol as it may change:
- the classical vision of recommendation evaluation, still often focused on RMSE,
- the way to improve the recommender systems to achieve their tasks.

### 4.5.1 What correlation between RMSE and recommendations' quality?

Where is the widely implicitly assumed correlation between RMSE and Precision? If we compare for instance the global RMSE of the Default Predictor (a simple sum of means) and the global RMSE of Gravity or a KNN model, we would conclude that the Default Predictor will not perform well in term of precision, compared to Gravity or KNN. It is the contrary that we observe on the Netflix data.

Where is the widely implicitly assumed correlation between RMSE and the quality of the recommendation? If our assumption that the Average Measure of Impact we defined in this chapter captures well the added value of the Help to Discover task, then Gravity would show an AMI significantly greater than that of the Default Predictor. We do not observe that on the Netflix data. For the AMI, there is no significant improvement of Gravity compared with the Default Predictor, whereas Gravity is a state-of-the-art algorithm for RMSE optimization.

---

- [10] But in this case we would certainly face another performance decreasing as another approximation of the neighborhood would be done.

There is no evidence of strong correlation between RMSE improvement and recommendations' quality improvement.

The reason might be simple: remember that today the best algorithms still have a RMSE of about 0.84 for datasets with rating scale between 1 on 5: the average absolute error is still about 21%, and improving the error rate of 10% leads to an absolute improvement of only 2%: the improvement in accuracy may be restricted on heavy users popular items without any positive effect on global ranking or quality of the recommendation.

If there is no correlation between RMSE and Precision and no correlation between RMSE and the impact of the recommendation, evaluating recommender systems only on RMSE is questionable. However many articles dealing with recommender systems papers use only the RMSE to evaluate their algorithms.

RMSE is not a wrong measure, but it is only one indicator on one recommendation task: Help to Decide. This task, Help to Decide, is not the crucial one if we remember the overview of our state of the art. The tasks Help to Compare, Help to Discover and Help to Explore are more important in the industry.

As no evidence of strong correlations between RMSE and other indicators for the other recommender tasks can be found, RMSE (or MAE) cannot be used alone to prove the efficiency of a recommender system.

## 4.5.2 When should we personalize the recommendations?

Can we use a recommender system to drive the demand down the long tail? Actually, yes, but a switch-based, segment-aware hybrid recommender would be preferable for this purpose. For instance, it seems illusory to make personalized recommendation of unpopular items to light users as personalized algorithms do not perform well on this segment.

It is assumed that recommender systems have to improve usages on weak segments, especially usages of light users on the long tail. In fact, personalized algorithms seem to be efficient especially with heavy users and popular items. Remember that with this segment a KNN model can reach a precision greater that 97% for 10 recommended items. At the opposite, when the information about users and items is missing, very simple predictors are as efficient as powerful algorithms.

We showed that a Default Predictor mainly based on the robust mean of the items' rating achieves more than 92% of precision on all segment, so on the light-user-unpopular-item segment. This clearly means that selecting the top rated of the unpopular items (not often rated) is a strategy good enough to make recommendation to light users.

Golden nuggets may be items not often rated but with a good ratings' average. With our definition of the recommendation's impact, our Default Predictor performs well, gathering a lot of positive impacts (AMI) on the unpopular items segments.

But pushing an almost static list Top N items of an almost static catalog is just worth once. Otherwise users may switch off. Many recommendations are done during a selection in a catalog, or given a short list. In these cases, where a Top-N ranking is insufficient and global ranking is mandatory, we saw that the personalized algorithms are significantly better than the Default Predictor . For the "Help to Compare" task, the Default Predictor has only a 68% of predictive preference accuracy whereas KNN can reach 77% and more.

### 4.5.3  Which algorithm use in a recommender engine?

We saw that today's two mainstream algorithms, Gravity and KNN are in fact complementary. No algorithm dominates another one for all the recommendation tasks and for all the segments, and in some cases a Default Predictor based on the means of items' rating can be the good strategy.

KNN seems better to model heavy user and popular item segments. A KNN or similarity-based approach is mandatory for the item-to-item recommendations. Native KNN is better to the task "Help to explore" as it produces better item-item similarity matrices.

Gravity seems better to model heavy user unpopular segment and light user popular item segment. Gravity is better to the global ranking task, the "Help to Compare".

Finally, Gravity could be used to emulate a KNN search, without significantly losing performances. This could be very useful for fast KNN search process in a low dimensional space.

# Conclusion

*"Divide each difficulty into as many parts as is feasible and necessary to resolve it."*

René Descartes.

## Main results of this work

We have proposed a new approach to analyze the performances and the added value of automatic Recommender Systems in an industrial context.

First, we have defined 4 core functions for these systems, which are:

- Help users to Decide
- Help users to Compare
- Help users to Discover
- Help users to Explore

We have proposed some vocabulary normalization and a possible new classification of the automatic recommender systems. We have also reviewed the main prerequisites of the recommender systems. We showed that K-Nearest-Neigbors methods (KNN) and more precisely item-item similarity-based engines can implement the four aforementioned core functions and are compliant with the prerequisites.

We have built an operational hybrid recommender engine, called Reperio, based on KNN approach and studied its accuracy performances, with the classical RMSE measure, and its speed performances. We studied the effect of the similarity measures on KNN techniques and show that this effect is quite strong, although often neglected in the literature.

We carried out several experiments to compare the performances of KNN methods with thematic, collaborative, and hybrid approaches. We showed that thematic filtering and hybrid filtering can be useful for critical cases such as cold start with short users' profiles. We also showed that beyond a critical mass of information, collaborative filtering performs better than thematic filtering and hybrid filtering.

Finally, using our conclusions of the first chapter, we proposed a general off-line protocol to evaluate a recommender system taking into account industrial and marketing needs: we crossed our 4 core functions with 4 users×items segments:

- Heavy-users and popular items
- Heavy users and unpopular items
- Light users and popular items
- Light users and unpopular items

and we compared two major state of the art methods, KNN and Gravity, with 2 baselines methods used as reference. We showed that the two major methods are complementary as they perform differently across the different segments. We showed that we could use the two algorithm together, for instance to produce a KNN matrix based on the dimension reduction done by Gravity, with still a correct quality. Gravity could be used to implement a similarity function between items to support the "Help to Explore" function, and Gravity could be used as a component for fast KNN search.

We proposed a new measure, the Average Measure of Impact, to deal with the usefulness and the trust of the recommendations. Using the precision measure, and the AMI, we showed that there is no clear evidence of correlation between the RMSE and the quality of the recommendation.

We have demonstrated the utility of our protocol as it may change
- the classical vision of the recommendation evaluation, often focused on the RMSE/MAE measures as they are assumed correlated with the system overall performances,
- and the way to improve the recommender systems to achieve their tasks.

When designing a recommender engine 's general recommendation strategy, we have to think about the impact of the recommender: recommending popular items to heavy users might be not so useful. On the other hand, it can be illusory to make personalized recommendations of unpopular and unknown items to light and unknown users.

A possible simple strategy could be:
- Rely on robust Default Predictors, for instance based on robust item's means to try to push unknown golden nuggets to unknown users.
- Use personalized algorithms to recommend popular items to light users.
- Finally use personalized algorithms to recommend unpopular items of the long tails for heavy "connoisseurs" users.

# Future works

An interesting work would be to deepen the notion of recommendations' impact and to cross the off-line performance measures with users' interviews, and A/B testing protocols.

In Chapter 4, the results about algorithms' impact performances are relative to the basic strategy we used to generate the recommendations: global scoring of the entire catalog.

The study and comparison of more elaborate recommendation strategies to optimize the impact of the recommendation should be carried out. Many improvements could certainly be found with well-adapted strategies for users' and each items' segments. The segmentation itself could be improved. Time-aware segmentation, with short term and long term profiling should be investigated both for items, and for users.

As we show that using Gravity's dimension reduction to perform a KNN search does not decrease too much the quality of a similarity matrix.

We are currently implementing a Fast Matrix Factorization model to pre-process huge databases before computing similarities in our Reperio Framework. This mode will be an intermediate mode between the centralized native mode of Reperio-C and the Embedded mode of Reperio-E, running on Android$^{TM}$ devices.

The typical use cases of the different working modes of Reperio are summarized in the following table. See (Meyer and Fessant, 2011) for details about this approach. This is still a work in progress.

The first complete method we could test is the process described in Figure C-1.

|  | **Centralized native mode** | **Centralized virtual mode** | **Embedded mode** |
|---|---|---|---|
| Typical example of item dataset | Video On Demand catalog | Cultural products on e-commerce website | Electronic Program Guides (TV programs) |
| Size of item dataset | Large | Very large | Small |
| Catalog's update frequency | Low | Low | High |
| Typical adapted filtering mode | Collaborative, multi user scoring mode | Collaborative, multi user scoring mode | Content-based, mono user scoring mode |
| Size of items' description (metadata or ratings) | Large, several hundreds | Very large, several thousands | Small, several tens |
| Reperio's similarity computation mode | Pre-computed into a similarity matrix, on native data | Online, based on a matrix factor decomposition | Online, on native data |
| Pre-computed model used | Pre-computed similarity matrix | Matrix factor decomposition | No pre-computed model |
| Adapted Similarity used | Extended Pearson | Pearson | Jaccard using attribute weightings |
| Typical Item-to-item function implementation | Exact K nearest neighbors extract from the similarity matrix | Approximate K nearest neighbors by random sampling | Restricted to items explicitly selected as interesting, or to a time-slice. |

**Table C-0-1: The near-future of Reperio: 3 possible working modes - illustration**

Another line of work will be the switch-mode, segment-aware hybrid recommender systems using the best algorithm for the adapted task and the adapted segment. It is easy to see the possible improvement brought by such systems as each global performance measure would be the sum of the best measures for each segment.

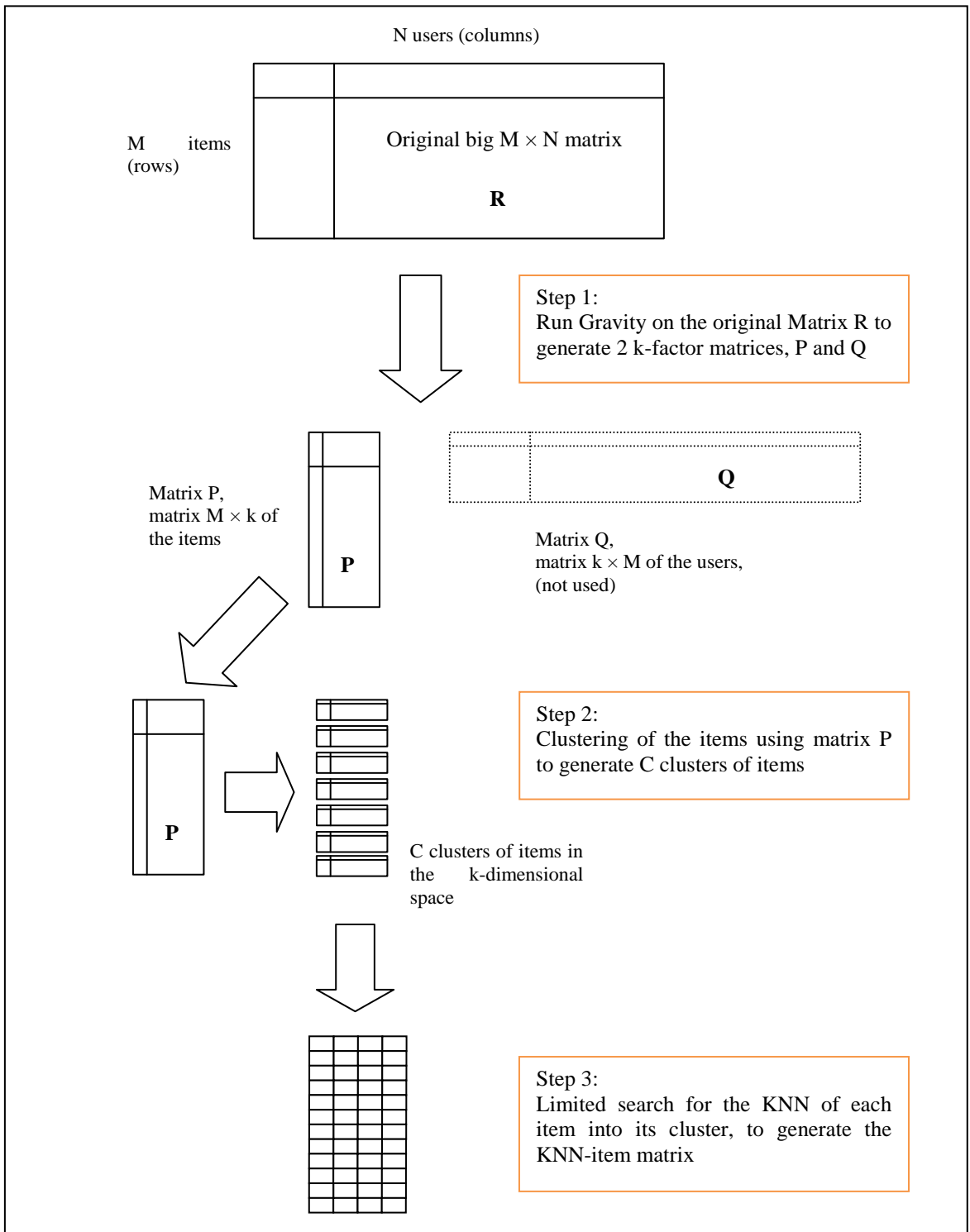

**Figure C-1 A possible solution for fast KNN item-item matrix generation**

Page 162

**Résumé**

Cette thèse traite des systèmes de recommandation automatiques. Les moteurs de recommandation automatique sont des systèmes qui permettent, par des techniques de data mining, de recommander automatiquement à des clients, en fonction de leurs consommations passées, des produits susceptibles de les intéresser. Ces systèmes permettent par exemple d'augmenter les ventes sur des sites web marchands : le site Amazon a une stratégie marketing en grande partie basée sur la recommandation automatique. Amazon a popularisé l'usage de la recommandation automatique par la célèbre fonction de recommandation que nous qualifions d'item-to-items, le fameux : " les personnes qui ont vu/acheté cet articles ont aussi vu/acheté ces articles.

La contribution centrale de cette thèse est d'analyser les systèmes de recommandation automatiques dans le contexte industriel, incluant les besoins marketing, et de croiser cette analyse avec les travaux académiques. Cette thèse comporte 4 parties :

- Une analyse des fonctions cœurs et des pré-requis des systèmes de recommandation dans un contexte industriel: nous identifions 4 fonctions cœur dans les systèmes de recommandation : Aide à la Décision, Aide à la Comparaison, Aide à l'Exploration, Aide à la Découverte. L'implémentation de ces fonctions a des implications dans les choix algorithmiques au cœur des systèmes de recommandations.

- Un état de l'art, qui présente les principales techniques utilisées dans les systèmes de recommandation automatique: les deux méthodes algorithmiques les plus utilisées, les méthodes à K-plus-proches-voisins et les méthodes de factorisation rapide de matrices sont détaillées. L'état de l'art présente aussi les méthodes purement thématiques, les techniques d'hybridation, et les mesures de performance classiques pour évaluer les systèmes. Cet état de l'art donne ensuite un panorama de plusieurs systèmes de recommandation, du monde académique, ou des acteurs industriels connus ( Amazon, Google...).

- Une analyse des performances et des implications d'un système de recommandation industriel développé au cours de cette thèse: ce système, Reperio, est un moteur hybride utilisant une technique de K-Plus-Proches Voisins (KPPV). Nous étudions les performances des méthodes KPPV, notamment l'impact des fonctions de similarités utilisées. Puis nous étudions les performances de Reperio dans le cas critique du démarrage à froid.

- Une méthodologie d'analyse des performances des systèmes de recommandation en contexte industriel : cette méthodologie permet d'évaluer la plus-value des méthodes algorithmiques ou des stratégies de recommandation sur l'ensemble des fonctions cœurs. Pour cela nous reprenons les 4 fonctions que nous avons définies et nous les croisons avec 4 segments clés de l'analyse des performances du système de recommandation : gros clients et items fréquents, gros clients et items peu fréquents, petits clients et items fréquents, petits clients et items peu fréquents. Nous montrons que les systèmes de recommandation devraient redéfinir leurs enjeux : il est illusoire de recommander de manière personnalisée des items peu populaires à des utilisateurs peu connus par exemple. Pour augmenter les usages, les stratégies efficaces seraient plutôt de recommander des items peu fréquents à des gros utilisateurs, et de recommander des items fréquents aux petits utilisateurs. Le paradigme de la "Long Tail", pour les items peu fréquents et pour les petits utilisateurs, devrait être revu.

# Abstract

This thesis deals with automatic recommendation systems. Automatic recommendation systems are systems that allow, through data mining techniques, to recommend automatically to users, based on their past consumption, items that may interest them. These systems allow for example to increase sales on e-commerce websites: the Amazon site has a marketing strategy based mainly on the recommendation. Amazon has popularized the use of automatic recommendation based on the recommendation function that we call item-to-items, the famous "people who have seen / bought this product have also seen / bought these articles".

The central contribution of this thesis is to analyze the automatic recommendation systems in the industrial context, including marketing needs, and to cross this analysis with academic works. This thesis consists of four parts:

- An analysis of the core functions and the prerequisites for recommender systems in an industrial context: we identify four core functions for recommendation systems: Help do Decide, Help to Compare, Help to Explore, Help to Discover. The implementation of these functions has implications for the choices at the heart of algorithmic recommender systems.

- A state of the art, which deals with the main techniques used in automated recommendation system: the two most commonly used algorithmic methods, the K-Nearest-Neighbor methods (KNN) and the fast factorization methods are detailed. The state of the art presents also purely content-based methods, hybridization techniques, and the classical performance metrics used to evaluate the recommender systems. This state of the art then gives an overview of several systems, both from academia and industry (Amazon, Google ...).

- An analysis of the performances and implications of a recommendation system developed during this thesis: this system, Reperio, is a hybrid recommender engine using KNN methods. We study the performance of the KNN methods, including the impact of similarity functions used. Then we study the performance of the KNN method in critical uses cases in cold start situation.

- A methodology for analyzing the performance of recommender systems in industrial context: this methodology assesses the added value of algorithmic strategies and recommendation systems according to its core functions. For this we take the four functions we have defined and we cross with four key segments of the performance analysis of the recommendation system: heavy users and popular items, heavy users and unpopular items, light users and popular items, light users and unpopular items. We show that the recommendation systems should redefine their challenges: it is unrealistic to recommend a personalized unpopular items to users who are not known, for example. To increase usage, effective strategies are rather: always recommend unpopular items only to heavy users, always recommend to light users only the popular items. The long tail paradigm, both for infrequent items and infrequent users, should be revisited.